# Application of Nonlinear Autoregressive with Exogenous Input (NARX) neural network in macroeconomic forecasting, national goal setting and global competitiveness assessment

**Liyang Tang**[1, *]

[1]Research Bureau, People's Bank of China, Beijing 100800, China; tliyang@pbc.gov.cn

[*]Correspondence: tliyang@pbc.gov.cn; Tel.: +86-10-13581917180

**Abstract:**

As the trove of economic big data released by statistical agencies, private and public surveys, and other sources every day becomes continuously available in real time, making high-quality economic forecasts requires tracking a large and complex set of data, however, the limited deductive and inductive abilities of the human brain or the inadequacies of current econometric approaches and other modeling methods greatly limit the above full use of the trove of economic big data. But with the rapid development of neural network methods and their cross-field applications in recent years, certain types of neural networks such as the Nonlinear Autoregressive with Exogenous Input (NARX) neural network may be considered as a general method of making full use of the trove of economic data to make time series predictions. This paper selects the NARX neural network as the method of this study through literature review, and constructs specific NARX neural networks under specific application scenarios involving macroeconomic forecasting, national goal setting and global competitiveness assessment, where after this study focuses on analyzing how different settings for exogenous inputs from the trove of economic big data affect the prediction performance of NARX neural networks. Next, through case studies on China, US and Eurozone, this study wants to explore and summarize how those limited & partial exogenous inputs or abundant & comprehensive exogenous inputs, a small set of most relevant exogenous inputs or a large set of exogenous inputs covering all major aspects of the macro economy, whole area related exogenous inputs or both whole area and subdivision area related exogenous inputs specifically affect the forecasting performance of NARX neural networks for specific macroeconomic indicators or indices. And next, through the case study on Russia this paper similarly explores how the limited & most relevant exogenous inputs set or the abundant & comprehensive exogenous inputs set specifically influences the fitting performance and prediction performance of those specific NARX neural networks for national goal setting. Finally, comparative studies on the application of NARX neural networks for the forecasts of Global Competitiveness Indices (GCIs) of various economies are conducted, in order to explore whether the specific NARX neural network trained on the basis of the GCI related data of some economies can make sufficiently accurate predictions about GCIs of other economies, and whether the specific NARX neural network trained on the basis of the data of some type of economies can give more accurate predictions about GCIs of the same type of economies than those of different type of economies. Based on all of the above successful application, this paper provides policy recommendations on applying fully trained NARX neural networks that are assessed as qualified to assist or even replace the deductive and inductive abilities of the human brain in a variety of appropriate tasks.

**Keywords:** Nonlinear Autoregressive with Exogenous Input (NARX) neural network; macroeconomic forecasting; goal setting; competitiveness assessment; leading business and economic indicators and indices

# 1. Introduction

The trove of economic big data released by statistical agencies, private and public surveys, and other sources is parsed every day by not only governments and authorities, but also economists and market analysts. The governments and authorities usually parse the trove of economic big data in order to assist decision-making, and making forecasts on the basis of the trove of economic big data is also essential to governments and authorities in informing their policy decisions and communicating their economic outlook to the public. Economists usually parse the trove of economic big data to assess the health of the economy and serve specific research topics. Market analysts usually strive to parse the most valuable part of the trove of economic big data, in order to understand where the economy currently is and to forecast in which direction it is going, which can provide direct and indirect market value.

However, the full use of the trove of economic big data is not that easy. For example, although monitoring a large number of different economic indicators and indices to detect early signals, which not only enables one to exploit different sampling frequencies and different timing of macroeconomic data releases, but also mitigates the risk of overweighting idiosyncratic fluctuations as well as measurement errors, can enhance timeliness and accuracy in assessing the health of the economy, separating meaningful signals from noise for each indicator or index and integrating the extracted information (sometimes even contradictory) into a single conclusion are always difficult tasks with no fixed solution (Bok et al., 2017). Take another example, as the trove of economic big data becomes continuously available in real time, making high-quality economic forecasts to help understand where the economy is now requires tracking a large and complex set of data, since information about different aspects and sectors of the economy can be considered as imperfect measures of latent economic situations, the more systematic and comprehensive the information, the higher the quality of prediction, however, the limited deductive and inductive abilities of the human brain or the inadequacies of current econometric methods and modelling methods (such as can't provide economic forecasts at high frequency) greatly limit the above full use of the trove of economic big data, then forecasters could either only track a small number of key and comprehensive indicators of economic activity to make predictions, or use complicated mathematical models to form projections and apply subjective adjustments to pure model-generated forecasts, or use a combination of a suite of models and a fair amount of expert judgment to generate forecasts, all of the above workarounds are just stopgaps and problem-specific solutions, and cannot be regarded as general methods of making full use of the trove of economic big data to make predictions (Ghoddusi et al., 2019; Ozbayoglu et al., 2020).

With the development of a large number of machine learning methods and their cross-field applications, representative machine learning methods such as neural network methods have been well equipped to exploit the trove of economic big data more effectively and efficiently than traditional econometric or modeling methods, in other words, certain types of neural networks may be considered as a general method of making full use of the trove of economic big data to make predictions (Bok et al., 2017; Storm et al., 2019). Compared with current econometric approaches and other modeling methods, neural network methods that focus on making predictions have the following advantages: first, neural networks are highly flexible and may be helpful in settings where other flexible models have computational problems due to the size of the dataset or the number of economic variables that need to be considered; second, neural network approaches can automatically extract the most relevant features for a prediction task, and are potentially capable of deriving more complex features from the raw data that are missed by other approaches; third, neural network methods can be useful in addressing problems with large number of explanatory variables, and they are especially crucial and almost irreplaceable when the number of explanatory variables exceeds the number of observations in large datasets; fourth, the predictive ability of neural networks in complex and high-dimensional settings can be used to improve causal estimates under some application scenarios (Storm et al., 2019).

Since neural network methods that are specifically developed for making predictions have already demonstrated great potential in improving prediction, then **the idea of this study is to first determine both the specific type of neural network and the specific application scenarios that have important economic significance but have not been systematically and deeply studied, and then apply this particular type of neural network to these specific economic scenarios, in order to focus on analyzing how different settings for exogenous inputs from the trove of economic big data affect the prediction performance of neural networks.**

The remainder of this paper is structured as follows. The literature review in the next section focuses on the systematic and comprehensive search for application of neural networks in various subfields of economics and finance from the most representative academic and policy studies, which are used to provide inspiration to either the determination of specific application scenarios or the selection of specific neural network methods in our study. Section 3 selects and introduces the Nonlinear Autoregressive with Exogenous Input (NARX) neural network, and constructs specific NARX neural networks under various specific application scenarios including macroeconomic forecasting, national goal setting and global competitiveness assessment, so that the subsequent studies can focus on analyzing how different settings for exogenous inputs affect the prediction performance of NARX neural networks. Section 4 carries out the application of NARX neural networks for macroeconomic forecasting in the form of case studies on China, US and Eurozone, in order to explore and summarize how those limited & partial exogenous inputs or abundant & comprehensive exogenous inputs, a small set of most relevant exogenous inputs or a large set of exogenous inputs covering all major aspects of the macro economy, whole area related exogenous inputs or both whole area and subdivision area related exogenous inputs specifically affect the forecasting performance of NARX neural networks for specific macroeconomic indicators or indices. Section 5 similarly explores how the limited & most relevant exogenous inputs set or the abundant & comprehensive exogenous inputs set specifically influences the fitting performance and prediction performance of those specific NARX neural networks for national goal setting, through the case study on Russia. Section 6 conducts comparative studies on the application of NARX neural networks for the forecasts of Global Competitiveness Indices (GCIs) of various economies, in order to simply and intuitively explore whether the specific NARX neural network trained on the basis of the GCI related data of some economies can make sufficiently accurate predictions about GCIs of other economies, and whether the specific NARX neural network trained on the basis of the data of some type of economies can give more accurate predictions about GCIs of the same type of economies than those of different type of economies. Section 7 offers policy recommendations on applying fully trained NARX neural networks that are assessed as qualified to assist or even replace the deductive and inductive abilities of the human brain in a variety of appropriate tasks. Section 8 concludes.

## 2. Literature Review

In this section we systematically and comprehensively search for application of neural networks in both empirical and normative fields of economics and finance from the most representative academic and policy studies, which could be used to provide enlightenment for this study not only in the selection of specific neural network methods, but also in the determination of specific application scenarios such as economic & financial forecasting, government decision-making and policy evaluation that have not been systematically and deeply studied.

If we have to pick not only the hottest but also the most representative application areas for neural network methods among various economic and financial subfields, that must be the various subfields of energy economics and finance. More specifically, the application of neural networks covers areas such as energy prices prediction, energy demand or consumption forecasting, structure of energy system, energy policy analysis, model calibration, energy trading strategies, and data management. Neural network methods can provide superior performance for forecasting energy prices because they have higher flexibility in handling energy commodity price series that typically demonstrate complex features such as non-linearity, lag-dependence, non-stationarity, and volatility clustering (Cheng et al., 2018; Ghoddusi et al., 2019). Among literatures about energy price predictions, a vast majority of papers focus on either crude oil price prediction (Cheng et al., 2018; Ding, 2018; Godarzi et al., 2014; Huang and Wang, 2018; Jammazi and Aloui, 2012; Moshiri and Foroutan, 2006; Safari and Davallou, 2018; Wang and Wang, 2016; Yu et al., 2008; Yu et al., 2017; Zhao et al., 2017) or electricity price prediction (Aggarwal et al., 2009; Bento et al., 2018; Conejo et al., 2005; Dudek, 2016; Lago et al., 2018; Panapakidis and Dagoumas, 2016; Peng et al., 2018; Singh et al., 2017; Wang et al., 2017; Weron, 2014; Yang et al., 2017); papers on predicting natural gas prices (Čeperić et al., 2017; Nguyen and Nabney, 2010) and carbon prices (Fan et al., 2015; Sun et al., 2016) are much less frequent, while major papers on predicting coal prices are hard to be found, even though coal is a major energy source. Since government agencies and financial & trading institutions are all interested in having a realistic forecast of energy consumption portfolio in the future, a vast majority of literatures make long-range predictions of aggregate and sectoral energy demand based on neural network methods (Debnath and Mourshed, 2018; Geem and Roper, 2009; Kaytez et al., 2015; Liu et al., 2016;

Sözen and Arcaklioglu, 2007; Sözen et al., 2007; Xiao et al., 2018), among different types of energy, electricity demand forecasting is the most traditional domains for neural network methods (Ardakani and Ardehali, 2014; Azadeh et al., 2008; Lai et al., 2008; Pao, 2006), and neural network is much better suited for short-term electricity demand forecast than the national level forecast, since the electricity sector can provide a large number of high-frequency observations on a large set of potential input variables (Anderson et al., 2011; Bassamzadeh and Ghanem, 2017; Liu et al., 2014), however, there are less neural network related literatures on predicting natural gas demand (Azadeh et al., 2010; Panapakidis and Dagoumas, 2017; Szoplik, 2015), on predicting transport energy demand (Geem, 2011; Limanond et al., 2011; Murat and Ceylan, 2006), and on predicting coal demand (Hu et al., 2008; Ning, 2003; Yang et al., 2014). Among other relatively minor application areas of the neural network approach, there are relatively abundant studies on structure of energy system (Ermis et al., 2007; Fang et al., 2013; Farajzadeh and Nematollahi, 2018; Skiba et al., 2017; Sözen, 2009; Zhang et al., 2016), and researches on energy policy analysis are also relatively easy to be found (Dagoumas et al., 2017; Mahmoud and Alajmi, 2010; Skiba et al., 2017), while studies on model calibration (Sun et al., 2011), on energy trading strategies (Moreno, 2009), and on data management (Abdella and Marwala, 2005; Nelwamondo et al., 2007) are the least frequent.

    If the merits and limitations of neural network methods are compared horizontally among all machine learning methods that have already been applied to some subfield of energy economics and finance, then accuracy in general, speed of classification, dealing with binary/continuous attributes, and attempts for incremental learning are all considered to be the outstanding merits of most neural network methods, while relatively low speed of learning, relatively low tolerance to missing values, relatively low tolerance to irrelevant attributes, relatively low tolerance to redundant attributes, relatively low tolerance to data noise, dealing with danger of overfitting, lack of explanation ability or statistical inference, black box nature that leads to difficulties in understanding how the results were obtained compared to other more transparent methods, and no general rules for model parameter handling are considered as the limitations of some neural network methods in some situations but not always (Ghoddusi et al., 2019). Among all kinds of neural networks, the NARX neural network usually has a higher level of accuracy, superior

generality and practicability for time series prediction because of various advantages such as the ability to keep a memory of past events to predict future trends (Wang and Wang, 2016), the capacity to provide better forecasting accuracy than traditional time series analysis, especially in the multi-step ahead short-term forecast (Cheng et al., 2018), its special structure that allows the algorithm to automatically and efficiently model a high degree of complexity not only in the interaction among various exogenous inputs, but also in the possible relationships between inputs and outputs (Hatcher and Yu, 2018; LeCun et al., 2015), the ability to provide a high degree of flexibility to address features as including a large number of exogenous variables in time-series models, in other words, the ability to accept almost any number of exogenous inputs, which relieves the neural network builder from the task of picking a small number of informative input variables (Zhao et al., 2017), the flexibility of outputs depending on the problem under study, in other words, the flexible and changeable application of the same type of NARX neural network in various problems such as regression problems, classification problems or ranking problems (Ghoddusi et al., 2019), and no special data preprocessing required, for example the detrending, seasonal adjustment, or decomposition of the time series, since the NARX neural network can consider those characteristics as additional features of the data and incorporate them into the final forecasting algorithm (Dudek, 2016).

 Subfields of finance such as algorithmic trading, risk assessment, fraud detection, portfolio management, asset pricing and derivatives market (such as options, futures, and forward contracts), cryptocurrency and blockchain studies, financial sentiment analysis and behavioral finance, and financial text mining are also among the hottest application areas for neural network methods where specific neural networks are developed to provide real-time working solutions for the financial industry (Ozbayoglu et al., 2020). For the sake of simplicity, we only state and compare the major conclusions from those representative reviews of neural network-related studies in each of the above subfields as follows. Algorithmic trading is defined as buy-sell decisions made solely by algorithmic models, since financial time series forecasting is highly coupled with algorithmic trading, most of the algorithmic trading studies have been concentrated on the prediction of stock or index prices, meanwhile, the various types of recurrent neural networks (RNNs) have been the most preferred neural networks in this subfield, since various types of trade indicators and technical indicators that provide either real-time information or historical information can be simply treated as the exogenous inputs into RNNs to help improve the quality of forecasts (Hu et al., 2015; Ozbayoglu et al., 2020; Sezer et al., 2019). Risk assessment studies that adopt neural networks as the method identify the riskiness of any given asset, firm, person, product, and bank, and form various research topics such as bankruptcy prediction, credit scoring, credit evaluation, loan/insurance underwriting, bond rating, loan application, consumer credit determination, corporate credit rating, mortgage choice decision, financial distress prediction, and business failure prediction, among various machine learning methods, a great number of researches have turned their attention to specific neural networks such as the deep neural networks (DNNs) for higher accuracy (Chen et al., 2015; Fethi & Pasiouras, 2010; Kirkos & Manolopoulos, 2004; Kumar & Ravi, 2007; Lahsasna et al., 2010; Lin et al., 2012; Marqués et al., 2013; Ozbayoglu et al., 2020; Ravi et al., 2008; Sun et al., 2014; Verikas et al., 2009). Financial fraud is one of the areas where the governments and authorities are desperately trying to find a permanent solution, that's one important reason why fraud detection is also among the hottest application areas for neural network methods, most of the studies in this area can be considered as anomaly detection and are generally classification problems between fraud and non-fraud, and those neural networks that specialize in accurate classification are usually adopted by these studies, for example the convolutional neural networks (CNNs) (Kirkos et al., 2007; Ngai et al., 2011; Ozbayoglu et al., 2020; Phua et al., 2010; Sharma & Panigrahi, 2013; Wang, 2010; West & Bhattacharya, 2016; Yue et al., 2007). Portfolio management, which is the process of choosing various assets within the portfolio for a predetermined period, covers those closely related and even interchangeable areas such as portfolio optimization, portfolio selection, and portfolio allocation, since portfolio management is actually an optimization problem, identifying the best possible course-of-action for selecting the best-performing assets for a given period, as a result, a lot of neural networks that specialize in solving optimization problems, such as RNNs and CNNs, have been developed for this purpose (Li & Hoi, 2012; Metaxiotis & Liagkouras, 2012; Ozbayoglu et al., 2020). Asset pricing and derivatives market (options, futures, forward contracts) are almost destined to be one of the application areas for neural network methods, since accurate pricing or valuation of an asset is a fundamental study area in finance, among a vast number of special neural networks developed for banks, corporates, real estate, derivative products, etc., specially designed RNNs and DNNs are usually proved to be more capable of assisting the asset pricing researchers or valuation experts (Ozbayoglu et al., 2020). Since price forecasting and trading systems dominate the area of cryptocurrency and blockchain studies, then application of neural network methods in this area is similar to that in the area of algorithmic trading, which we've already mentioned above (Ozbayoglu et al., 2020). Since emotion or investor sentiment is among the most important components of behavioral finance, neural network methods are increasingly applied to financial sentiment analysis, especially for trend forecasting, since most of the researches in this area are focused on financial forecasting and based on text mining, then specially constructed RNNs, DNNs, and CNNs are used most often in these researches (Kearney & Liu, 2014; Ozbayoglu et al., 2020). With the rapid spreading of social media and real-time streaming news, instant text-based information retrieval has become available, as a result, financial text mining studies have been more and more popular in recent years, in fact text mining is usually accompanied by the semantic analysis and forecast application of the extracted information, such as those financial sentiment analysis coupled with text mining for forecasting, which we've already mentioned above, and text mining without sentiment analysis for forecasting, which we haven't talked about yet, and various types of specially designed RNNs are the most widely and frequently used neural networks in both types of studies (Kumar & Ravi, 2016; Li, 2011; Loughran & McDonald, 2016; Mitra & Mitra, 2012; Mittermayer & Knolmayer, 2006; Ozbayoglu et al., 2020).

Some newly developed representative and interesting application areas for neural network methods and even machine learning methods in a broader sense in recent years that can offer inspiration to either the determination of specific application scenarios or the selection of specific neural network methods in our study are also briefly described as follows.

Recent literatures that can provide inspiration for application scenario determination are presented as follows. Bandiera et al. (2020) use an unsupervised machine learning algorithm to measure the behavior of CEOs in large samples via a survey that collects high-frequency, high-dimensional diary data, this algorithm uncovers two distinct behavioral types that are leader type and manager type, and its ability of accurate classification is proved to be much higher than other traditional methods. Sabahi & Parast (2020) use predictive analytics by proposing a machine learning approach to predict individuals' project performance, which is considered as a new measurement system for predicting performance. Mittal et al. (2019) monitor the impact of economic crisis on crime in India through machine learning methods, which are proved to be capable of automatically predicting the factors that affect the crimes effectively and efficiently. Sansone (2019) helps high schools obtain more precise predictions of student dropout through exploiting the available high-dimensional data from 9th grade jointly with machine learning tools, which are proved to perform much better than the parsimonious early warning systems as implemented in many high schools. Erel et al. (2018) demonstrate that machine learning algorithms can even assist firms in their decisions on nominating corporate directors, specifically speaking, machine learning holds promise for understanding the process by which governance structures are chosen, and has potential to help real-world firms improve their governance. Kleinberg et al. (2018) take bail decisions as a good test case, and prove that although machine learning can be valuable when it is used to improve human decision making, realizing this value requires integrating these tools into an economic framework: being clear about the link between predictions and decisions; specifying the scope of payoff functions; and constructing unbiased decision counterfactuals. Handel & Kolstad (2017) implement machine learning-based models to assess treatment effect heterogeneity, which are considered as new methods applied to health behaviors related new data based on wearable technologies to understand population health. Chalfin et al. (2016) demonstrate that studying the nature of production functions in social policy applications requires not an estimate of a causal effect, but rather a prediction, then there can be large social welfare gains from using machine learning tools to predict worker productivity, since these research works can be used to help improve productivity. McBride & Nichols (2016) use out-of-sample validation and machine learning to retool poverty targeting, since poverty targeting tools have become common tools for beneficiary targeting and poverty assessment where full means tests are costly.

All the above representative and interesting literatures in recent years show that **lots of different implementations of neural network methods and even machine learning methods in a broader sense are constantly emerging, and the broad interest in combining economic analysis and neural network methods is always continuing, however, the playfield for neural** **network methods in subfields of economics and finance is wide open, and a lot of research opportunities, such as macroeconomic forecasting, national goal setting and global competitiveness assessment that are all specific scenarios selected in this study, still exist and haven't been extensively covered and fully explored.**

Recent literatures that can offer inspiration for neural network selection are introduced as follows. Athey & Imbens (2019) highlight newly developed methods at the intersection of machine learning and econometrics, which typically perform better than either off-the-shelf machine learning or more traditional econometric methods when applied to particular classes of problems, such as causal inference problems, optimization problems, performance evaluation problems, and policy effect estimation problems. Storm et al. (2019) review application of machine learning methods in agricultural and applied economics, and find that a large number of machine learning methods have already demonstrated great potential in improving prediction and computational power in agricultural and applied economic analysis, while economists still have a vital role in addressing the shortcomings of machine learning methods when used for quantitative economic analysis. Kasy (2018) suggests an approach based on maximizing posterior expected social welfare, combining insights from both optimal policy theory as developed in the field of public finance, and machine learning using Gaussian process priors, in fact this study tells us how to use (quasi-)experimental evidence when choosing policies such as optimal taxation and insurance. Athey & Imbens (2017) argue that the use of machine learning can help buttress the credibility of policy evaluation, and believe that further developed literatures in the area of causality and policy evaluation can help researchers avoid unnecessary functional form and other modeling assumptions, and increase the credibility of policy analysis. Athey (2017) argues that there are a number of gaps between making a prediction and making a decision when applying machine-learning prediction methods, and underlying assumptions need to be understood in order to optimize data-driven decision-making. Kleinberg et al. (2015) argue that an important class of policy problems does not require causal inference but instead requires predictive inference, and newly developed machine learning methods are particularly useful for addressing these prediction problems, specifically, this study uses an example from health policy to illustrate the large potential social welfare gains from improved prediction.

All the above representative and interesting literatures in recent years argue that **although novel neural network approaches and even machine learning approaches in a broader sense have led to important breakthroughs in various subfields of economics and finance, uniting data-driven machine learning methods with the amassed theoretical disciplinary knowledge of economics and finance still remains a central challenge for the application of neural network methods and even machine learning methods in a broader sense, in this respect, economists have a vital role not only in addressing the shortcomings of neural network methods when used for quantitative economic analysis, but also in combining results of neural network methods with theoretical knowledge to answer economic questions.**

From all of the above literature reviews, **neural network methods hold significant potential for capturing complex spatial and temporal relationships, and are the most widely used, effective supervised machine learning approaches currently available, among many neural network architectures, the three most relevant for economists are convolutional neural networks (CNNs), recurrent neural networks (RNNs) and deep neural networks (DNNs), since CNNs are well placed to process grid-like data such as 2D or 3D data, RNNs are an alternative to CNNs for processing sequential data or time series data, handling dynamic relationships and long-term dependencies, and DNNs are the basis for the first two, then this study selects a special type of RNNs, the Nonlinear Autoregressive with Exogenous Input (NARX) neural network, according to the specific application scenarios involving macroeconomic forecasting, national goal setting and global competitiveness assessment.**

## 3. Method

### 3.1. Introduction of General Nonlinear Autoregressive with Exogenous Input (NARX) Neural Network

Since the Nonlinear Autoregressive with Exogenous Input (NARX) neural network is a special type of Recurrent Neural Network (RNN), here we first introduce the ideas behind RNN. The advantage of RNNs is that they could use their reasoning about previous events in a task to inform later ones, since they are networks with loops in them, allowing information to persist.

Both the basic and unrolled structures of a general RNN are shown in Figure 1. In the basic structure, a chunk of neural network A looks at some inputs $x_t$ and outputs values $y_t$ in each period t, a loop in each period t allows information to be passed from one step of the network to the next (Olah, 2015). In the unrolled structure, a general RNN can be thought of as multiple copies of the module A, each passing a message to a successor, this repeating module A usually has a very simple structure such as a single tanh layer (here tanh is a hyperbolic function that is the ratio of sinh to cosh, which is expressed as $\tanh(x)=(e^x-e^{-x})/(e^x+e^{-x})$), this chain-like nature reveals that RNNs are intimately related to sequences and lists, they're the natural architecture of neural network to use for time series data (Olah, 2015).

As a special type of RNN, NARX neural network is further specifically designed to serve time series prediction, it usually provides better predictions than the above general input-output RNN model (with $\{x_t, x_{t-1},…\}$ as inputs and get $\{y_t, y_{t-1},…\}$ as outputs), because it uses the additional information contained in the series of interest $\{y_{t-1}, y_{t-2},…\}$ that has already been output before period t, and both the basic and unrolled structures of a general NARX neural network as an update and improvement to a general RNN in Figure 1 are comparatively shown in Figure 2. Switching to the perspective of econometrics, since a general NARX neural network would like to nonlinearly predict future values $y_t$ of a time series $\{y_t, y_{t-1},…\}$ from not only past values of that time series $\{y_{t-1}, y_{t-2},…\}$, but also past values of a second time series $\{x_{t-1}, x_{t-2},…\}$, then it could also be considered as an update and improvement to the classic autoregressive (AR) model which specifies that the output variable $y_t$ depends linearly on its own previous values $\{y_{t-1}, y_{t-2},…\}$ and on a stochastic term (an imperfectly predictable term), that's just the reason why this form of prediction is called nonlinear autoregressive with exogenous (external) input, or NARX (Number23, 2019). Since the externally determined inputs series $\{x_t, x_{t-1},…\}$ influence the series of interest $\{y_t, y_{t-1},…\}$ based on the above modeling ideas, the exogenous inputs series $\{x_t, x_{t-1},…\}$ are also called the driving series of a general NARX model (Wikipedia, 2019).

A general NARX neural network model can be mathematically stated as follows:

$y_t = F(y_{t-1}, …, y_{t-d}; x_{t-1}, …, x_{t-d})+\varepsilon_t$

Here t represents for period t; d is the exogenously designated time delays; $\{y_t, y_{t-1},…\}$ is the series of interest; $\{x_t, x_{t-1},…\}$ is the exogenous inputs series or driving series, here it's worth mentioning that each $x_t$ could contain several variables based on time t, in other words, {$x_t$, $x_{t-1}$,…} could be made up of several exogenously determined time series; function F represents for a general NARX neural network; moreover, the above model contains the error term $\varepsilon_t$ (also called noise of the above NARX model), which signifies that although knowledge of both past values of the exogenous inputs series {$x_{t-1}$, $x_{t-2}$,…} and past values of the series of interest {$y_{t-1}$, $y_{t-2}$,…} help predict the current value of the series of interest $y_t$, they will not enable $y_t$ to be predicted exactly (Number23, 2019).

Based on the above simplest model expression, the performance of a general NARX neural network model can be simply and directly evaluated by the following mean squared error (MSE) and coefficient of determination ($R^2$):

MSE := {$\sum_{t=T0}^{T1}$[$y_t$ - F($y_{t-1}$, …, $y_{t-d}$; $x_{t-1}$, …, $x_{t-d}$)]^2}/(T1-T0+1)

$R^2$ := 1 - {$\sum_{t=T0}^{T1}$[$y_t$ - F($y_{t-1}$, …, $y_{t-d}$; $x_{t-1}$, …, $x_{t-d}$)]^2}/[$\sum_{t=T0}^{T1}$F($y_{t-1}$, …, $y_{t-d}$; $x_{t-1}$, …, $x_{t-d}$)^2]

Here [$T_0$, $T_1$] is the entire time interval for the specific time series prediction problem; $T_0$ is equal to the starting time of exogenous inputs series {$x_t$, $x_{t-1}$,…} or target series {$y_t$, $y_{t-1}$,…} plus the time delays parameter d that represents the minimum number of previous periods required to make the earliest prediction in the target series; $T_1$ is equal to the end time of exogenous inputs series {$x_t$, $x_{t-1}$,…} or target series {$y_t$, $y_{t-1}$,…} (Beale et al., 2014; MathWorks, 2020). Then the optimum configuration of a general NARX neural network model that gives the lowest MSE and highest $R^2$ for the training dataset is usually considered to have the optimal performance (Islam & Morimoto, 2015).

In addition to the set of exogenous inputs series {$x_t$, $x_{t-1}$,…}, target series {$y_t$, $y_{t-1}$,…}, and the time delays parameter d which we're going to discuss in the next subsection when setting specific NARX neural networks for various application scenarios, other user-defined parameters in the NARX modeling comprise the selection of some target timesteps for training, validation, or testing, and the selection of number of hidden neurons in the NARX neural network. Here the target timesteps for training are used to directly be presented to the NARX neural network during training, in order to help adjust the model according to its intermediate error term series {$\varepsilon_t$, $\varepsilon_{t-1}$,…}; the target timesteps for validation are adopted not only to measure NARX neural network generalization, but also to halt training when model generalization stops improving; the target timesteps for testing are ruled out during the training process, so they have no effects on training, and their main function is to provide an independent measure of prediction performance of the NARX neural network during and after training (Beale et al., 2014; MathWorks, 2020). In our subsequent study, all target timesteps along with associated inputs series and target series are randomly divided into the training set, the validation set, and the testing set for all applications of NARX neural networks in macroeconomic forecasting, national goal setting and global competitiveness assessment, with 70% be incorporated into the training set, 10% be put into the validation set, and the last 20% be included in the testing set.

    Based on the conclusion from the literature that any multi-dimensional nonlinear mapping of any continuous function can be carried out by a two-layer model with a suitable chosen number of neurons in its hidden layer (Cybenko, 1989), the standard NARX network is set as a two-layer feedforward network, with the tangent sigmoid transfer function (also called the logistic function, which is expressed as tangent-sigmoid$(x)=\tanh(x):=(e^x-e^{-x})/(e^x+e^{-x})$) in the hidden layer and the linear transfer function in the output layer, then only the selection of number of hidden neurons in the hidden layer is closely related to the nonlinearity degree of the model, the learning ability of past information from both the exogenous inputs series and target series, the forecasting ability of the target series, and the training efficiency and complexity of the model (Beale et al., 2014; Lee & Sheridan, 2018; MathWorks, 2020). Specifically speaking, if the number of hidden neurons is too small, the model cannot have the necessary degree of nonlinearity for highly nonlinear forecasting problems, the necessary learning ability or past information processing ability, and the sufficient forecasting ability at the target timesteps for testing, although the training efficiency of the model is guaranteed to be high and the training complexity is usually low; on the contrary, if the number of hidden neurons is too large, not only the complexity of the NARX neural network structure (which is especially important for the neural network implemented by hardware) will greatly increase, which brings about the model is more likely to fall into the local minimum rather than the global minimum in the training process, but also the training speed of the NARX neural network will become quite slow and the training efficiency is too difficult to be guaranteed, sometimes the above situation even leads to the forecasting problem no longer solvable under the constraints of existing software and hardware resources, although the nonlinearity degree of the

model could be guaranteed to be enough for catching nonlinear dynamics of the system for nonlinear forecasting problems, besides, increasing the number of hidden neurons cannot always and necessarily improve model accuracy and generalizability, which means either the learning ability of past information or the forecasting ability of the target series are not guaranteed to be improved with the increase of the number of hidden neurons, moreover, even when the learning & forecasting abilities improve as the number of hidden neurons increase, the marginal learning & forecasting abilities improvement decreases with the increase of the number of hidden neurons (Asgari, 2014; Beale et al., 2014; Islam & Morimoto, 2015; MathWorks, 2020).

Considering the above impacts of number of hidden neurons on the various characteristics and performance of a general NARX neural network, and since it is desirable for us to adopt the simplest possible network structure to carry out macroeconomic forecasting, national goal setting and global competitiveness assessment (Lee & Sheridan, 2018), we always start with 1 hidden neuron and gradually increase the number of hidden neurons until there is very limited improvement in further reducing the mean square error (MSE) of the model (more concretely, the average squared difference between output series and target series of the model), or further increasing the coefficient of determination ($R^2$) of the neural network. In the subsequent preliminary experiments for various applications of NARX neural networks in macroeconomic forecasting, national goal setting and global competitiveness assessment, in most cases after choosing Bayesian Regularization as the NARX training algorithm (which is going to be discussed in the next paragraph), setting the number of hidden neurons to 2 or sometimes 3 can be enough for the specific NARX neural network to not only yield the best performance, but also best balance the nonlinearity degree of the model for those specific nonlinear forecasting problems, the learning ability and the forecasting ability, and the training efficiency and complexity of this specific NARX neural network.

After exogenous inputs series $\{x_t, x_{t-1},…\}$ and target series $\{y_t, y_{t-1},…\}$ are fully prepared, meanwhile the time delays parameter d, target timesteps for training, validation, or testing, and number of hidden neurons in the hidden layer of a specific NARX neural network are all explicitly set, before training and testing the NARX neural network we still need to make a choice about the NARX training algorithm at the end. There are three commonly used NARX training algorithms involving Levenberg-Marquardt Optimization, Bayesian Regularization, and Scaled Conjugate Gradient Optimization. All algorithms are briefly described and especially their advantages and disadvantages are highlighted as follows:

1. Levenberg-Marquardt Optimization is the training algorithm that is usually considered to be a good balance between model performance, training efficiency and computational resource consuming, the training under this algorithm automatically stops when the average squared difference between output series and target series at those target timesteps for validation almost stops declining (Beale et al., 2014; Guzman et al., 2017; Lee et al., 2016; MathWorks, 2020);

2. Bayesian Regularization is modified to include the regularization technique on the basis of Levenberg-Marquardt Optimization, since this algorithm minimizes a combination of squared errors & weights and determines the correct combination to produce a NARX neural network that generalizes well (more precisely, Bayesian Regularization training algorithm updates weight and bias values according to a proper combination of gradient descent with momentum, gradient descent with adaptive learning rate, and gradient descent momentum & adaptive learning rate), and training under this algorithm stops according to adaptive weight minimization (regularization), then it can result in good generalization for difficult, small or noisy datasets, however, this algorithm typically takes more training time and is more dependent on computational resources (Beale et al., 2014; Eugen, 2012; Guzman et al., 2017; Islam & Morimoto, 2015; MathWorks, 2020);

3. Scaled Conjugate Gradient Optimization is considered as the time saving and computational resource saving algorithm, it's especially recommended for time wasting and computational resource wasting problems with extremely large datasets, because it uses gradient calculations which are more memory efficient than the Jacobian calculations used by the above two algorithms, the training under this algorithm also automatically stops when the average squared difference between output series and target series at those target timesteps for validation almost stops improving, which is the same as that under Levenberg-Marquardt Optimization (Beale et al., 2014; MathWorks, 2020).

Since those subsequent applications of NARX neural networks in macroeconomic forecasting, national goal setting and global competitiveness assessment mostly work on difficult and noisy datasets, also since this study focuses more on the forecasting performance of the model and relatively less on the training time efficiency and computational resource utilization efficiency of the neural network, the Bayesian Regularization training algorithm is ultimately chosen for all the subsequent applications of NARX neural networks in the following sections.

## 3.2. Specific Nonlinear Autoregressive with Exogenous Input (NARX) Neural Networks for Macroeconomic Forecasting, National Goal Setting and Global Competitiveness Assessment

The selection of exogenous inputs series $\{x_t, x_{t-1},…\}$ is usually considered as the most important part in the process of constructing a specific NARX neural network, whether the selection is appropriate or not is directly related to the applicability & generalization ability of the specific NARX neural network and the accuracy of its forecasting. Since when the training of NARX neural network does not have the problem of overfitting, previous studies show that the greater the relationship between exogenous inputs series $\{x_t, x_{t-1},…\}$ and target series $\{y_t, y_{t-1},…\}$, the more accurate the forecasting value given by the trained NARX neural network (Fan & Yu, 2015; Wang & Hou, 2015), then specific exogenous inputs series for specific target series in the construction of a particular NARX neural network are selected based on the following preference order.

1. The exogenous inputs series have similar statistical meaning to the target series;
2. The functional relationship between the exogenous inputs series and the target series is linear;
3. The exogenous inputs series are highly linearly correlated with the target series from an empirical perspective, although there is no obvious & direct functional relationship between the two from a theoretical perspective;
4. The exogenous inputs series and the target series have low degree nonlinear functional relationship, which is closer to linear rather than nonlinear, for example they have the quasi linear function relations;

5. Although the exogenous inputs series and the target series have high degree nonlinear functional relationship, this functional relationship is explicit and can be expressed analytically, such as they have polynomial function relations;

6. There is an arbitrary nonlinear functional relationship between the exogenous inputs series and the target series, and the steps of function operation from the exogenous inputs series to the target series are relatively short;

7. There is still an arbitrary nonlinear function relation between the exogenous inputs series and the target series, but the steps of function operation from the exogenous inputs series to the target series are relatively long, and function operations are relatively complicated;

8. The exogenous inputs series are highly nonlinearly correlated with the target series from an empirical perspective, although there is no obvious & direct functional relationship between them from a theoretical perspective.

Based on the above preference order for the selection of exogenous inputs series, and in consideration of the fact that those leading business and economic indicators & indices (LBEIs) from the corresponding country's business and economic surveys usually contain the most abundant information that is useful for the forecasting of the relevant target series, the specific NARX neural network for macroeconomic forecasting can be mathematically stated as follows on the basis of adjustment & improvement to a general NARX neural network:

$MEI_t = F(MEI_{t-1}, \ldots, MEI_{t-d}; LBEIs_{t-1}, \ldots, LBEIs_{t-d}) + \varepsilon_t$

Here t represents for period t; $MEI_t$ represents for the value of a specific macro-economic indicator or index (MEI) in period t; $LBEIs_t$ represent for the values of selected leading business and economic indicators & indices (LBEIs) in period t; d is the exogenously designated time delays, since 1-year past values of both exogenous inputs series $\{LBEIs_{t-1}, LBEIs_{t-2},\ldots\}$ and target series $\{MEI_{t-1}, MEI_{t-2},\ldots\}$ are considered sufficient to capture the underlying level of the specific macro-economic indicator or index $MEI_t$ in the future period t and its recent growth rate, d is set to 12 or 4 if the specific NARX neural network is used to make monthly forecast or quarterly forecast respectively; function F represents for the specific NARX neural network for the forecasting of the specific macroeconomic target series; the error term $\varepsilon_t$ is used to calculate the mean squared error (MSE) and coefficient of determination ($R^2$) mentioned in the previous subsection in order to evaluate the fitting or prediction performance of this specific NARX neural network. Figures 3 and Figure 4 give the example of the specific NARX neural network for monthly frequency forecasting of the target series and quarterly frequency forecasting of the target series respectively.

The only difference between the specific NARX neural network for macroeconomic forecasting and the specific NARX neural network for national goal setting is that the latter requires not only the inclusion of LBEIs in exogenous inputs series, but also the inclusion of major and conventional macroeconomic statistical indicators & indices in exogenous inputs series. Thus the specific NARX neural network for national goal setting can be mathematically expressed as follows on the basis of minor modifications to the specific NARX neural network for macroeconomic forecasting:

$NG_t = F(NG_{t-1}, \ldots, NG_{t-d}; LBEIs_{t-1}, \ldots, LBEIs_{t-d}; MCMSIs_{t-1}, \ldots, MCMSIs_{t-d}) + \varepsilon_t$

Here $NG_t$ represents for the value of a specific national goal (NG) in period t; $LBEIs_t$ still represent for the values of selected leading business and economic indicators & indices (LBEIs) in period t; additional added $MCMSIs_t$ represent for the values of selected major and conventional macroeconomic statistical indicators & indices (MCMSIs) in period t; the exogenously designated time delays d is still set to 12 or 4, depending on whether the specific NARX neural network is used to set monthly goal or quarterly goal; function F represents for the specific NARX neural network for the forecasting of the specific national goal series; the error term $\varepsilon_t$ is still used for the evaluation of fitting or prediction performance of this specific NARX neural network.

Modeling ideas behind the specific NARX neural network for global competitiveness assessment are almost the same as those behind the specific NARX neural network for national goal setting, except that only those LBEIs and MCMSIs collected or estimated for all countries in the training sample could be incorporated into the exogenous inputs series, meanwhile those selected LBEIs and MCMSIs should cover more aspects besides the country's macro economy, since the global competitiveness assessment is in fact the prediction of the Global Competitiveness Index (GCI), which is a systematic and comprehensive assessment of the corresponding country's institutions, infrastructure, macroeconomic environment, health and primary education, higher education and training, goods market efficiency, labour market efficiency, financial market development, technological readiness, market size, business sophistication, and innovation by World Economic Forum (WEF). Based on the above preparation, the specific NARX neural network for global competitiveness assessment can be mathematically stated as follows on the basis of minor adjustments to the specific NARX neural network for national goal setting:

$$GCIs_t = F(GCIs_{t-1}, \ldots, GCIs_{t-d}; LBEIs_{t-1}, \ldots, LBEIs_{t-d}; MCNSIs_{t-1}, \ldots, MCNSIs_{t-d}) + \varepsilon_t$$

Here $GCIs_t$ represent for the values of all sample countries' global competitiveness indices in period t; $LBEIs_t$ represent for the values of all sample countries' selected leading business and economic indicators & indices (LBEIs) that cover more aspects besides macro economy in period t; $MCNSIs_t$ represent for the values of all sample countries' selected major and conventional national statistical indicators & indices (MCNSIs) that cover more aspects besides macro economy in period t; since any country's GCI is yearly released by WEF, and 2-year past values of a particular country's GCI are considered sufficient to capture the underlying level of that country's GCI in the next period and its recent growth rate, then the exogenously designated time delays d is set to 2; function F represents for the specific NARX neural network for the forecasting of the specific GCI series; the error term $\varepsilon_t$ is still used for the evaluation of fitting or prediction performance of this specific NARX neural network. Figure 5 gives an example of a specific NARX neural network for yearly frequency forecasting of GCIs.

In addition, the training for all of the above specific NARX neural networks can be made more efficient if the values of exogenous inputs series and target series are all normalized into the interval [−1, 1], which simplifies the problem of the outliers for the NARX neural network (Islam & Morimoto, 2015). Based on the above considerations, most statistical indicators and indices adopted in exogenous inputs series & target series are converted to year-on-year growth type indicators and indices when the conversion is possible and necessary.

## 4. Application of NARX Neural Networks for Macroeconomic Forecasting

In this section, we're going to carry out the application of NARX neural networks for macroeconomic forecasting in the form of case studies on China (at the national level), US (at the national level) and Eurozone (at the regional level). Through comparing the fitting and prediction performance of NARX neural networks with different settings for exogenous inputs, we're going to explore and summarize how those limited & partial exogenous inputs or abundant & comprehensive exogenous inputs specifically affect the forecasting performance of NARX neural networks for specific macroeconomic indicators or indices.

## 4.1. Application of NARX Neural Networks for Macroeconomic Forecasting: A Case Study on China

In this subsection, we apply specific NARX neural networks for macroeconomic forecasting to forecast the following macroeconomic indicators and indices of China, all of which are regularly predicted by China's most famous financial database WIND, while WIND's predictions are in fact the forecast average of major financial institutions in China and abroad:

- Year-on-Year Growth of Constant Price GDP
- Year-on-Year Growth of Value-Added of Industrial Enterprises above Designated Size (IVA)
- Year-on-Year Growth of Consumer Price Index (CPI)
- Year-on-Year Growth of Ex-factory Price Index of Industrial Products or Producer Price Index (PPI)
- Year-on-Year Growth of Investment in Fixed Assets (FAI)
- Year-on-Year Growth of Total Retail Sales of Consumer Goods (TRSCG)
- Year-on-Year Growth of Value of Exports
- Year-on-Year Growth of Value of Imports
- Average Exchange Rate of USD/CNY
- Year-on-Year Growth of M2
- Year-on-Year Growth of RMB Loans of Financial Institutions
- 1-Year Time Deposit Rate (Lump-Sum Deposit and Withdrawal)
- 6-Month to 1-Year Short-Term Loan Interest Rate

The precondition of the above application is to select all available leading business and economic indicators & indices (LBEIs) related to each target indicator or index waiting for prediction, therefore, before presenting and comparing the forecast results of each macroeconomic indicator or index, we first list all their related LBEIs that are treated as exogenous inputs for those specific NARX neural networks.

Year-on-year growth of constant price GDP from the second quarter of 1992 to the fourth quarter of 2019 is predicted based on the following exogenous inputs.

- Macro-economic climate indices: involving the coincident index, the leading index and the lagging index, which are all released by the National Bureau of Statistics of China and converted from monthly to quarterly;
- Diffusion indices from business survey of 5000 principal industrial enterprises: including overall operation situation index, utility of equipment capacity index, inventory of manufactured products index, domestic orders index, orders of export products index, capital turnover index, reflow of corporate sales income index, conditions of bank loans index, enterprise profit capability index, sales price of products index, and fixed assets investment index, which are all quarterly indices released by the People's Bank of China;
- Consumer confidence indices: containing consumer confidence index, consumer satisfaction index and consumer expectation index, which are all quarterly indices released by the National Bureau of Statistics of China.

Figure 6 not only displays the predicted year-on-year growth of constant price GDP of China as output series of the specific NARX neural network, with all the above related LBEIs taken as input series, but also shows the forecasts of China's most famous financial database WIND (which are first released as late as the first quarter of 2010) that are taken as a comparison object for the above predicted results. The upper subgraph in Figure 6 displays the specific NARX neural network's outputs, reference targets for training neural network, comparative targets from WIND database and forecast errors versus time, while the lower subgraph uses this neural network's outputs as the benchmark and shows the gaps between targets/comparative targets and outputs versus time. The time on the horizontal axis of both upper and lower subgraphs corresponds to each quarter from the second quarter of 1993 to the fourth quarter of 2019, which is also the time span of the exogenous inputs series or reference target series for training neural network minus the initial four quarters as the initial time delays for forecast. It is indicated that which time points were selected for training and testing neural network in both subgraphs.

From the outputs at those time points used for training neural network in Figure 6, we find that the average deviation of those training outputs from the training targets is even greater than the average deviation of comparative targets from the training targets, which means that the trained NARX neural network does not provide a fit to the training targets as good as comparative forecasts from WIND database. From the outputs at other time points adopted for testing neural network in Figure 6, although the deviations of very few test outputs from test targets are smaller than the deviations of the corresponding comparative targets from those test targets, the deviations of most test outputs from test targets are much greater than the deviations of the corresponding comparative targets from test targets, which signifies that the trained NARX neural network does not give a more accurate prediction that is closer to the true value in the future than the comparative prediction from WIND database. **The main reason for both poor fitting performance and poor prediction performance of the specific NARX neural network for predicting year-on-year growth of constant price GDP could be that the above selected LBEIs only provide limited and partial information for the forecasts of year-on-year growth of constant price GDP of China.**

Year-on-year growth of value-added of industrial enterprises above designated size (IVA) from January 2005 to February 2020 is forecasted on the basis of the following exogenous inputs.

- China purchasing managers' indices (PMIs) of the manufacturing industry: involving PMI overall index, PMI on production, PMI on new orders, PMI on new export orders, PMI on backlog of orders, PMI on stocks of finished goods, PMI on quantity of purchases, PMI on imports, PMI on ex-factory price, PMI on prices of purchased materials, PMI on inventory of raw materials, PMI on employment, and PMI on speed of supplier deliveries, which are all monthly indices released by the National Bureau of Statistics of China.

After still adopting the forecasts (which are first released as late as January 2008) by WIND database as the comparative forecasts in Figure 7, and omitting a description of Figure 7 similar to that of Figure 6, we directly find that the deviations of most training outputs from the training targets are smaller than the deviations of the corresponding comparative targets from the training targets, which means that the trained NARX neural network provides a better fit to the training targets than comparative forecasts from WIND database, however, the deviations of most test outputs from test targets are greater than the deviations of the corresponding comparative targets from test targets, which signifies that the trained NARX neural network provides less accurate forecasts than the comparative forecasts by WIND database. **The main reason for good fitting performance but poor prediction performance of the specific NARX neural network for forecasting year-on-year growth of IVA could still be that the above selected LBEIs are only limited and partial exogenous inputs around the forecasts of year-on-year growth of IVA of China.**

Year-on-year growth of consumer price index (CPI) from May 2015 to February 2020 is predicted based on all of the following exogenous inputs.

- Both the initial value (play the role of leading index) and final value (play the role of coincident index) of CICC cyclical momentum index (CMI) on prices, which are all monthly released by China International Capital Corporation (CICC);
- PMI on ex-factory price, non-manufacturing PMI on selling price, non-manufacturing PMI of construction industry on selling price, and non-manufacturing PMI of service industry on selling price, which are all monthly indices released by the National Bureau of Statistics of China;
- Consumer confidence indices: containing consumer confidence index, consumer satisfaction index and consumer expectation index, which are all monthly indices released by the National Bureau of Statistics of China;
- 9M macro indices: including the inflation index, the monetary condition index, and the monetary policy index, which are all released by the Nine Martingale Investment Management LP of China and converted from daily to monthly;
- Indices from national survey of urban depositors: involving index of future price expectation, index of future price expectation on proportion of choosing rises, and index of future price expectation on proportion of choosing falls, which are all released by the People's Bank of China and converted from quarterly to monthly.

Based on Figure 8 with a description similar to that of Figure 6, we find that the deviations of all training outputs from the training targets are so small that they can almost be ignored, which means that the trained NARX neural network provides an extremely accurate fit to the training targets that is much better than the comparative targets from WIND database, however, the deviations of most test outputs from test targets are much greater than the deviations of the corresponding comparative targets from test targets, which signifies that the trained NARX neural network could only give less accurate forecasts than the comparative forecasts from WIND database. **The main reason for extremely good fitting performance but relatively poor prediction performance of the specific NARX neural network for predicting year-on-year growth of CPI could still be that the above selected LBEIs could only provide relatively limited and partial information for the forecasts of year-on-year growth of CPI of China.**

Year-on-year growth of ex-factory price index of industrial products or producer price index (PPI) from May 2015 to February 2020 is forecasted on the basis of all the following exogenous inputs.

- Both the initial value (play the role of leading index) and final value (play the role of coincident index) of CICC CMI index on prices, which are all monthly released by China International Capital Corporation (CICC);
- PMI on ex-factory price, PMI on prices of purchased materials, PMI of large enterprises on main raw material purchase price, PMI of medium-sized enterprises on main raw material purchase price, PMI of small enterprises on main raw material purchase price, non-manufacturing PMI on selling price, non-manufacturing PMI of construction industry on selling price, and non-manufacturing PMI of service industry on selling price, which are all monthly indices released by the National Bureau of Statistics of China;
- Iron and steel PMI on purchasing price of raw materials, which is monthly released by China Federation of Logistics & Purchasing (CFLP);
- LanGe steel circulation PMI on selling price, and LanGe steel circulation PMI on purchase cost, which are monthly released by LanGe Steel Platform of China;
- Emerging industries PMI on purchase price, which is calculated according to the press finishing by WIND database;
- China enterprises development indices: including micro-sized enterprise operating index

on cost, micro-sized enterprise operating index of agriculture, forestry, animal husbandry and fishery industry on cost, micro-sized enterprise operating index of manufacturing industry on cost, micro-sized enterprise operating index of construction industry on cost, micro-sized enterprise operating index of transport industry on cost, micro-sized enterprise operating index of wholesale and retail industry on cost, micro-sized enterprise operating index of accommodation and catering industry on cost, micro-sized enterprise operating index of service industry on cost, micro-sized enterprise operating index of North China on cost, micro-sized enterprise operating index of Northeast China on cost, micro-sized enterprise operating index of East China on cost, micro-sized enterprise operating index of Central South China on cost, micro-sized enterprise operating index of Southwest China on cost, and micro-sized enterprise operating index of Northwest China on cost, which are all monthly indices released by the Postal Savings Bank of China;

- Chinese business conditions indices (BCIs): including BCI on labor costs, BCI on overall costs, BCI on consumer prices, and BCI on producer prices, which are all monthly indices released by the Cheung Kong Graduate School of Business (CKGSB);

- 9M macro indices: including the inflation index, the monetary condition index, and the monetary policy index, which are all released by the Nine Martingale Investment Management LP of China and converted from daily to monthly;

- Indices from entrepreneur poll: involving raw material purchase price index, raw material purchase price index on proportion of choosing rises, raw material purchase price index on proportion of choosing equals, and raw material purchase price index on proportion of choosing falls, which are all released by the People's Bank of China and converted from quarterly to monthly.

Compared with the forecasts of year-on-year growth of CPI in Figure 8, we find that the forecasts of year-on-year growth of PPI in Figure 9 are more accurate, not only the deviations of all training outputs from the training targets are small enough to be ignored, which signifies that the trained NARX neural network provides a sufficiently accurate fit to the training targets that is much better than the comparative targets from WIND database, but also the deviations of the majority of test outputs from test targets are smaller than the deviations of the corresponding comparative targets from test targets, which means that the trained NARX neural network gives more accurate forecasts than the comparative forecasts by WIND database. **Compared with the relatively poor prediction performance for year-on-year growth of CPI, the above selected LBEIs provide more abundant and comprehensive information for the forecasts of year-on-year growth of PPI than those selected LBEIs for the forecasts of year-on-year growth of CPI, that could be the main reason why the specific NARX neural network for forecasting year-on-year growth of PPI has extremely good fitting performance and relatively good prediction performance for year-on-year growth of PPI of China.**

Year-on-year growth of investment in fixed assets (FAI) from March 2012 to February 2020 is predicted based on all of the following exogenous inputs.

- Non-manufacturing PMIs of construction industry: involving non-manufacturing PMI of construction industry, non-manufacturing PMI of construction industry on new orders, non-manufacturing PMI of construction industry on new export orders, non-manufacturing PMI of construction industry on expected operational activities, non-manufacturing PMI of construction industry on input price, non-manufacturing PMI of construction industry on selling price, and non-manufacturing PMI of construction industry on employment, which are all monthly indices released by the National Bureau of Statistics of China;

- Iron and steel PMIs: including iron and steel PMI, iron and steel PMI on production, iron and steel PMI on raw material purchasing volume, iron and steel PMI on inventory of raw materials, iron and steel PMI on new orders, iron and steel PMI on stocks of finished goods, and iron and steel PMI on purchasing price of raw materials, which are all monthly indices released by China Federation of Logistics & Purchasing (CFLP);

- LanGe steel circulation PMIs: containing LanGe steel circulation PMI, LanGe steel circulation PMI on sales, LanGe steel circulation PMI on selling price, LanGe steel circulation PMI on total orders, LanGe steel circulation PMI on export orders, LanGe steel circulation PMI on domestic orders, LanGe steel circulation PMI on purchase cost, LanGe steel circulation PMI on speed of arrival, LanGe steel circulation PMI on inventory level, LanGe steel circulation PMI on financing environment, LanGe steel circulation PMI on employees, LanGe steel circulation PMI on trend judgement, and LanGe steel circulation PMI on willingness of purchase, which are all monthly indices released by LanGe Steel Platform of China;

- Chinese business conditions index (BCI) on investment, which is monthly released by the Cheung Kong Graduate School of Business (CKGSB).

From Figure 10 with a description similar to that of Figure 6, we find that the deviations of all training outputs from the training targets are so small that could almost be ignored, which means that the trained NARX neural network provides a sufficiently precise fit to the training targets that is much better than the comparative targets from WIND database, however, the deviations of all test outputs from test targets are much greater than the deviations of the corresponding comparative targets from test targets, which signifies that the trained NARX neural network gives much less accurate forecasts than the comparative forecasts by WIND database. **The main reason for sufficiently good fitting performance but relatively poor forecast performance of the specific NARX neural network for forecasting year-on-year growth of FAI could still be that the above selected LBEIs could only provide relatively limited and partial information for the forecasts of year-on-year growth of FAI of China.**

Year-on-year growth of total retail sales of consumer goods (TRSCG) from January 2009 to February 2020 is forecasted on the basis of all the following exogenous inputs.

- Both the initial value (play the role of leading index) and final value (play the role of coincident index) of CICC CMI index on domestic demand, which are all monthly released by China International Capital Corporation (CICC);
- PMI on ex-factory price, non-manufacturing PMI on expected operational activities, non-manufacturing PMI on selling price, non-manufacturing PMI of service industry on expected operational activities, and non-manufacturing PMI of service industry on selling price, which are all monthly indices released by the National Bureau of Statistics of China;
- Caixin China PMIs: involving Caixin China services PMI on business activity, and Caixin China composite PMI on output, which are sponsored by China Caixin Media and are monthly compiled and distributed by IHS Markit;

- Consumer confidence indices: including consumer confidence index, consumer satisfaction index, consumer expectation index, and consumer confidence index on consumption willingness, which are all monthly indices released by the National Bureau of Statistics of China;
- Chinese business conditions indices (BCIs): including BCI on sales, and BCI on consumer prices, which are all monthly indices released by the Cheung Kong Graduate School of Business (CKGSB).

Since both fitting performance and prediction performance of the specific NARX neural network for forecasting year-on-year growth of TRSCG in Figure 11 are almost the same as those for predicting year-on-year growth of FAI in Figure 10, we directly summarize that **the sufficiently good fitting performance but relatively poor prediction performance of the specific NARX neural network for predicting year-on-year growth of TRSCG could still be due to the relatively limited and partial information contained in the above selected LBEIs for the forecasts of year-on-year growth of TRSCG of China.**

The forecasts of year-on-year growth of value of exports are compared with the forecasts of year-on-year growth of value of imports because of their similarity and proximity. Year-on-year growth of value of exports from December 2012 to February 2020 is predicted on the basis of the following abundant and comprehensive exogenous inputs.

- Both the initial value (play the role of leading index) and final value (play the role of coincident index) of CICC CMI index on external demand, which are all monthly released by China International Capital Corporation (CICC);
- China PMIs of the manufacturing industry involving PMI on new export orders, PMI of large enterprises on new export orders, PMI of medium-sized enterprises on new export orders, and PMI of small enterprises on new export orders, China non-manufacturing PMIs including non-manufacturing PMI on new export orders, non-manufacturing PMI of construction industry on new export orders, and non-manufacturing PMI of service industry on new export orders, which are all monthly indices released by the National Bureau of Statistics of China;
- LanGe steel circulation PMI on export orders, which is monthly released by LanGe Steel

Platform of China;
- Export leading indicators (ELIs): containing export leading indicator, export managers' index (EMI), EMI on new export orders, EMI on export confidence, EMI on cost of export enterprises, EMI of large enterprises, EMI of medium-sized enterprises, EMI of small enterprises, business ratio of decrease in amount of new orders year-on-year, business ratio of increase in amount of new orders year-on-year, business ratio of no increase and decrease in amount of new orders year-on-year, business ratio of not optimistic for confidence in the export situation, business ratio of optimism for confidence in the export situation, business ratio of increase in comprehensive cost of exports year-on-year, business ratio of decrease in comprehensive cost of exports year-on-year, business ratio of no increase and decrease in comprehensive cost of exports year-on-year, business ratio of rising labor costs (as one type of comprehensive cost of exports) year-on-year, business ratio of rising exchange rate costs (as one type of comprehensive cost of exports) year-on-year, and business ratio of rising raw material costs (as one type of comprehensive cost of exports) year-on-year, which are all monthly indicators released by China Customs;
- Maritime silk road trade index (STI) on overall exports, which is monthly released by Ningbo Shipping Exchange of China;
- Indices from entrepreneur poll: involving export order index, export order index on proportion of choosing rises, export order index on proportion of choosing equals, and export order index on proportion of choosing falls, which are all released by the People's Bank of China and converted from quarterly to monthly.

As a contrast, year-on-year growth of value of imports from January 2013 to February 2020 is predicted only on the basis of the following limited and partial exogenous inputs.
- China PMIs of the manufacturing industry: involving PMI on imports, PMI of large enterprises on imports, PMI of medium-sized enterprises on imports, and PMI of small enterprises on imports, which are all monthly indices released by the National Bureau of Statistics of China;
- Maritime silk road trade index (STI) on overall imports, which is released by Ningbo

Shipping Exchange of China.

Through comparing fitting performance and prediction performance between the specific NARX neural network for forecasting year-on-year growth of value of exports in Figure 12 and the neural network for predicting year-on-year growth of value of imports in Figure 13, we find that the trained NARX neural network for predicting either year-on-year growth of value of exports or year-on-year growth of value of imports provides a general level of fit to the training targets that is no worse than the comparative targets from WIND database, however, the trained NARX neural network for forecasting year-on-year growth of value of exports gives accurate forecasts that are at least as good as the comparative forecasts by WIND database, while the trained NARX neural network for predicting year-on-year growth of value of imports could only give less accurate forecasts than the comparative forecasts by WIND database. **The main reason for significantly different forecast performance between the specific NARX neural network for forecasting year-on-year growth of value of exports and the NARX neural network for predicting year-on-year growth of value of imports has been proved to be that the export-related abundant and comprehensive exogenous inputs provide much more abundant and comprehensive information for the forecasts of year-on-year growth of value of exports than the import-related limited and partial exogenous inputs for the forecasts of year-on-year growth of value of imports.**

The forecasts of average exchange rate of USD/CNY, year-on-year growth of M2, year-on-year growth of RMB loans of financial institutions, 1-year time deposit rate (lump-sum deposit and withdrawal), and 6-month to 1-year short-term loan interest rate are put together for comparison and analysis, because all of them are important statistical indicators for China's financial system. Firstly, all the relevant exogenous inputs for each of the above five financial indicators are listed as follows.

All relevant exogenous inputs for the forecasts of average exchange rate of USD/CNY from November 2014 to February 2020 are as follows.

- CNH exchange rates: involving 1-week USD/CNY non-delivery forward (NDF), 1-month USD/CNY NDF, 2-month USD/CNY NDF, 3-month USD/CNY NDF, 6-month USD/CNY NDF, 9-month USD/CNY NDF, 1-year USD/CNY NDF, overnight USD/CNH delivery forward (DF), 1-week USD/CNY DF, 2-week USD/CNY DF, 1-month USD/CNY DF, 2-month USD/CNY DF, 3-month USD/CNY DF, 6-month USD/CNY DF, 9-month USD/CNY DF, and 1-year USD/CNY DF, which are all released by the UK-based money broker ICAP and converted from daily to monthly;

- CNY forward/swap quote: including 1-week USD/CNY bid, 1-month USD/CNY bid, 3-month USD/CNY bid, 6-month USD/CNY bid, 9-month USD/CNY bid, 1-year USD/CNY bid, 1-week USD/CNY ask, 1-month USD/CNY ask, 3-month USD/CNY ask, 6-month USD/CNY ask, 9-month USD/CNY ask, and 1-year USD/CNY ask, which are all released by China Foreign Exchange Trade System (CFETS) and converted from daily to monthly.

Year-on-year growth of M2 from October 1999 to February 2020 is forecasted on the basis of all the following exogenous inputs.

- 9M macro indices: including the inflation index, the monetary condition index, and the monetary policy index, which are all released by the Nine Martingale Investment Management LP of China and converted from daily to monthly;
- Government economic target for year-on-year growth of M2, which is released by the National Development and Reform Commission of China and converted from yearly to monthly.

All of the exogenous inputs for the prediction of year-on-year growth of RMB loans of financial institutions from January 2009 to February 2020 are listed as follows.

- Indices from bankers survey: involving loan demand climate index, loan demand climate index of manufacturing industry, loan demand climate index of non-manufacturing industry, loan demand climate index of infrastructure construction, loan demand climate index of large enterprises, loan demand climate index of medium-sized enterprises, loan demand climate index of small enterprises, and loan approval index, which are all released by the People's Bank of China and converted from quarterly to monthly;
- China enterprises development indices: including micro-sized enterprise operating index on financing, micro-sized enterprise operating index of agriculture, forestry, animal husbandry and fishery industry on financing, micro-sized enterprise operating index of manufacturing industry on financing, micro-sized enterprise operating index of construction industry on financing, micro-sized enterprise operating index of transport industry on financing, micro-sized enterprise operating index of wholesale and retail industry on financing, micro-sized enterprise operating index of accommodation and catering industry on financing, micro-sized enterprise operating index of service industry on financing, micro-sized enterprise operating index of North China on financing, micro-sized enterprise operating index of Northeast China on financing, micro-sized enterprise operating index of East China on financing, micro-sized enterprise operating index of Central South China on financing, micro-sized enterprise operating index of Southwest China on financing, and micro-sized enterprise operating index of Northwest China on financing, which are all monthly indices released by the Postal Savings Bank of China;

- Chinese business conditions index (BCI) on financing, which is monthly released by the Cheung Kong Graduate School of Business (CKGSB);
- 9M macro indices: including the monetary condition index, and the monetary policy index, which are all released by the Nine Martingale Investment Management LP of China and converted from daily to monthly.

1-year time deposit rate (lump-sum deposit and withdrawal) from January 2004 to February 2020 is forecasted based on all the following relevant exogenous inputs.

- The offer prices for interbank deposit rates: involving the deposit price for 7-day interbank deposit rate, the deposit price for 14-day interbank deposit rate, the deposit price for 1-month interbank deposit rate, the deposit price for 2-month interbank deposit rate, the deposit price for 3-month interbank deposit rate, the deposit price for 6-month interbank deposit rate, the deposit price for 1-year interbank deposit rate, the deposit prices by state-owned commercial banks for the above various types of interbank deposit rates, the deposit prices by joint-stock commercial banks for the above various types of interbank deposit rates, the deposit prices by urban commercial banks for the above various types of interbank deposit rates, and the deposit prices by rural commercial banks for the above various types of interbank deposit rates, which are the arithmetic average value of the offer price of big QQ Groups from each industry and converted from daily to monthly;

- Major interest rate data: including 3-month treasury bonds interest rate, 1-year treasury bonds interest rate, 10-year treasury bonds interest rate, which are all released by China Foreign Exchange Trade System (CFETS) and converted from daily to monthly;

- Index of willingness to save on proportion of choosing more savings from national survey of urban depositors, which is released by the People's Bank of China and converted from quarterly to monthly;

- 9M macro indices: including the monetary condition index, and the monetary policy index, which are all released by the Nine Martingale Investment Management LP of China and converted from daily to monthly.

6-month to 1-year short-term loan interest rate from January 2004 to February 2020 is predicted on the basis of all the following relevant exogenous inputs.

- The withdraw prices for interbank deposit rates: involving the withdraw price for overnight interbank deposit rate, the withdraw price for 7-day interbank deposit rate, the withdraw price for 14-day interbank deposit rate, the withdraw price for 1-month interbank deposit rate, the withdraw price for 2-month interbank deposit rate, the withdraw price for 3-month interbank deposit rate, the withdraw price for 6-month interbank deposit rate, the withdraw price for 1-year interbank deposit rate, the withdraw prices by state-owned commercial banks for the above various types of interbank deposit rates, the withdraw prices by joint-stock commercial banks for the above various types of interbank deposit rates, and the withdraw prices by urban commercial banks for the above various types of interbank deposit rates, which are the arithmetic average value of the withdraw price of big QQ Groups from each industry and converted from daily to monthly;

- Loan prime rates (LPRs): including 1-year LPR released by the National Interbank Funding Center of China, 1-year LPR released by Bank of Communications, 1-year LPR released by Shanghai Pudong Development Bank, 1-year LPR released by China Everbright Bank, 1-year LPR released by China Minsheng Banking, 1-year LPR released by Agricultural Bank of China, and 1-year LPR released by Bank of China, all LPRs are converted from daily to monthly;

- 9M macro indices: including the monetary condition index, and the monetary policy index, which are all released by the Nine Martingale Investment Management LP of China and converted from daily to monthly.

Since the formation mechanism of all the above five financial indicators is usually less complicated than that of the previous macroeconomic indicators, and their influencing factors are also less numerous and simpler than those of the previous macroeconomic indicators, then the fitting performance and prediction performance of the specific NARX neural networks for forecasting any of them can be good enough even if only based on a small number of relevant exogenous inputs, when taking the comparative forecasts by WIND database as criteria for performance evaluation.

From Figure 14 to Figure 18, we find that the deviations of all training outputs from the training targets are so small that could almost be ignored in Figure 14, Figure 16, Figure 17 and Figure 18, while only the deviations of most training outputs from the training targets are not small enough to be ignored in Figure 15, which means that all of the trained NARX neural networks for forecasting average exchange rate of USD/CNY, year-on-year growth of RMB loans of financial institutions, 1-year time deposit rate, and 6-month to 1-year short-term loan interest rate provide sufficiently accurate fits to the training targets that are much better than the comparative forecasts by WIND database, while the trained NARX neural network for forecasting year-on-year growth of M2 could only provide insufficiently accurate fits to the training targets that are at most slightly better than the comparative forecasts by WIND database. **The main reason for the above significantly different fitting performance of the specific NARX neural networks for forecasting different financial indicators of China is that the M2-related limited and partial exogenous inputs are the least numerous and provide the minimum amount of and the least systematic and comprehensive information for the fitting of year-on-year growth of M2, compared with the other financial indicators related more abundant and comprehensive exogenous inputs for the fitting of the corresponding financial indicators.**

Moving from fitting performance to prediction performance in Figure 14 to Figure 18, we find that the deviations of most test outputs from test targets are smaller than the deviations of the corresponding comparative forecasts by WIND database from test targets in Figure 14, the deviations of about half of the test outputs from test targets are at least as small as the deviations of the corresponding comparative forecasts by WIND database from test targets in Figure 16, and the deviations of only a very small percentage of the test outputs from test targets are at least as small as the deviations of the corresponding comparative forecasts by WIND database from test targets in Figure 17 and Figure 18, while the deviations of the majority of test outputs from test targets are greater than the deviations of the corresponding comparative forecasts by WIND database (which are first released as late as December 2007) from test targets in Figure 15, which signifies that the prediction performance of the trained NARX neural network for forecasting average exchange rate of USD/CNY is the best, the prediction performance of the trained NARX neural network for forecasting year-on-year growth of RMB loans of financial institutions is the second best, and the prediction performance of the trained NARX neural networks for forecasting 1-year time deposit rate and 6-month to 1-year short-term loan interest rate are the third best, while the prediction performance of the trained NARX neural network for forecasting year-on-year growth of M2 is the worst, when taking the comparative forecasts by WIND database as criteria for performance evaluation. **The main reason for the above significantly different prediction performance of the specific NARX neural networks for forecasting different financial indicators of China is that the exchange rate related abundant and comprehensive exogenous inputs provide the maximum amount of and the most relevant, systematic and comprehensive information for the forecasts of average exchange rate of USD/CNY, the RMB loans related abundant and comprehensive exogenous inputs provide the information of the second best quality for the forecasts of year-on-year growth of RMB loans of financial institutions, and either the deposit rate related or loan interest rate related abundant and comprehensive exogenous inputs provide the information of the third best quality for the forecasts of 1-year time deposit rate or 6-month to 1-year short-term loan interest rate respectively, while the M2-related limited and partial exogenous inputs are the least numerous and provide the minimum amount of and the least systematic and comprehensive information for the forecasts of year-on-year growth of M2.**

As a summary of the case study on China, it is found that **the significantly different fitting performance and prediction performance among the specific NARX neural networks for forecasting different macroeconomic or financial indicators of China are closely related with and most likely caused by whether the forecast target related abundant and comprehensive exogenous inputs provide the sufficient amount of and the sufficiently relevant, systematic and comprehensive information for the forecasts of the corresponding macroeconomic or financial indicators, or the forecast target related limited and partial exogenous inputs only provide the insufficient amount of and the insufficiently relevant, incomplete and partial information for the forecasts of the corresponding macroeconomic or financial indicators**, it is also found that **if the formation mechanism of some simpler indicators (for example financial indicators) is less complicated than that of other more complex indicators (for example macroeconomic indicators), and their influencing factors are also less numerous and simpler than those of more complex indicators, then the fitting performance and prediction performance of the specific NARX neural networks for forecasting simpler indicators can be as good as or even better than those of the specific NARX neural networks for forecasting more complex indicators, even if only based on fewer number of but the most relevant exogenous inputs that can provide the sufficient amount of and the sufficiently relevant, systematic and comprehensive information for the forecasts of the corresponding simpler indicators**.

## 4.2. Application of NARX Neural Networks for Macroeconomic Forecasting: Case Study on US

In this subsection, we apply specific NARX neural networks for macroeconomic forecasting to predict the main macroeconomic indicators and indices of US that are regularly predicted at a lower time frequency (yearly) by International Monetary Fund (IMF) and released in its World Economic Outlook (WEO). Here in the case study on US **we take another way of selecting available leading business and economic indicators & indices (LBEIs) related to target indicators and indices waiting for prediction, first we are going to build a large set of LBEIs covering all major aspects of the macro economy of US (such as national accounts,** **production, government policy and governance performance, labor market, inflation, domestic trade, foreign trade, business environment, financial market, etc.), and then this large set of LBEIs is adopted to predict all the following major macroeconomic indicators and indices of US, rather than only using those relevant LBEIs to predict different indicators or indices.**

- Year-on-Year Growth of Current Price GDP or Nominal GDP with Seasonal Adjustment (Converted from Quarterly to Monthly)
- Year-on-Year Growth of Chain Linked 2012 Price GDP or Real GDP with Seasonal Adjustment (Converted from Quarterly to Monthly)
- Year-on-Year Growth of General Government Revenue (Converted from Quarterly to Monthly)
- Year-on-Year Growth of General Government Expense (Converted from Quarterly to Monthly)
- Unemployment Rate with Seasonal Adjustment from Current Population Survey
- Year-on-Year Growth of Urban Consumer Price Index (CPI) with Seasonal Adjustment
- Year-on-Year Growth of Exports of Goods, Services & Income with Seasonal Adjustment (Converted from Quarterly to Monthly)
- Year-on-Year Growth of Imports of Goods, Services & Income with Seasonal Adjustment (Converted from Quarterly to Monthly)
- Balance of Current Account (CA) with Seasonal Adjustment as a Percentage of Nominal GDP with Seasonal Adjustment (Converted from Quarterly to Monthly)

All available LBEIs that cover all major aspects of the macro economy of US and can be incorporated into the large exogenous inputs set are listed as follows.

- PMIs: including overall PMI, PMI on production, PMI on new orders, PMI on backlog of orders, PMI on supplier deliveries, PMI on inventories, PMI on employment, PMI on prices, PMI on new export orders, PMI on imports, and PMI on customers inventories, which are all monthly indices released by Institute for Supply Management (ISM);
- Non-Manufacturing Indices (NMIs): involving overall NMI, NMI on business activity, NMI on new orders, NMI on backlog of orders, NMI on supplier deliveries, NMI on new export orders, NMI on imports, NMI on inventory change, NMI on inventory sentiment, NMI on prices, and NMI on employment, which are also monthly indices released by ISM;

- Manufacturing operating rate and non-manufacturing operating rate, which are still monthly indices released by ISM;
- Sentix Economic Indicators: containing overall sentix economic indicator, sentix economic indicator on current situation from headline opinions, sentix economic indicator on expectation from headline opinions, sentix economic indicator on current situation from private investors' opinions, sentix economic indicator on expectation from private investors' opinions, sentix economic indicator on current situation from institutional investors' opinions, and sentix economic indicator on expectation from institutional investors' opinions, which are all monthly indicators released by the Sentix-project as one of the leading surveys of investor opinion;
- Economic Optimism Indices: including overall economic optimism index, economic optimism index for 6-month outlook, economic optimism index for personal financial outlook, and economic optimism index for federal policies, which are all monthly indices released by Investor's Business Daily (IBD) and Techno Metrica Market Intelligence (TIPP);
- Chicago Fed National Activity Index (CFNAI): involving overall CFNAI, CFNAI on production & income, CFNAI on employment, unemployment & hours, CFNAI on personal consumption & housing, CFNAI on sales, orders & inventories, 3-month moving average CFNAI, and diffusion index for CFNAI, which are all monthly indices released by US Federal Reserve Bank of Chicago;
- State Coincident Indices: containing overall state coincident index, 1-month diffusion index for state coincident index, 3-month diffusion index for state coincident index, and state coincident indices for all states, which are all monthly indices released by US Federal Reserve Bank of Philadelphia;
- State Leading Indices: including overall state leading index, and state leading indices for all states, which are all monthly indices released by US Federal Reserve Bank of

Philadelphia;

- National Federation of Independent Business (NFIB) Small Business Optimism Indices (SBOIs): involving overall SBOI, SBOIs on outlook for expansion, SBOIs on outlook for general business condition, SBOIs on earnings, SBOIs on sales, SBOIs on prices, SBOIs on employment, SBOIs on compensation, SBOIs on credit conditions, SBOIs on inventories, SBOIs on capital outlay, and SBOIs on most pressing problems, which are all monthly indices with seasonal adjustments and released by US National Federation of Independent Business (NFIB) Research Foundation;

- CEO Economic Outlook Index, which is a quarterly index released by US Business Roundtable and converted from quarterly to monthly;

- Recession Probability Indices: containing recession probability, National Bureau of Economic Research (NBER) recorded recession, recession probability based on 10-year treasury yield curve, recession probability based on 3-month treasury yield curve, and recession probability based on yield spread curve, which are all monthly indices released by US Federal Reserve Bank of New York;

- Financial Stress Index, which is a weekly index released by US Federal Reserve Bank of St. Louis and converted from weekly to monthly;

- Bank Lending Practice Indices from Senior Loan Officer Opinion Survey: including indices on demand for commercial & industrial loans, indices on reason for strong demand for commercial & industrial loans, indices on reason for weak demand for commercial & industrial loans, indices on lending policies for large & middle market firms, indices on lending policies for small firms, indices on lending policies' reason for credit tightening, indices on lending policies' reason for credit easing, indices on demand for consumer loans, indices on lending policies for individual credit cards, indices on lending policies for new and used autos, indices on lending policies for other consumer loans except credit cards & autos, indices on demand for commercial real estate loans, indices on lending policies for commercial real estate loans, indices on revolving home equity, and indices on residential mortgage loans, which are all quarterly indices released by US Federal Reserve Board and converted from quarterly to monthly;

- Smoothed CEIC leading indicator, which is a monthly indicator released by the world famous CEIC Database;
- Country Governance Indicators: involving country governance indicator on government effectiveness, country governance indicator on regulatory quality, country governance indicator on rule of law, country governance indicator on control of corruption, country governance indicator on voice and accountability, and country governance indicator on political stability and absence of violence/terrorism, which are all annual indicators estimated by World Bank (WB) and converted from yearly to monthly.

From Figure 19 to Figure 27 with descriptions similar to that of Figure 6, first we focus on and compare the fitting performance of the specific NARX neural networks for forecasting different macroeconomic indicators and indices of US. It is found that the deviations of all training outputs from the training targets are small enough to be ignored in almost all of the above figures except Figure 21, while the deviations of training outputs from the training targets at those time points where the training targets have risen or fallen sharply are not small enough to be ignored in Figure 21, which means that all of the trained NARX neural networks for forecasting other macroeconomic indicators and indices of US except year-on-year growth of general government revenue provide sufficiently accurate fits to the training targets that are much better than the comparative forecasts by IMF (although the annual time frequency is low, the prediction accuracy is sufficient), while the trained NARX neural network for forecasting year-on-year growth of general government revenue can only provide insufficiently accurate fits to the training targets that are assessed to be slightly better than the comparative forecasts by IMF. **The main reasons for the above significantly different fitting performance between the specific NARX neural network for forecasting year-on-year growth of general government revenue and the specific NARX neural networks for predicting other macroeconomic indicators and indices of US are not as singular as in case studies on China, these differences are not only caused by the fact that the large set of LBEIs or the large exogenous inputs set contains smaller number of LBEIs associated with general government revenue of US than LBEIs associated with other macroeconomic indicators and indices of US, and provides smaller amount of and less systematic and comprehensive information for the fitting of year-on-year growth of general** **government revenue than for the fitting of other macroeconomic indicators and indices of US, but also caused by the fact that the true values of year-on-year growth of general government revenue at those time points where the training targets have risen or fallen sharply are extremely difficult to be fitted by the specific NARX neural network, and this type of volatility is particularly large and discontinuous when applying the specific NARX neural network to forecast year-on-year growth of general government revenue of US among all macroeconomic indicators and indices of US.**

**Taking the forecasts of year-on-year growth of real GDP as an example, an additional comparison between the case study on China and the case study on US suggests that the large set of LBEIs covering all major aspects of the macro economy of US really helps to make the fitting performance of the specific NARX neural network for forecasting year-on-year growth of real GDP much better than a small set of LBEIs closely related only to China's GDP, as illustrated by the comparison of both the upper left subgraph and the coefficient of determination $R^2$ above that subgraph between Figure 28 and Figure 29.**

Next our concerns transfer from fitting performance to prediction performance of the specific NARX neural networks for forecasting different macroeconomic indicators and indices of US in Figure 19 to Figure 27. It is found that the deviations of the majority of test outputs from test targets are smaller than the deviations of the corresponding comparative forecasts by IMF from test targets in Figure 25 and Figure 26, and the deviations of the majority of test outputs from test targets are at least as small as the deviations of the corresponding comparative forecasts by IMF from test targets in Figure 19, Figure 20, Figure 22, Figure 23, Figure 24 and Figure 27, while the deviations of the majority of test outputs from test targets are greater than the deviations of the corresponding comparative forecasts by IMF from test targets in Figure 21, which signifies that the prediction performance of the trained NARX neural networks for forecasting year-on-year growth of exports of goods, services & income and year-on-year growth of imports of goods, services & income are the best, the prediction performance of the trained NARX neural networks for forecasting year-on-year growth of nominal GDP, year-on-year growth of real GDP, year-on-year growth of general government expense, unemployment rate, year-on-year growth of CPI, and balance of current account as a percentage of nominal GDP are the second best, while the prediction performance of the trained NARX neural network for forecasting year-on-year growth of general government revenue is the worst, when taking the comparative forecasts by IMF as criteria for prediction performance evaluation. **The main reasons for the above significantly different prediction performance of the specific NARX neural networks for forecasting different macroeconomic indicators and indices of US are also not that singular as in case studies on China, these differences are not only caused by the fact that the large set of LBEIs or the large exogenous inputs set provides the maximum amount of and the most relevant, systematic and comprehensive information for the forecasts of year-on-year growth of exports of goods, services & income and year-on-year growth of imports of goods, services & income, and provides the information of the second best quality for the forecasts of year-on-year growth of nominal GDP, year-on-year growth of real GDP, year-on-year growth of general government expense, unemployment rate, year-on-year growth of CPI, and balance of current account as a percentage of nominal GDP, while only provides the minimum amount of and the least systematic and comprehensive information for the forecasts of year-on-year growth of general government revenue, but also caused by the fact that the true values of year-on-year growth of general government revenue at those time points where the training targets have risen or fallen sharply are extremely difficult to be accurately predicted by the specific NARX neural network, and the fluctuation of time series of year-on-year growth of general government revenue is the largest and most discontinuous among time series of all macroeconomic indicators and indices of US, and also caused by the fact that the monthly frequency fluctuations of time series of both year-on-year growth of exports of goods, services & income and year-on-year growth of imports of goods, services & income usually keep the monthly frequency test targets the furthest from the annual frequency comparative forecasts by IMF among time series of all macroeconomic indicators and indices of US, which make the monthly frequency forecasts of year-on-year growth of exports of goods, services & income and year-on-year growth of imports of goods, services & income the easiest to be much more accurate than the annual frequency forecasts by IMF.**

  **And an additional comparison between the case study on China and the case study on US similarly suggests that the large set of LBEIs covering all major aspects of the macro**

economy of US really makes the prediction performance of the specific NARX neural network for forecasting year-on-year growth of real GDP much better than a small set of LBEIs closely related only to China's GDP, as demonstrated by the comparison of both the upper right subgraph and the coefficient of determination $R^2$ above that subgraph between Figure 28 and Figure 29.

As a summary of the case study on US, it is found that **the significantly different fitting performance and prediction performance among the specific NARX neural networks for forecasting different macroeconomic indicators and indices of US are not only most likely caused by the amount of relevant, systematic and comprehensive information for the forecasts of the corresponding macroeconomic indicator or index provided by the large set of LBEIs covering all major aspects of the macro economy of US, but also closely related with whether those large and discontinuous fluctuations of time series of some macroeconomic indicator or index make the forecasts of the specific NARX neural network extremely difficult to be sufficiently accurate at those time points where the time series of the corresponding macroeconomic indicator or index have risen or fallen sharply, and also closely related with the monthly frequency fluctuations of time series of some macroeconomic indicator or index that keep the monthly frequency test targets far away from the annual frequency comparative forecasts by IMF usually make the monthly frequency forecasts of the specific NARX neural network easier to be much more accurate than the annual frequency comparative forecasts by IMF. An additional comparison between the case study on China and the case study on US suggests that the large set of LBEIs covering all major aspects of a country's macro economy really helps to make the prediction performance of the specific NARX neural network for forecasting a certain macroeconomic indicator or index of that country much better than a small set of LBEIs closely related only to that macroeconomic indicator or index.**

## 4.3. Application of NARX Neural Networks for Macroeconomic Forecasting: Case Study on Eurozone

Since the large set of LBEIs used for the forecasts of main macroeconomic indicators and indices of US not only just contains national level LBEIs, but also involves a small number of state level LBEIs, such as state coincident indices for all states of US, and state leading indices for all states of US, then in this subsection, through the case study on Eurozone we want to take our research one step further and answer the following further question:

**Whether adding those subdivision area related LBEIs on the basis of the whole area or global area related LBEIs can help to make the prediction performance of the specific NARX neural network for forecasting some macroeconomic indicator or index of the whole area or global area better?**

All major macroeconomic indicators and indices of Eurozone waiting for prediction are listed as follows.

- Year-on-Year Growth of Current Price GDP or Nominal GDP with Seasonal & Working Days Adjustment (Converted from Quarterly to Monthly)
- Year-on-Year Growth of Chain Linked 2015 Price GDP or Real GDP with Seasonal & Working Days Adjustment (Converted from Quarterly to Monthly)
- General Government Revenue as a Percentage of Nominal GDP (Converted from Quarterly to Monthly)
- General Government Expenditure as a Percentage of Nominal GDP (Converted from Quarterly to Monthly)
- Unemployment Rate with Seasonal Adjustment
- Year-on-Year Growth of Harmonised Index of Consumer Prices (HICP)
- Year-on-Year Growth of Exports with Seasonal & Working Days Adjustment
- Year-on-Year Growth of Imports with Seasonal & Working Days Adjustment
- Balance of Current Account (CA) as a Percentage of Nominal GDP (Converted from Quarterly to Monthly)

Before applying the specific NARX neural networks to forecast the above macroeconomic indicators and indices of Eurozone, we need to build two large sets of LBEIs covering all major aspects of the macro economy of Eurozone (such as national accounts, production, industrial situation, labor market, inflation, domestic trade, foreign trade, business environment, financial market, etc.), one only contains those whole Eurozone related LBEIs, and we simply call it the overall regional exogenous inputs set, the other includes both those whole Eurozone related LBEIs and those country-specific LBEIs, and it is called the overall regional & country-specific exogenous inputs set. Both whole Eurozone related LBEIs and country-specific LBEIs are listed together as follows.

- Survey of Professional Forecasters (SPF) Indicators: involving SPF inflation for current calendar year, SPF inflation for next calendar year, SPF inflation for long term, SPF real GDP growth for current calendar year, SPF real GDP growth for next calendar year, SPF real GDP growth for long term, SPF unemployment rate for current calendar year, SPF unemployment rate for next calendar year, and SPF unemployment rate for long term, which are all quarterly indicators released by European Central Bank (ECB) and converted from quarterly to monthly;
- Business Climate Indicator with seasonal adjustment, which is a monthly indicator released by European Commission's Directorate-General for Economic and Financial Affairs;
- Economic Sentiment Indices with seasonal adjustment: including economic sentiment index for Euro Area 19, and economic sentiment indices for all 19 countries in Euro Area, which are all monthly indices with seasonal adjustment and released by European Commission's Directorate-General for Economic and Financial Affairs;
- Monthly Industrial Confidence Indicators (ICIs) with seasonal adjustment: containing overall ICI for Euro Area 19, overall ICIs for all 19 countries in Euro Area, ICI on production trend observed in recent months for Euro Area 19, ICIs on production trend observed in recent months for all 19 countries in Euro Area, ICI on assessment of order book for Euro Area 19, ICIs on assessment of order book for all 19 countries in Euro Area, ICI on assessment of export order book for Euro Area 19, ICIs on assessment of export order book for all 19 countries in Euro Area, ICI on assessment of stocks of finished products for Euro Area 19, ICIs on assessment of stocks of finished products for all 19 countries in Euro Area, ICI on production expectations for the months ahead for Euro Area 19, ICIs on production expectations for the months ahead for all 19 countries in Euro Area, ICI on selling price expectations for the months ahead for Euro Area 19, ICIs on selling price expectations for the months ahead for all 19 countries in Euro Area, ICI on employment expectations for the months ahead for Euro Area 19, ICIs on employment expectations for the months ahead for all 19 countries in Euro Area, ICIs on consumer goods for all 19 countries in Euro Area, ICIs on durable consumer goods for all 19 countries in Euro Area, ICIs on non-durable consumer goods for all 19 countries in Euro Area, ICIs on intermediate goods for all 19 countries in Euro Area, and ICIs on investment goods for all 19 countries in Euro Area, which are all monthly indicators with seasonal adjustment and released by European Commission's Directorate-General for Economic and Financial Affairs;

- Quarterly Industrial Confidence Indicators (ICIs) with seasonal adjustment: involving ICI on no factors limiting the production for Euro Area 19, ICI on demand factors limiting the production for Euro Area 19, ICI on labour factors limiting the production for Euro Area 19, ICI on equipment factors limiting the production for Euro Area 19, ICI on financial factors limiting the production for Euro Area 19, ICI on other factors limiting the production for Euro Area 19, ICI on assessment of current production capacity for Euro Area 19, ICIs on assessment of current production capacity for all 19 countries in Euro Area, ICI on duration of production assured by current order book level for Euro Area 19, ICIs on duration of production assured by current order book level for all 19 countries in Euro Area, ICI on new orders in recent months for Euro Area 19, ICIs on new orders in recent months for all 19 countries in Euro Area, ICI on export expectations for the months ahead for Euro Area 19, ICIs on export expectations for the months ahead for all 19 countries in Euro Area, ICI on competitive position in domestic market for Euro Area 19, ICIs on competitive position in domestic market for all 19 countries in Euro Area, ICI on competitive position inside EU for Euro Area 19, ICIs on competitive position inside EU for all 19 countries in Euro Area, ICI on competitive position outside EU for Euro Area 19, and ICIs on competitive position outside EU for all 19 countries in Euro Area, which are all quarterly indicators with seasonal adjustment released by European Commission's Directorate-General for Economic and Financial Affairs, and converted from quarterly to monthly;

- Service Confidence Indicators (SCIs) with seasonal adjustment: including overall SCI for Euro Area 19, overall SCIs for all 19 countries in Euro Area, SCI on business situation development over the past 3 months for Euro Area 19, SCIs on business situation development over the past 3 months for all 19 countries in Euro Area, SCI on evolution of the demand over the past 3 months for Euro Area 19, SCIs on evolution of the demand over the past 3 months for all 19 countries in Euro Area, SCI on expectation of the demand over the next 3 months for Euro Area 19, SCIs on expectation of the demand over the next 3 months for all 19 countries in Euro Area, SCI on evolution of the employment over the past 3 months for Euro Area 19, SCIs on evolution of the employment over the past 3 months for all 19 countries in Euro Area, SCI on expectations of the employment over the next 3 months for Euro Area 19, SCIs on expectations of the employment over the next 3 months for all 19 countries in Euro Area, SCI on expectations of the prices over the next 3 months for Euro Area 19, and SCIs on expectations of the prices over the next 3 months for all 19 countries in Euro Area, which are all monthly indicators with seasonal adjustment and released by European Commission's Directorate-General for Economic and Financial Affairs;
- Sentix Economic Indicators: containing overall sentix economic indicator, sentix economic indicator on current situation from headline opinions, sentix economic indicator on expectation from headline opinions, sentix economic indicator on current situation from private investors' opinions, sentix economic indicator on expectation from private investors' opinions, sentix economic indicator on current situation from institutional investors' opinions, and sentix economic indicator on expectation from institutional investors' opinions, which are all monthly indicators released by the Sentix-project as one of the leading surveys of investor opinion;
- Smoothed CEIC leading indicator, which is a monthly indicator released by the world famous CEIC Database;
- Business Environment Indices: involving business extent of disclosure index, depth of credit information index, distance to frontier score, strength of legal rights index, labour tax and contributions as a percentage of commercial profits, profit tax as a percentage of commercial profits, and other taxes payable by businesses as a percentage of commercial profits, which are all annual indices released by World Bank (WB) and converted from yearly to monthly.

Figure 30, Figure 32, Figure 34, Figure 36, Figure 38, Figure 40, Figure 42, Figure 44 and Figure 46 show those specific NARX neural networks' forecast results for the above 9 major macroeconomic indicators and indices of Eurozone on the basis of the overall regional exogenous inputs set, as a contrast, Figure 31, Figure 33, Figure 35, Figure 37, Figure 39, Figure 41, Figure 43, Figure 45 and Figure 47 present the corresponding specific NARX neural networks' forecast results for the above 9 major macroeconomic indicators and indices of Eurozone on the basis of the overall regional & country-specific exogenous inputs set. After still adopting the comparative forecasts by IMF as criteria for both fitting performance and prediction performance evaluation, the differences of fitting performance and prediction performance of those specific NARX neural networks for forecasting major macroeconomic indicators and indices of Eurozone based on different exogenous inputs sets can be simply and intuitively expressed as follows.

**Both the fitting performance and prediction performance of the specific NARX neural network for forecasting any one of the 9 major macroeconomic indicators and indices of Eurozone based on the overall regional & country-specific exogenous inputs set are far better than those based on the overall regional exogenous inputs set.** To be more specific, the deviations of most training outputs from the training targets are not small enough to be ignored in Figure 30, Figure 32, Figure 34, Figure 36, Figure 38, Figure 40, Figure 42, Figure 44 and Figure 46 (all of them show those specific NARX neural networks' forecast results on the basis of the overall regional exogenous inputs set), while the deviations of almost all training outputs from the training targets are small enough to be ignored in Figure 31, Figure 33, Figure 35, Figure 37, Figure 39, Figure 41, Figure 43, Figure 45 and Figure 47 (all of them show those specific NARX neural networks' forecast results on the basis of the overall regional & country-specific exogenous inputs set), which means that all of the trained NARX neural networks based on the overall regional & country-specific exogenous inputs set provide sufficiently accurate fits to the training targets that are much better than those of the trained NARX neural networks based on the overall regional exogenous inputs set. It's similarly found that the deviations of the majority of test

outputs from test targets are smaller than or at most as small as the deviations of the corresponding comparative forecasts by IMF from test targets in Figure 31, Figure 33, Figure 35, Figure 37, Figure 39, Figure 41, Figure 43, Figure 45 and Figure 47 (all of them show forecast results based on the overall regional & country-specific exogenous inputs set), while the deviations of the majority of test outputs from test targets are greater than or at least as large as the deviations of the corresponding comparative forecasts by IMF from test targets in Figure 30, Figure 32, Figure 34, Figure 36, Figure 38, Figure 40, Figure 42, Figure 44 and Figure 46 (all of them show forecast results based on the overall regional exogenous inputs set), which signifies that all of the trained NARX neural networks based on the overall regional & country-specific exogenous inputs set give more accurate predictions that are closer to the true values in the future than the comparative forecasts by IMF, while those trained NARX neural networks based on the overall regional exogenous inputs set give less accurate predictions that are further away from the true values in the future than the comparative forecasts by IMF.

**The main reason for the above significantly different fitting performance and prediction performance of those specific NARX neural networks for forecasting 9 major macroeconomic indicators and indices of Eurozone can only fall on the differences between the overall regional exogenous inputs set and the overall regional & country-specific exogenous inputs set. Since not only those whole Eurozone related LBEIs contained in the overall regional exogenous inputs set, but also those country-specific LBEIs involved in the overall regional & country-specific exogenous inputs set can help to provide relevant, systematic and comprehensive information for the forecasts of major macroeconomic indicators and indices of Eurozone, then the overall regional & country-specific exogenous inputs set is surely to provide the larger amount of relevant, more systematic and more comprehensive information for the forecasts of major macroeconomic indicators and indices of Eurozone than the overall regional exogenous inputs set.**

As a summary of the case study on Eurozone, we can answer the question raised at the beginning of this subsection, it is suggested that **not only those whole area or global area related LBEIs, but also those subdivision area related LBEIs can provide relevant, systematic and comprehensive information for the forecasts of major macroeconomic** **indicators and indices of the whole area or global area, then using both types of LBEIs can help to make the fitting performance and prediction performance of the specific NARX neural network for forecasting some macroeconomic indicator or index of the whole area or global area much better than only adopting the former-those whole area or global area related LBEIs.**

## 5. Application of NARX Neural Networks for National Goal Setting: A Comparative Case Study on Russia

We have already discussed in subsection 3.1 that the only difference between the specific NARX neural network for macroeconomic forecasting and the specific NARX neural network for national goal setting is that **available exogenous inputs required by the latter not only comprise those available leading business and economic indicators & indices (LBEIs) that are relevant to national goals waiting for prediction, as suggested in subsection 4.1, 4.2 and 4.3, but also contain those major and conventional macroeconomic statistical indicators & indices (MCMSIs) that are also related to these national goals**. In addition, the relevance for these national goals here can be understood in a broad sense, then selecting all available LBEIs and MCMSIs covering all major aspects of both the macro economy and the market or industry segmentation is also suggested, in order to further promote the prediction performance of the specific NARX neural networks for the forecasts of major national goals.

Since we still want to simply and intuitively explore **how those limited & partial exogenous inputs or abundant & comprehensive exogenous inputs specifically influence the fitting performance and prediction performance of those specific NARX neural networks for national goal setting**, we still build two exogenous inputs sets for each of those selected national goals of Russia, one only contains a small number of MCMSIs that are most relevant to the specific national goal, and we simply call it the limited & most relevant exogenous inputs set, the other involves all available LBEIs and MCMSIs covering all major aspects of both the macro economy and the market or industry segmentation of Russia (such as national accounts, flow of funds, production, government and public finance, demographic characteristics and labour market, domestic trade and household situation, inflation, foreign trade, balance of payments, monetary data, interest and foreign exchange rates, investment, financial market, etc.), and we simply call it the abundant & comprehensive exogenous inputs set, here the abundant & comprehensive exogenous inputs set is the same for all selected national goals of Russia.

We apply the specific NARX neural network for national goal setting to the 14 most important national goals of Russia, among them 13 national goals set by Ministry of Economic Development (MED) of the Russian Federation are listed as follows.

- Target for Year-on-Year Growth of Real GDP (Converted from Yearly to Monthly)
- Target for GDP Deflator (Converted from Yearly to Monthly)
- Target for Year-on-Year Growth of Labor Productivity (Converted from Yearly to Monthly)
- Target for Unemployment Rate (Converted from Yearly to Monthly)
- Target for Year-on-Year Growth of Real Wage (Converted from Yearly to Monthly)
- Target for Year-on-Year Growth of Real Disposable Income (Converted from Yearly to Monthly)
- Target for Year-on-Year Growth of Paid Services Rendered to Population (Converted from Yearly to Monthly)
- Target for Year-on-Year Growth of Year Average CPI (Converted from Yearly to Monthly)
- Target for Year-on-Year Growth of Exports (Converted from Yearly to Monthly)
- Target for Year-on-Year Growth of Imports (Converted from Yearly to Monthly)
- Target for Current Account (Unit Billion USD, Converted from Yearly to Monthly)
- Target for Capital Inflow (+) /Outflow (-) (Unit Billion USD, Converted from Yearly to Monthly)
- Target for Year Average Foreign Exchange Rate RUB/USD (Converted from Yearly to Monthly)

In addition, one more national goal that is set by Central Bank of the Russian Federation (CBRF) is as follows.

- Target for Inflation (Converted from Yearly to Monthly)

The limited & most relevant exogenous inputs sets for each of the above 14 national goals are specifically listed as follows.

The limited & most relevant exogenous inputs set around the target for year-on-year growth of real GDP only contains the following unique statistical indicator:

- Year-on-year growth of 2016 constant price GDP, which is quarterly released by Russia Federal State Statistics Service (RFSSS) and converted from quarterly to monthly.

The limited & most relevant exogenous inputs set around the target for GDP deflator includes the following single statistical indicator:

- Year-on-year growth of GDP deflator, which is quarterly released by RFSSS and converted from quarterly to monthly.

The limited & most relevant exogenous inputs set around the target for year-on-year growth of labor productivity involves the following unique statistical indicator:

- Year-on-year growth of labour productivity, which is yearly released by RFSSS and converted from yearly to monthly.

The limited & most relevant exogenous inputs set around the target for unemployment rate includes the following two statistical indicators:

- Extended history unemployment rate that is monthly released by the world famous CEIC Database;
- Unemployment rate with seasonal & working days adjustment that is monthly released by RFSSS.

The limited & most relevant exogenous inputs set around the target for year-on-year growth of real wage involves the following four statistical indicators and indices:

- Year-on-year growth of labour force demand, nominal wages index (same month previous year=100), and real wages index (same month previous year=100), which are all monthly released by RFSSS;
- Year-on-year growth of official minimal wages per month that is monthly released by Ministry of Labour and Social Security of the Russian Federation (MLSSRF).

The limited & most relevant exogenous inputs set around the target for year-on-year growth of real disposable income contains the following three statistical indices:

- Real household income index (same quarter previous year=100), real disposable income index (same quarter previous year=100), and household income per capita index (same quarter previous year=100), which are all quarterly released by RFSSS and converted from quarterly to monthly.

The limited & most relevant exogenous inputs set around the target for year-on-year growth of paid services rendered to population only contains the following unique statistical indicator:

- Paid services rendered to population (same month previous year=100), which is monthly released by RFSSS.

The limited & most relevant exogenous inputs set around the target for inflation involves the following four statistical indicators and indices:

- Median value of inflation expectations that is monthly released by CBRF;
- Year-on-year growth of CPI (2010=100), core inflation for CPI (same month previous year=100), and PPI (same month previous year=100) calculated in accordance with new All-Russia classification of products by kinds of economic activities (OKVED2), which are all monthly released by RFSSS.

The limited & most relevant exogenous inputs set around the target for year-on-year growth of year average CPI includes the following two statistical indices:

- Year-on-year growth of CPI (2010=100), and core inflation for CPI (same month previous year=100), which are all monthly released by RFSSS.

The limited & most relevant exogenous inputs set around the target for year-on-year growth of exports involves the following three statistical indicators and indices:

- Year-on-year growth of exports that is monthly released by Russia Federal Customs Service (RFCS);
- Export volume index (same month previous year=100), and export unit value index (same month previous year=100), which are all monthly released by Ministry of Economic Development of the Russian Federation (MEDRF).

The limited & most relevant exogenous inputs set around the target for year-on-year growth of imports contains the following three statistical indicators and indices:

- Year-on-year growth of imports that is monthly released by RFCS;
- Import volume index (same month previous year=100), and import unit value index (same month previous year=100), which are all monthly released by MEDRF.

The limited & most relevant exogenous inputs set around the target for current account (unit billion USD) involves the following four statistical indicators:

- Current account with seasonal adjustment (unit million USD), analytical presentation of current account (unit million USD), and neutral presentation of current account (unit million USD), which are all quarterly released by CBRF and converted from quarterly to monthly;
- Analytical presentation of current account (unit billion USD), which is monthly released by CBRF.

The limited & most relevant exogenous inputs set around the target for capital inflow (+) /outflow (-) (unit billion USD) contains the following three statistical indicators:

- Capital account (unit million USD), financial account (unit million USD), and net capital inflow (+) /outflow (-) of private sector (unit billion USD), which are all quarterly released by CBRF and converted from quarterly to monthly.

The limited & most relevant exogenous inputs set around the target for year average foreign exchange rate RUB/USD includes the following three statistical indicators and indices:

- Monthly average exchange rate RUB/USD, which is monthly released by CBRF;
- Nominal effective exchange rate index for RUB (2010=100), and real effective exchange rate index for RUB, which are all monthly released by Bank for International Settlements (BIS).

All available LBEIs and MCMSIs contained in the identical Abundant & Comprehensive Exogenous Inputs Set prepared for all selected national goals of Russia are specifically listed as follows, according to the module to which they belong.

**Module 1. National Accounts Indicators & Indices**

Year-on-year growth of nominal GDP, year-on-year growth of final consumption in nominal GDP, year-on-year growth of final consumption by households, year-on-year growth of final consumption by government, year-on-year growth of final consumption by non-profit institutions serving households, year-on-year growth of gross capital formation, year-on-year growth of gross fixed capital formation, year-on-year growth of exports, year-on-year growth of imports, year-on-year growth of gross value added, year-on-year growth of taxes on products, year-on-year growth of subsides on products, and year-on-year growth of net taxes on products;

Year-on-year growth of 2016 constant price GDP, year-on-year growth of 2016 constant price final consumption in 2016 constant price GDP, year-on-year growth of 2016 constant price final consumption by households, year-on-year growth of 2016 constant price final consumption by government, year-on-year growth of 2016 constant price final consumption by non-profit institutions serving households, year-on-year growth of 2016 constant price gross capital formation, year-on-year growth of 2016 constant price gross fixed capital formation, year-on-year growth of 2016 constant price exports, year-on-year growth of 2016 constant price imports, year-on-year growth of 2016 constant price gross value added, year-on-year growth of 2016 constant price taxes on products, year-on-year growth of 2016 constant price subsides on products, and year-on-year growth of 2016 constant price net taxes on products;

Year-on-year growth of average earning of employees, year-on-year growth of net taxes on products and import, year-on-year growth of gross profit of economy and gross mixed income, and year-on-year growth of GDP deflator;

All of the above indicators in this module are quarterly released by RFSSS and converted from quarterly to monthly.

Year-on-year growth of GDP per capita, year-on-year growth of gross national income, year-on-year growth of current transfers receivable from rest of world, year-on-year growth of current transfers payable to rest of world, year-on-year growth of gross national disposable income, year-on-year growth of gross savings, year-on-year growth of gross savings by non-financial corporations, year-on-year growth of gross savings by financial corporations, year-on-year growth of gross savings by general government, year-on-year growth of gross savings by households, and year-on-year growth of gross savings by non-profit institutions serving households, which are all yearly released by RFSSS and converted from yearly to monthly.

2016 constant price output index for key economic activities (same month previous year=100), which is monthly released by RFSSS.

**Module 2. Flow of Funds Indicators & Indices**

Year-on-year growth of financial balance of the total economy, year-on-year growth of financial assets of the total economy, year-on-year growth of financial liabilities of the total economy, year-on-year growth of financial balance of financial corporations, year-on-year growth of financial assets of financial corporations, year-on-year growth of financial liabilities of financial corporations, year-on-year growth of financial balance of non-financial corporations, year-on-year growth of financial assets of non-financial corporations, year-on-year growth of financial liabilities of non-financial corporations, year-on-year growth of financial balance of general government, year-on-year growth of financial assets of general government, year-on-year growth of financial liabilities of general government, year-on-year growth of financial balance of households & non-profit institutions serving households, year-on-year growth of financial assets of households & non-profit institutions serving households, year-on-year growth of financial liabilities of households & non-profit institutions serving households, year-on-year growth of financial balance of rest of the world, year-on-year growth of financial assets of rest of the world, and year-on-year growth of financial liabilities of rest of the world, which are all quarterly released by CBRF and converted from quarterly to monthly.

**Module 3. Production Indicators & Indices**

Year-on-year growth of industrial production index, year-on-year growth of industrial production index with seasonal adjustment, year-on-year growth of agricultural production index, and year-on-year growth of orders value of all industries, which are all monthly released by RFSSS.

**Module 4. Government and Public Finance Indicators & Indices**

Year-on-year growth of consolidated government revenue, year-on-year growth of consolidated government expenditure, year-on-year growth of consolidated government budget balance, year-on-year growth of consolidated government financing, year-on-year growth of consolidated internal government financing, and year-on-year growth of consolidated external government financing, which are all monthly released by Russia Federal Treasury (RFT);

Year-on-year growth of consolidated government tax revenue, which is monthly released by Federal Tax Service of Russia (FTSR);

Year-on-year growth of domestic debt, which is monthly released by Ministry of Finance of the Russian Federation (MFRF).

**Module 5. Demographic and Labour Market Indicators & Indices**

Year-on-year growth of period average population, year-on-year growth of working age population, year-on-year growth of life expectancy at birth, and year-on-year growth of labour productivity, which are all yearly released by RFSSS and converted from yearly to monthly;

Extended history unemployment rate, which is monthly released by the world famous CEIC Database;

Economic activity rate among age 15 and above, employment rate among age 15 and above, unemployment rate among age 15 and above, unemployment rate among age 15 to 72, unemployment rate with seasonal & working days adjustment, year-on-year growth of labour force demand, year-on-year growth of period average nominal wages, nominal wages index (same month previous year=100), real wages index (same month previous year=100), and year-on-year growth of average weekly hours worked per employee among age 15 and above for main job, which are all monthly released by RFSSS;

Year-on-year growth of official minimal wages per month, which is monthly released by MLSSRF.

**Module 6. Domestic Trade and Household Survey Indicators & Indices**

Year-on-year growth of retail trade turnover, year-on-year growth of wholesale trade turnover, year-on-year growth of trade turnover, paid services rendered to population (same month previous year=100), year-on-year growth of paid personal services, and year-on-year growth of public catering turnover, which are all monthly released by RFSSS;

Retail trade confidence indicator (RTCI), RTCI on actual economic situation, RTCI on forecast economic situation, RTCI on actual storage stock, wholesale trade confidence indicator, consumer confidence index, consumer expectation index (CEI) on current economic situation, CEI on changes occurred in economic situation, CEI on expecting changes in economic situation for short term perspective, CEI on expecting changes in price, CEI on expecting changes in number of unemployed, CEI on financial situation of household, CEI on changes occurred in financial situation of household, CEI on expecting changes in financial situation of household, CEI on conditions favourable for major purchase, CEI on conditions favourable for money savings, real household income index (same quarter previous year=100), real disposable income index (same quarter previous year=100), household income per capita index (same quarter previous year=100), and year-on-year growth of average household income per capita per month, which are all quarterly released by RFSSS and converted from quarterly to monthly;

Year-on-year growth of retail trade turnover per capita, year-on-year growth of household expenditures & savings, and year-on-year growth of household expenditure per capita, which are all yearly released by RFSSS and converted from yearly to monthly;

Median value of inflation expectations, which is monthly released by CBRF;

Year-on-year growth of household disposable money income, which is yearly released by CBRF and converted from yearly to monthly.

### Module 7. Inflation Indicators & Indices

Year-on-year growth of CPI (2010=100), CPI of food (same month previous year=100), CPI of non food (same month previous year=100), CPI of food & non food (same month previous year=100), CPI of services (same month previous year=100), core inflation for CPI (same month previous year=100), and PPI (same month previous year=100) calculated in accordance with new All-Russia classification of products by kinds of economic activities (OKVED2), which are all monthly released by RFSSS.

### Module 8. Foreign Trade Indicators & Indices

Year-on-year growth of exports, year-on-year growth of imports, and year-on-year growth of foreign trade turnover, which are all monthly released by RFCS;

Foreign trade balance (unit million USD), which is monthly released by CBRF;

Export volume index (same month previous year=100), import volume index (same month previous year=100), export unit value index (same month previous year=100), and import unit value index (same month previous year=100), which are all monthly released by MEDRF.

### Module 9. Balance of Payments Indicators & Indices

Current account with seasonal adjustment (unit million USD), capital account (unit million USD), financial account (unit million USD), neutral presentation of current account (unit million USD), neutral presentation of goods & services in current account (unit million USD), neutral presentation of primary & secondary income in current account (unit million USD), neutral presentation of capital account (unit million USD), neutral presentation of balance from current & capital account (unit million USD), neutral presentation of financial account (unit million USD), neutral presentation of reserve assets (unit million USD), assets for foreign claims and liabilities (unit million USD), liabilities for foreign claims and liabilities (unit million USD), net international investment position for foreign claims and liabilities (unit million USD), assets for international investment position (unit million USD), liabilities for international investment position (unit million USD), reserve assets for international investment position (unit million USD), external debt (unit million USD), external debt of general government (unit million USD), external debt of monetary authorities (unit million USD), external debt of banks (unit million USD), external debt of other sectors (unit million USD), short term external debt (unit million USD), long term external debt (unit million USD), foreign currency external debt (unit million USD), domestic currency external debt (unit million USD), net capital inflow (+) /outflow (-) of private sector (unit billion USD), inward foreign direct investments (unit million USD), and outward foreign direct investments (unit million USD), which are all quarterly released by CBRF and converted from quarterly to monthly;

Analytical presentation of current account (unit billion USD), analytical presentation of capital account (unit billion USD), analytical presentation of balance from current & capital account (unit billion USD), analytical presentation of financial account except reserve assets (unit billion USD), and analytical presentation of change in reserve assets (unit billion USD), which are all monthly released by CBRF.

**Module 10. Monetary Indicators & Indices**

Year-on-year growth of monetary base, year-on-year growth of currency issued, year-on-year growth of M2 supply, year-on-year growth of M1 supply, year-on-year growth of M0 supply, year-on-year growth of broad money M2x supply, year-on-year growth of broad money M2x supply with seasonal adjustment, year-on-year growth of official reserve assets, required reserve ratio for national currency liabilities to non-residents, required reserve ratio for foreign currency liabilities to non-residents, required reserve ratio for national currency liabilities to individuals, and required reserve ratio for foreign currency liabilities to individuals, which are all monthly released by CBRF;

Year-on-year growth of number of personal banking cards, year-on-year growth of number of personal transactions through banking cards, year-on-year growth of amount of personal transactions through banking cards, year-on-year growth of number of corporate's banking cards, year-on-year growth of number of corporate's transactions through banking cards, and year-on-year growth of amount of corporate's transactions through banking cards, which are all quarterly released by CBRF and converted from quarterly to monthly.

**Module 11. Interest and Foreign Exchange Rates Indicators & Indices**

Month end policy rate, central bank bonds rate, 10-year GKO-OFZ government bond zero coupon yield as key long term interest rate, central bank repo rate, 1 to 3-month short term interest rate, short term national currency deposit rate for households, short term national currency lending rate for households, short term national currency lending rate for non-financial institutions, short term national currency lending rate for mortgage loans, short term foreign currency deposit rate for households, short term foreign currency lending rate for households, short term foreign currency lending rate for non-financial institutions, short term foreign currency lending rate for mortgage loans, up to 1-year deposit rate for individuals, over 1-year deposit rate for individuals, up to 1-year deposit rate for non-financial organisations, over 1-year deposit rate for non-financial organisations, up to 1-year lending rate for individuals, over 1-year lending rate for individuals, up to 1-year lending rate for non-financial organisations, over 1-year lending rate for non-financial organisations, monthly average exchange rate RUB/USD by central bank, monthly average exchange rate RUB/EUR by central bank, period end exchange rate of RUB against dual-currency basket of USD & EUR by central bank, and period end exchange rate RUB/SDR by central bank, which are all monthly released by CBRF;

Nominal effective exchange rate index for RUB (2010=100), and real effective exchange rate index for RUB, which are all monthly released by Bank for International Settlements (BIS);

Period average official foreign exchange rate RUB/USD, period average official foreign exchange rate RUB/EUR, period average official foreign exchange rate RUB/SDR, nominal effective exchange rate index for RUB (2010=100), and real effective exchange rate index for RUB based on CPI (2010=100), which are all monthly released by IMF.

**Module 12. Investment Indicators & Indices**

Year-on-year growth of fixed capital investment, and year-on-year growth of financial investment, which are all quarterly released by RFSSS and converted from quarterly to monthly;

Year-on-year growth of number of registered new enterprises, year-on-year growth of number of registered Russian ownership new enterprises, year-on-year growth of number of registered foreign ownership new enterprises, year-on-year growth of number of registered joint Russian & foreign ownership new enterprises, year-on-year growth of number of liquidated enterprises, year-on-year growth of number of Russian ownership liquidated enterprises, year-on-year growth of number of foreign ownership liquidated enterprises, year-on-year growth of number of joint Russian & foreign ownership liquidated enterprises, balance (profit less loss) of big & medium enterprises (same period previous year=100), balance (profit less loss) for scientific research & development activities of big & medium enterprises (same period previous year=100), year-on-year growth of profit amount of big & medium enterprises, year-on-year growth of profit amount for scientific research & development activities of big & medium enterprises, year-on-year growth of loss amount of big & medium enterprises, and year-on-year growth of loss amount for scientific research & development activities of big & medium enterprises, which are all monthly released by RFSSS.

**Module 13. Financial Market Indicators & Indices**

Month end Russian Trading System (RTS) equity market index, RTS market capitalization index, RTS turnover rate index for classia & standard markets, Moscow Exchange (MOEX) Russia index, Moscow Interbank Currency Exchange (MICEX) 10 index, blue chip index, blue chip market capitalization index, and blue chip turnover rate index, which are all monthly released by Moscow Exchange;

Year-on-year growth of deposit certificates issued by credit institutions, year-on-year growth of saving certificates issued by credit institutions, year-on-year growth of bonds issued by credit institutions, and year-on-year growth of derivatives at fair value issued by credit institutions, which are all monthly released by CBRF;

Year-on-year growth of insurance premium of all contracts except mandatory medical, and year-on-year growth of insurance benefit of all contracts except mandatory medical, which are all quarterly released by CBRF and converted from quarterly to monthly;

Russia share price index, which is monthly released by IMF.

**Module 14. Leading Business & Economic Indicators and Indices (LBEIs)**

Russian Economic Barometer Indices: containing expected percentage of enterprises with rising sales prices in the next 3 months, expected percentage of enterprises with rising purchasing prices in the next 3 months, expected percentage of enterprises with rising wages in the next 3 months, expected percentage of enterprises with rising employment in the next 3 months, expected percentage of enterprises with rising production in the next 3 months, expected percentage of enterprises with rising equipment purchase in the next 3 months, expected percentage of enterprises with improving financial situation in the next 3 months, expected percentage of enterprises with rising orders in the next 3 months, expected percentage of enterprises with rising debt to banks in the next 3 months, actual percentage of enterprises with rising sales prices over 1 month, actual percentage of enterprises with rising purchasing prices over 1 month, actual percentage of enterprises with rising wages over 1 month, actual percentage of enterprises with rising employment over 1 month, actual percentage of enterprises with rising production over 1 month, actual percentage of enterprises with rising orders over 1 month, actual percentage of enterprises with rising stocks over 1 month, actual percentage of enterprises with rising sales/purchasing prices ratio over 1 month, actual percentage of enterprises with rising equipment purchase over 1 month, actual percentage of enterprises with improving credit conditions over 1 month, enterprises indicator (EI) for actual capacity utilisation rate (normal monthly level=100), EI for actual labour utilisation rate (normal monthly level=100), EI for actual stocks (normal monthly level=100), EI for actual orders (normal monthly level=100), EI for enterprises debt to banks (normal monthly level=100), percentage of enterprises in good or normal financial situation, percentage of enterprises not buying equipment for 2 months and more, interest rates on RUB bank loans that attract enterprises in the next 3 months, percentage of enterprises without debt to banks & not expected to have debt to banks in the next 3 months, and percentage of enterprises not going to take banking loans in the next 3 months, which are all monthly indicators released by Institute of World Economy and International Relations, Russian Academy of Sciences (IMEMO RAS);

Vnesheconombank Monthly Monitoring Indices: including overall 1-month Semaphon index (Semaphon 1M), Semaphon 1M for macro economy, Semaphon 1M for industry, Semaphon 1M for assessments, overall 12-month Semaphon index (Semaphon 12M), Semaphon 12M for macro economy, Semaphon 12M for industry, Semaphon 12M for assessments, and Vnesheconombank monthly GDP index as a leading index of quarterly GDP, which are all monthly indices released by Vnesheconombank of Russia (VEB);

Higher School of Economics (HSE) Composite Indices: involving HSE leading index, HSE coinciding index, and HSE lagging index, which are all monthly indices released by Higher School Of Economics Russia (HSER);

Entrepreneur Confidence Indicators (ECIs): including ECI on mining & quarrying, ECI on manufacturing, and ECI on electricity, gas, steam & air conditioning supply, which are all monthly indicators released by RFSSS;

Smoothed CEIC leading indicator, which is a monthly indicator released by the world famous CEIC Database;

Business Environment Indices: involving business extent of disclosure index, depth of credit information index, distance to frontier score, strength of legal rights index, index for time to obtain an electrical connection, index for power outages in firms in a typical month, firms experiencing electrical outages as a percentage of firms, firms expected to give gifts in meetings with tax officials as a percentage of firms, labour tax and contributions as a percentage of commercial profits, profit tax as a percentage of commercial profits, and other taxes payable by businesses as a percentage of commercial profits, which are all annual indices released by World Bank (WB) and converted from yearly to monthly;

Country Governance Indicators: involving country governance indicator (CGI) on government effectiveness, CGI on regulatory quality, CGI on rule of law, CGI on control of corruption, CGI on voice and accountability, and CGI on political stability and absence of violence/terrorism, which are all annual indicators estimated by World Bank (WB) and converted from yearly to monthly.

Based on all the above preparations, those specific NARX neural networks' forecast results for the above 14 national goals of Russia on the basis of the corresponding Limited & Most Relevant Exogenous Inputs Sets are presented in Figure 48, Figure 50, Figure 52, Figure 54, Figure 56, Figure 58, Figure 60, Figure 62, Figure 64, Figure 66, Figure 68, Figure 70, Figure 72, and Figure 74 respectively, as a comparison, the corresponding specific NARX neural networks' forecast results for the above 14 national goals of Russia on the basis of the identical Abundant & Comprehensive Exogenous Inputs Set are shown in Figure 49, Figure 51, Figure 53, Figure 55, Figure 57, Figure 59, Figure 61, Figure 63, Figure 65, Figure 67, Figure 69, Figure 71, Figure 73, and Figure 75 respectively. Then the differences of fitting performance and prediction performance of those specific NARX neural networks for setting national goals of Russia based on different exogenous inputs sets can be simply and intuitively expressed as follows.

**The fitting performance of the specific NARX neural network for setting any one of the 14 major national goals of Russia based on the abundant & comprehensive exogenous inputs set is almost the same as that based on the limited & most relevant exogenous inputs set, in other words, at least there is no conspicuously visible difference between them. As a contrast, the prediction performance of the specific NARX neural network for setting any one of the 14 major national goals of Russia based on the abundant & comprehensive exogenous inputs set is always far better than that based on the limited & most relevant exogenous inputs set.**

More specifically, the deviations of all training outputs from the training targets are small enough to be ignored in Figure 48, Figure 52, Figure 56, Figure 58, Figure 62, Figure 64, Figure 66, Figure 68, Figure 70, Figure 72, and Figure 74 (all of them show those specific NARX neural networks' forecast results on the basis of the corresponding Limited & Most Relevant Exogenous Inputs Sets), and the deviations of all training outputs from the training targets are also small enough to be ignored in Figure 49, Figure 51, Figure 53, Figure 55, Figure 57, Figure 59, Figure 61, Figure 63, Figure 65, Figure 67, Figure 69, Figure 71, Figure 73, and Figure 75 (all of them show those specific NARX neural networks' forecast results on the basis of the identical Abundant & Comprehensive Exogenous Inputs Set), while the deviations of most training outputs from the training targets are not small enough to be ignored in Figure 50, Figure 54 and Figure 60 (all of them show those specific NARX neural networks' forecast results on the basis of the corresponding Limited & Most Relevant Exogenous Inputs Sets), which means that the majority of the trained NARX neural networks based on the corresponding Limited & Most Relevant Exogenous Inputs Sets can provide sufficiently accurate fits to the training targets that are close enough to the true values of most of the corresponding national goals issued by the authorities, except those for setting the target for GDP deflator, the target for unemployment rate, and the target for year-on-year growth of paid services rendered to population, while all of the trained NARX neural networks based on the identical Abundant & Comprehensive Exogenous Inputs Set provide sufficiently accurate fits to the training targets that are close enough to the true values of all the corresponding national goals issued by the authorities. **The main reason for the above poor fitting performance of the specific NARX neural network for setting the target for GDP deflator is that the GDP deflator goals issued by the authorities are usually much lower than the true values of GDP deflator in the future, since the authorities are usually too optimistic about the country's overall price level; the foremost reason for the poor fitting performance of the specific NARX neural network for setting the target for unemployment rate is that the extended history unemployment rate and the unemployment rate with seasonal & working days adjustment in the limited & most relevant exogenous inputs set are usually inconsistent on trends, then the conflicting information about trends provided by them is usually going to lower rather than promote the fitting performance of this specific NARX neural network; the primary cause for the poor fitting performance of the specific NARX neural network for setting the target for year-on-year growth of paid services rendered to population is that extracting the annual average trend as the national goal from the high frequency and excessive fluctuations of the single indicator-paid services rendered to population (same month previous year=100) is too difficult and almost infeasible for this specific NARX neural network, besides, even the moving average of this single indicator in the limited & most relevant exogenous inputs set is not entirely consistent with the annual trend of the national goal to be fitted.**

  The findings on prediction performance of those specific NARX neural networks for setting the 14 major national goals of Russia are simpler and more consistent than the above findings on fitting performance. The average deviations of all test outputs from test targets in Figure 48, Figure 50, Figure 52, Figure 54, Figure 56, Figure 58, Figure 60, Figure 62, Figure 64, Figure 66, Figure 68, Figure 70, Figure 72, and Figure 74 (all of them show forecast results on the basis of the corresponding Limited & Most Relevant Exogenous Inputs Sets) are far smaller than the average deviations of all test outputs from test targets in Figure 49, Figure 51, Figure 53, Figure 55, Figure 57, Figure 59, Figure 61, Figure 63, Figure 65, Figure 67, Figure 69, Figure 71, Figure 73, and Figure 75 (all of them show forecast results on the basis of the identical Abundant & Comprehensive Exogenous Inputs Set) respectively, which signifies that all of the trained NARX neural networks based on the identical Abundant & Comprehensive Exogenous Inputs Set give far more accurate predictions that are much closer to those major national goals issued by the authorities than those trained NARX neural networks based on the corresponding Limited & Most Relevant Exogenous Inputs Sets. **The main reason for the above significantly different prediction performance of those specific NARX neural networks for setting major national goals of Russia can only fall on the differences between the corresponding Limited & Most Relevant Exogenous Inputs Sets and the identical Abundant & Comprehensive Exogenous Inputs Set. Since not only a small number of MCMSIs that are most relevant to the specific national goal in the corresponding Limited & Most Relevant Exogenous Inputs Set, but also all available LBEIs and MCMSIs covering all major aspects of both the macro economy and the market or industry segmentation of Russia in the identical Abundant & Comprehensive Exogenous Inputs Set are capable of providing relevant, systematic and comprehensive information for the forecasts of the 14 major national goals of Russia, then the identical Abundant & Comprehensive Exogenous Inputs Set is always capable of providing the larger amount of, directly and indirectly relevant, more systematic and more comprehensive information for the settings of the 14 major national goals of Russia than those corresponding Limited & Most Relevant Exogenous Inputs Sets.**

As a summary of the comparative case study on Russia, it is found that **only a small number of MCMSIs that are most relevant to the specific national goal are usually enough to help the trained NARX neural network provide sufficiently accurate fits to the true values of this national goal issued by the authorities, however, they are still not enough to help this trained NARX neural network provide sufficiently accurate predictions that are close enough to the true values of this national goal, therefore, it's suggested that we adopt all available LBEIs and MCMSIs covering all major aspects of both the macro economy and the market or industry segmentation of this country, which are usually capable of providing the larger amount of, directly and indirectly relevant, more systematic and more comprehensive information for the setting of a major national goal, to support the trained NARX neural** network to give sufficiently accurate forecasts that are much closer to the true values of this national goal.

## 6. Application of NARX Neural Networks for Global Competitiveness Assessment: Comparative Case Studies on Global Competitiveness Index

The global competitiveness assessment is in fact the prediction of the Global Competitiveness Index (GCI) of a specific economy, through conducting comparative studies on the application of NARX neural networks for the forecasts of GCIs of various economies, we want to simply and intuitively explore the following two questions:

**Whether the specific NARX neural network trained on the basis of the GCI related data of some economies can make sufficiently accurate predictions about GCIs of other economies?**

**Whether the specific NARX neural network trained on the basis of the data of some type of economies can give more accurate predictions about GCIs of the same type of economies than those of different type of economies?**

Then our specific comparative research ideas are as follows:

**Step 1. Both China and US are chosen as the test economies, the former is the representative of the emerging and developing economies, while the latter is the representative of the advanced economies.**

**Step 2. The first specific NARX neural network for global competitiveness assessment is trained on the basis of the GCI related data of all advanced economies except US (it's generally assumed that this neural network's prediction performance for other advanced economies should be significantly better than its prediction performance for emerging and developing economies), then it is used to predict the GCIs of the two test economies, in order to verify whether the first specific NARX neural network's prediction of the GCI of US is much more accurate than that of the GCI of China.**

Here advanced economies except US involve Australia, Austria, Belgium, Canada, Cyprus, Czech Republic, Denmark, Finland, France, Germany, Greece, Hong Kong SAR (China), Iceland, Ireland, Israel, Italy, Japan, South Korea, Lithuania, Luxembourg, Malta, Netherlands, New

Zealand, Norway, Portugal, Singapore, Slovakia, Slovenia, Spain, Sweden, Switzerland, Taiwan (China), and United Kingdom.

**Step 3. The second specific NARX neural network for global competitiveness assessment is trained on the basis of the GCI related data of all emerging and developing Asian economies except China (it's generally assumed that this neural network's prediction performance for other emerging and developing economies should be significantly better than its prediction performance for advanced economies), then it is still used to predict the GCIs of the two test economies, in order to verify whether the second specific NARX neural network's prediction of the GCI of China is much more accurate than that of the GCI of US.**

Here emerging and developing Asian economies except China include Bangladesh, Bhutan, Brunei, Cambodia, India, Indonesia, Laos, Malaysia, Mongolia, Myanmar, Nepal, Philippines, Sri Lanka, Thailand, Timor-Leste, and Vietnam.

**Step 4. The third specific NARX neural network for global competitiveness assessment is trained on the basis of the GCI related data of both European Union 28 economies and emerging & developing European economies (it's generally assumed that this neural network's prediction performance for emerging and developing economies should be as good as its prediction performance for advanced economies), then it is also used to predict the GCIs of the two test economies, in order to verify whether the third specific NARX neural network's prediction of the GCI of China is as accurate as that of the GCI of US.**

Here European Union 28 economies involve Austria, Belgium, Bulgaria, Croatia, Cyprus, Czech Republic, Denmark, Estonia, Finland, France, Germany, Greece, Hungary, Ireland, Italy, Latvia, Lithuania, Luxembourg, Malta, Netherlands, Poland, Portugal, Romania, Slovakia, Slovenia, Spain, Sweden, and United Kingdom (before 2019); emerging and developing European economies include Albania, Bosnia and Herzegovina, Bulgaria, Croatia, Hungary, North Macedonia, Montenegro, Poland, Romania, Serbia, and Turkey; although overlap economies between European Union 28 economies and emerging & developing European economies lead to data duplication in the training data set, this situation could only improve the fitting performance of the specific NARX neural network on these overlap economies, but hardly affect the prediction performance of the neural network on economies outside the training data set.

It has already been mentioned in subsection 3.1 that the specific NARX neural network for global competitiveness assessment can be considered almost the same as the specific NARX neural network for national goal setting in the previous section, except that **those available leading business and economic indicators & indices (LBEIs) and those major and conventional macroeconomic statistical indicators & indices (MCMSIs) that are selected as exogenous inputs of the specific NARX neural network for global competitiveness assessment should cover more aspects besides the macro economy**, since the global competitiveness assessment is in fact the prediction of the Global Competitiveness Index (GCI), which is a systematic and comprehensive assessment of the corresponding economy's institutions, infrastructure, macroeconomic environment, health and primary education, higher education and training, goods market efficiency, labour market efficiency, financial market development, technological readiness, market size, business sophistication, and innovation by World Economic Forum (WEF).

In order to help the exogenous inputs cover other aspects besides economic performance, such as institutions, infrastructure, etc., and also taking into account that a single source of data can better ensure the consistency of data among economies, we mainly adopt authoritative statistical indicators & indices monthly, quarterly or yearly released by International Monetary Fund (IMF), supplemented by authoritative indicators & indices published by World Bank (WB). Moreover, in order to ensure the frequency & unit consistency of all exogenous inputs of the specific NARX neural networks for forecasting GCIs, not only all monthly or quarterly indicators are converted to yearly, but also all indicators with currency units are converted to dollar unit based on the year average exchange rate released by IMF.

Then all selected indicators & indices that are taken as systematic and comprehensive exogenous inputs of the specific NARX neural networks for forecasting GCIs are listed as follows, according to the module to which they belong.

**Module 1. National Accounts Indicators & Indices**

Current price GDP per capita, household consumption expenditure including non-profit institutions serving households (NPISHs) per capita, government consumption expenditure per capita, gross fixed capital formation per capita, changes in inventories per capita, exports of goods and services per capita, imports of goods and services per capita, gross national income per capita, gross national disposable income per capita, gross capital formation per capita, year-on-year growth of GDP deflator, gross value added of agriculture per capita at factor cost, gross value added of industry per capita at factor cost, and gross value added of manufacturing industry per capita at factor cost are involved, all their units are USD.

**Module 2. Production Indicators & Indices**

Only year-on-year growth of industrial production index as a key indicator estimated for all sample economies is included here.

**Module 3. Demographic and Labour Market Indicators & Indices**

Labour force (unit thousand person), year-on-year growth of employment index, unemployment (unit thousand person), unemployment rate, total population (unit person), female population as a percentage of total population, urbanization rate or urban population as a percentage of total population, labour force participation rate as a percentage of total population aged 15+, ratio of employment to total population aged 15+, ratio of unemployment to total labour force, male unemployment as a percentage of total male labour force, and female unemployment as a percentage of total female labour force collected or estimated for all sample economies are contained.

**Module 4. Inflation Indicators & Indices**

Only year-on-year growth of CPI and year-on-year growth of PPI collected or estimated for all sample economies are included.

**Module 5. Foreign Trade Indicators & Indices**

Value of exports per capita, value of imports per capita, and trade balance per capita are included, all their units are USD, and moreover, year-on-year growth of export volume index, year-on-year growth of import volume index, year-on-year growth of export value index, and year-on-year growth of import value index are also included.

**Module 6. Balance of Payments Indicators & Indices**

Per capita credit of current account, per capita debit of current account, per capita credit of goods in current account, per capita debit of goods in current account, per capita credit of services in current account, per capita debit of services in current account, per capita credit of primary income in current account, per capita debit of primary income in current account, per capita credit of secondary income in current account, per capita debit of secondary income in current account, per capita personal remittances received in current account, per capita personal remittances paid in current account, per capita credit of capital account, per capita debit of capital account, per capita assets of direct investment in financial account, per capita liabilities of direct investment in financial account, per capita assets of portfolio investment in financial account, per capita liabilities of portfolio investment in financial account, per capita financial derivatives & employee stock options in financial account, per capita assets of other investment in financial account, per capita liabilities of other investment in financial account, per capita official reserve assets in financial account, per capita assets for international investment position, and per capita liabilities for international investment position are included, units of all these indicators are USD.

**Module 7. Monetary Indicators & Indices**

Per capita international reserves as international liquidity, per capita gold as international liquidity, net domestic credit per capita collected or estimated for all sample economies are included, all their units are USD, and furthermore, ratio of bank non-performing loans to total gross loans is also included.

**Module 8. Interest and Foreign Exchange Rates Indicators & Indices**

End of period central bank policy rate, period average lending rate, and year-on-year growth of real effective exchange rate index for national currency based on CPI, which are collected or estimated for all sample economies.

**Module 9. Financial Market Indicators & Indices**

Only total value per capita of stocks traded in each economy's financial markets (unit USD) is included.

**Module 10. Telecommunication & Postal Sector Indicators & Indices**

Ratio of mobile cellular subscriptions to total population and ratio of fixed broadband internet subscribers to total population are included.

**Module 11. Tourism Sector Indicators & Indices**

Ratio of number of arrivals for international tourism to total population and ratio of number of departures for international tourism to total population are included.

### Module 12. Transportation Sector Indicators & Indices

Ratio of passengers carried by air transport to total population (unit person/person), ratio of freight by air transport to total population (unit ton-km/person), ratio of goods transported by railways to total population (unit ton-km/person), and ratio of passengers carried by railways to total population (unit person-km/person), which are collected or estimated for all sample economies.

### Module 13. Business Environment Indicators & Indices

Business extent of disclosure index, depth of credit information index, distance to frontier score, strength of legal rights index, index for time to obtain an electrical connection, index for power outages in firms in a typical month, firms experiencing electrical outages as a percentage of firms, firms expected to give gifts in meetings with tax officials as a percentage of firms, labour tax and contributions as a percentage of commercial profits, profit tax as a percentage of commercial profits, and other taxes payable by businesses as a percentage of commercial profits, which are assessed for all of the economies included in the sample.

### Module 14. Country Governance Indicators (CGIs)

CGI on government effectiveness, CGI on regulatory quality, CGI on rule of law, CGI on control of corruption, CGI on voice and accountability, and CGI on political stability and absence of violence/terrorism, which are assessed for all sample economies.

After 80% of all time points for all advanced economies except US were randomly selected for the training of the first specific NARX neural network and the remaining 20% for network performance testing, each sample economy's global competitiveness assessment results from this specific NARX neural network are presented in Figure 76. As we can see, the first specific NARX neural network can provide sufficiently accurate global competitiveness assessment results for each sample economy that are close enough to the actual values of global competitiveness index for the corresponding sample economy, no matter at those time points selected for neural network training or at other time points selected for network performance testing, which means we've already got a high quality NARX neural network trained by data of advanced economies except US. Subsequently, the first specific NARX neural network is applied to exogenous inputs of China and US that contain the same indicators & indices as those of the sample economies, and the global competitiveness assessment results for these two test economies that are directly output from this trained specific NARX neural network are shown in Figure 77. It is found that **the first specific NARX neural network trained by data of advanced economies except US can make sufficiently accurate predictions about GCIs of both China and US, while its predictions about GCIs of US (like the sample economies used for training neural network, they are all advanced economies) are more accurate than those of China (different from the sample economies, China is an emerging and developing economy), more interestingly, the first specific NARX neural network trained by data of advanced economies except US tends to underestimate GCIs of China, but does not overestimate or underestimate GCIs of US**. In fact, all of the above findings are also the answers to those two questions raised at the beginning of this section.

    Based on preparations similar to the first specific NARX neural network, each sample economy's global competitiveness assessment results from the second specific NARX neural network trained by data of emerging and developing Asian economies except China are presented in Figure 78. It is found that the second specific NARX neural network can provide sufficiently accurate global competitiveness assessment results for each sample economy that are close enough to the actual values of global competitiveness index for the corresponding sample economy, both at those time points selected for neural network training and at other time points selected for network performance testing, which means the second NARX neural network trained by data of emerging and developing Asian economies except China has been of high quality. Next, the second specific NARX neural network is applied to exogenous inputs of China and US, and the global competitiveness assessment results for these two test economies shown in Figure 79 are directly output from this trained specific NARX neural network. It is found that **the second specific NARX neural network trained by data of all emerging and developing Asian economies except China can make sufficiently accurate predictions about GCIs of both China and US, while its predictions about GCIs of China (like the sample economies used for training neural network, they are all emerging and developing economies) are more accurate than those of US (different from the sample economies, US is an advanced economy), more interestingly, the second specific NARX neural network trained by data of emerging and**

**developing Asian economies except China tends to slightly overestimate GCIs of China, while significantly underestimate GCIs of US**. Then the two questions raised at the beginning of this section are answered by the above findings.

Since most of the findings about the third specific NARX neural network are similar to those of the previous two specific NARX neural networks, we are going to briefly discuss these findings as follows. It is found from Figure 80 that the third specific NARX neural network trained by data of European Union 28 economies and emerging & developing European economies can provide sufficiently accurate global competitiveness assessment results for each sample economy that are close enough to the actual values of global competitiveness index for the corresponding sample economy, both at those time points selected for neural network training and at other time points selected for network performance testing, which means the third NARX neural network after training has already been of high quality. It is found in Figure 81 that **the third specific NARX neural network trained by data of both European Union 28 economies and emerging & developing European economies can make sufficiently accurate predictions about GCIs of both China and US, while its predictions about GCIs of China (like the emerging & developing European economies used for training neural network, they are all emerging and developing economies) are as accurate as those of US (like most of the European Union 28 economies used for training neural network, they are all advanced economies), more interestingly, the third specific NARX neural network trained by data of both European Union 28 economies and emerging & developing European economies usually tends to slightly overestimate GCIs of China, while slightly underestimate GCIs of US**. Then the two questions raised at the beginning of this section are answered by the above findings again.

As a summary of the comparative case study on Global Competitiveness Index (GCI), we can answer the two questions raised at the beginning of this section with enough certainty now: **the specific NARX neural network trained on the basis of the GCI related data of some economies are usually capable of making sufficiently accurate predictions about GCIs of other economies, as long as the number of sample economies used for training neural network is large enough and the amount of relevant, systematic and comprehensive information provided by their exogenous inputs is sufficiently large; the specific NARX** **neural network trained on the basis of the data of some type of economies usually can give more accurate predictions about GCIs of the same type of economies than those of different type of economies, as long as the differences between the two types of economies reflected in the indicators contained in their exogenous inputs are obvious and large enough, such as the differences between advanced economies and emerging & developing economies.**

## 7. Policy Recommendation

Unlike the human brain which is usually only capable of extracting information from a small number of statistics, and integrating these information inefficiently, unsteadily and unrepeatably to help improve the prediction of the object of interest, the NARX neural network specially constructed for the prediction of the object of interest has the following advantages over the human brain.

1. A NARX neural network has the potential to efficiently and effectively capture the dynamics of almost arbitrary nonlinear dynamic system, since it has been proved in the literature that any multi-dimensional nonlinear mapping of any continuous function can be carried out by a two-layer NARX neural network (one hidden layer and one output layer) with a suitable chosen number of neurons in its hidden layer, as a result, the ability of the NARX neural network to construct almost arbitrary nonlinear dynamic mapping, which takes complex multiple exogenous inputs and past values of the object of interest as independent variables and gives the predictions of the object of interest as dependent variables, is usually much higher than that of human brain.

2. During training, a NARX neural network is capable of automatically and efficiently extracting useful information from not only past values of the time series of both the object of interest and exogenous inputs (treated as independent variables), but also current values of the time series of the object of interest waiting to be fitted (treated as dependent variables), and also capable of automatically and adaptively building reasonable rules between these complex multiple independent variables and dependent variables, through the so called self-learning process, therefore, the fast self-learning ability of the NARX neural network during training is usually much higher than that of human brain.

3. A NARX neural network is good at discovering long-term dependencies between past values of both complex multiple exogenous inputs and the object of interest and current or even future values of the object of interest, in other words, when the gap between the time point where the relevant information appears and the time point where it is needed becomes extremely large, a reasonably constructed NARX neural network is still able to learn to connect the past information to the later output values over an extremely long period of time, which seems as if this NARX neural network has far greater capacity of long-term memory, and far more powerful ability of memory retrieval and memory connection construction than the average human brain.

4. The generalization ability of the specially constructed NARX neural network is usually much higher than that of human brain, in other words, the specially constructed NARX neural network is better at applying those reasonable rules mentioned above between complex multiple independent variables and dependent variables to new independent variables with unseen background and meaning, and sometimes even with unknown data accuracy and noise pollution, than the average human brain.

5. The fault tolerance ability of the specially constructed NARX neural network could be better or at least as good as that of human brain, to be more specific, even if a small number of hidden neurons in the hidden layer of a NARX neural network are damaged, for example the weights in these hidden neurons that are updated through the self-learning process are lost, the self-learning outcome of this NARX neural network after training can usually still be maintained, and its prediction performance is usually not significantly affected.

Of course, the application of the NARX neural network is still immature under some scenarios due to there are no good and general decision criteria for the setting of the architecture of the trained NARX neural network that affects its prediction performance. The most favorable behavior of a specially constructed NARX model is dependent upon the exogenously designated time delay that is especially important for the initial prediction, the selection of time points for training, validation, or testing that may lead to either different fitting performance for the known objects during training or different prediction performance for unknown objects outside of training, and the selection of number of hidden neurons in the hidden layer of this NARX neural network, which may result in the inadequate ability of the NARX neural network to construct highly nonlinear dynamic mapping when this number is too small, on the contrary, could bring about not only a real danger of overtraining the data and producing the false fitting of the known objects during training that does not lead to better forecasts for unknown objects outside of training, but also the excessively low training efficiency under the constraints of existing software and hardware resources, when this number is too large. In general, the determination of these architectural elements, in an optimal way, is a critical and difficult task for the construction of a specific NARX neural network, which has no fixed criteria and needs to be determined based on the specific requirements under the particular scenario.

In this study, **the successful application of specific NARX neural networks in macroeconomic forecasting, national goal setting and global competitiveness assessment demonstrate that the above five advantages of the specially constructed NARX neural networks over the human brain are obvious under the above three application scenarios, while all architectural elements of these specific NARX neural networks can be simply and directly determined as occasion demands**. As a consequence, it's strongly recommended that **government authorities and policy research institutions gradually turn to applying fully trained NARX neural networks that are assessed as qualified to assist or even replace the deductive and inductive abilities of the human brain in the macroeconomic forecasting tasks and the market or industry segmentation forecasting tasks, in the tasks for setting national or regional goals and market or industry targets, also in those tasks focusing on the global or regional or national comparative assessments of the general performance of individuals or the performance of individuals on particular subjects such as policy implementation and government governance**.

## 8. Conclusion

Although the trove of economic big data released by statistical agencies, private and public surveys, and other sources is parsed every day by governments and authorities, economists and market analysts, mostly for the purposes of economic forecasting, the full use of the trove of economic big data through current econometric approaches and other modeling methods is not that easy. However, with the rapid development of neural network methods and their cross-field applications in recent years, certain types of neural networks may be considered as a general method of making full use of the trove of economic data to make predictions.

After selecting the Nonlinear Autoregressive with Exogenous Input (NARX) neural network as the method of this study through systematic and comprehensive literature review, and constructing specific NARX neural networks under specific application scenarios involving macroeconomic forecasting, national goal setting and global competitiveness assessment, this study can concentrate on analyzing how different settings for exogenous inputs from the trove of economic big data affect the prediction performance of NARX neural networks.

The case study on China demonstrates that **the forecast target related abundant and comprehensive exogenous inputs, which provide the sufficient amount of and the sufficiently relevant, systematic and comprehensive information for the forecasts of the corresponding macroeconomic or financial indicators, can further improve the prediction performance of the specific NARX neural networks, compared with the forecast target related limited and partial exogenous inputs.**

The case study on US mainly demonstrates that **the large set of exogenous inputs covering all major aspects of a country's macro economy helps to make the prediction performance of the specific NARX neural network for forecasting a certain macroeconomic indicator or index of that country much better than a small set of exogenous inputs closely related only to that macroeconomic indicator or index.**

The case study on Eurozone argues that **not only those whole area related exogenous inputs, but also those subdivision area related exogenous inputs can provide relevant, systematic and comprehensive information for the forecasts of major macroeconomic indicators and indices of the whole area, then using both whole area and subdivision area related exogenous inputs can help to make the prediction performance of the specific NARX neural network for forecasting a certain macroeconomic indicator or index of the whole area much better than only adopting those whole area related exogenous inputs.**

In the case study on Russia, it is found that **only a small number of major and**

**conventional macroeconomic statistical indicators & indices (MCMSIs) that are most relevant to the specific national goal are usually enough to help the trained NARX neural network provide sufficiently accurate fits to the true values of this national goal issued by the authorities, however, they are still not enough to help this trained NARX neural network provide sufficiently accurate predictions that are close enough to the true values of this national goal**, therefore, this case study suggests that **we adopt all available leading business and economic indicators & indices (LBEIs) and MCMSIs covering all major aspects of both the macro economy and the market or industry segmentation of this country, which are usually capable of providing the larger amount of, directly and indirectly relevant, more systematic and more comprehensive information for the setting of this national goal, to support the trained NARX neural network to give sufficiently accurate forecasts that are much closer to the true values of this national goal.**

The case study on Global Competitiveness Index (GCI) not only demonstrates that **the specific NARX neural network trained on the basis of the GCI related data of some economies are usually capable of making sufficiently accurate predictions about GCIs of other economies, as long as the number of sample economies used for training neural network is large enough and the amount of relevant, systematic and comprehensive information provided by their exogenous inputs is sufficiently large**, but also demonstrates that **the specific NARX neural network trained on the basis of the data of some type of economies usually can give more accurate predictions about GCIs of the same type of economies than those of different type of economies, as long as the differences between the two types of economies reflected in the indicators contained in their exogenous inputs are obvious and large enough (take this case study as an example, the differences between advanced economies and emerging & developing economies)**.

Based on all of the above successful application of specific NARX neural networks in macroeconomic forecasting, national goal setting and global competitiveness assessment, it's strongly recommended that **government authorities and policy research institutions gradually turn to applying fully trained NARX neural networks that are assessed as qualified to assist or even replace the deductive and inductive abilities of the human brain in a variety of** **appropriate tasks**, such as the macroeconomic forecasting tasks and the market or industry segmentation forecasting tasks, the tasks for setting national or regional goals and market or industry targets, and those tasks focusing on the global or regional or national comparative assessments of the general performance of individuals or the performance of individuals on particular subjects such as policy implementation and government governance.

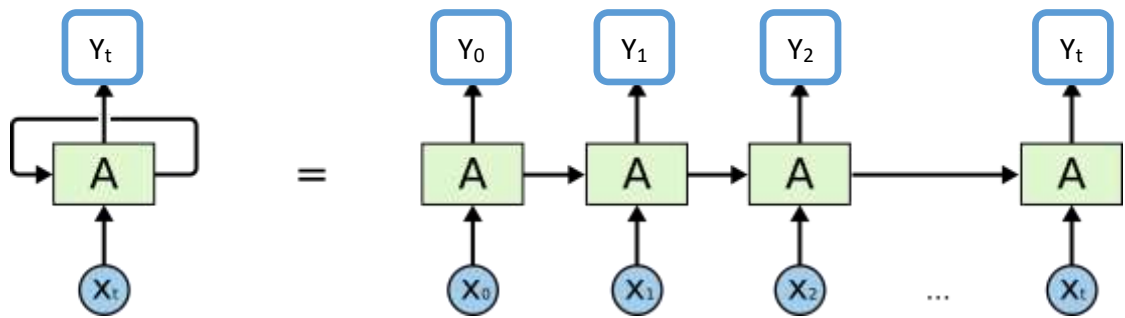

Source: colah's blog, available at http://colah.github.io/posts/2015-08-Understanding-LSTMs/

Notes: The left side of the equals sign is the basic structure of a general Recurrent Neural Network (RNN), while the right side of the equals sign is the unrolled structure of a general RNN.

Figure 1. Both the basic and unrolled structures of a general Recurrent Neural Network (RNN)

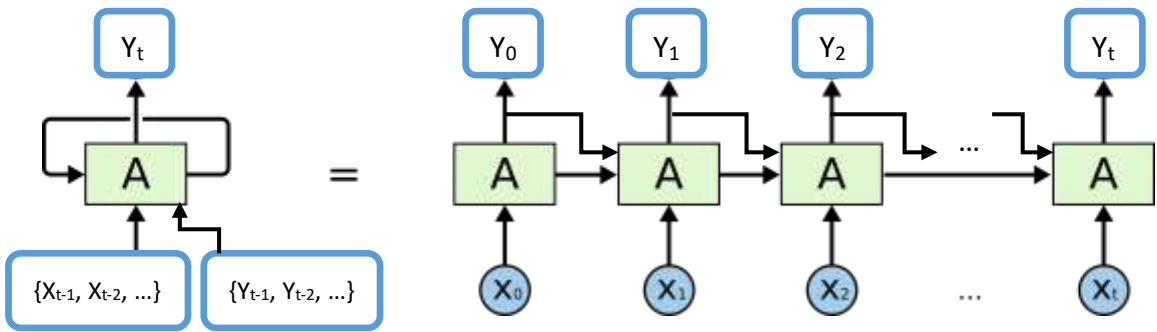

Source: author's drawing based on Figure 1 from colah's blog

Notes: The left side of the equals sign is the basic structure of a general NARX neural network, while the right side of the equals sign is the unrolled structure of a general NARX neural network.

Figure 2. Both the basic and unrolled structures of a general Nonlinear Autoregressive with Exogenous Input (NARX) neural network

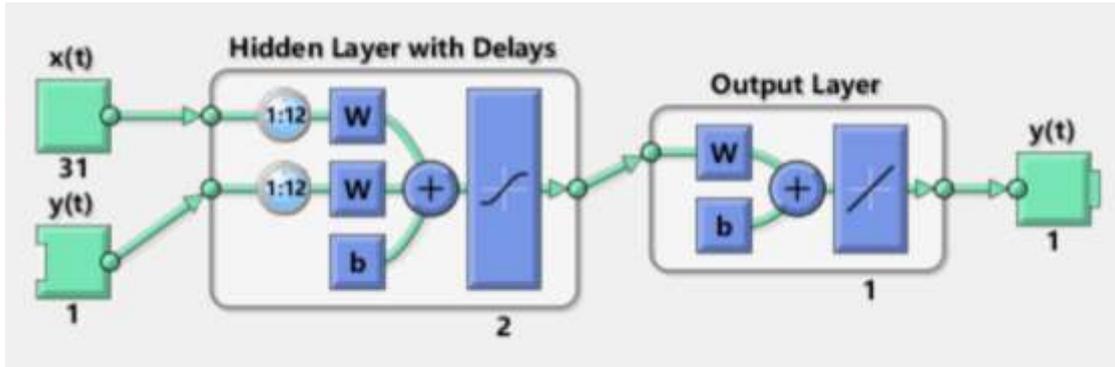

Source: Neural Net Time Series Toolbox of Matlab

Notes: The specific NARX neural network in the above figure is set to have 31 exogenous inputs series $\{x_t, x_{t-1},…\}$, 1 monthly frequency target series $\{y_t, y_{t-1},…\}$, 2 hidden neurons in the hidden layer, and 12-month or 1-year time delays for the dynamic forecasting of the target series.

Figure 3. Example of a specific NARX neural network for monthly frequency forecasting of the target series

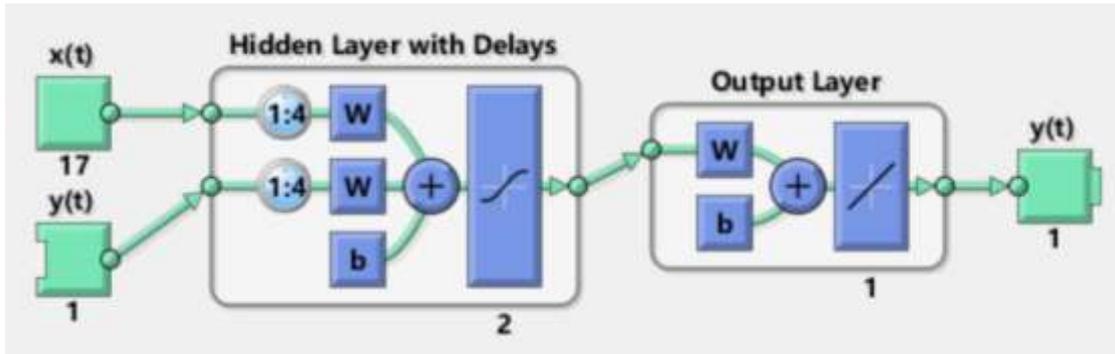

Source: Neural Net Time Series Toolbox of Matlab

Notes: The specific NARX neural network in the above figure is set to have 17 exogenous inputs series $\{x_t, x_{t-1},…\}$, 1 quarterly frequency target series $\{y_t, y_{t-1},…\}$, 2 hidden neurons in the hidden layer, and 4-quarter or 1-year time delays for the dynamic forecasting of the target series.

Figure 4. Example of a specific NARX neural network for quarterly frequency forecasting of the target series

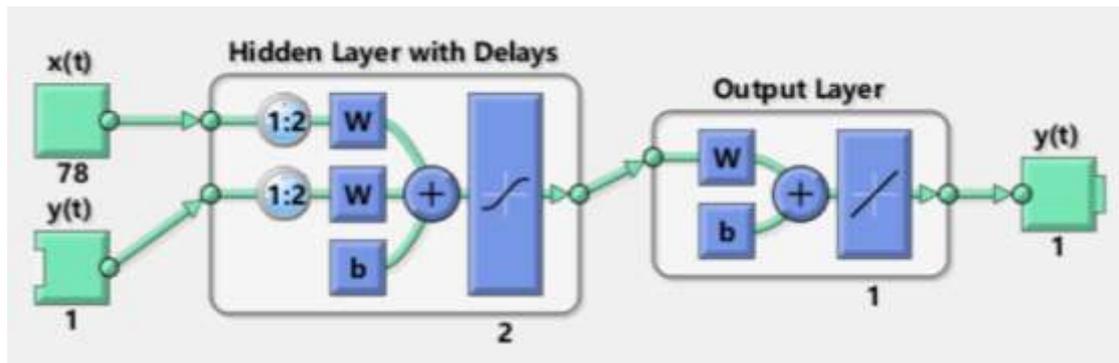

Source: Neural Net Time Series Toolbox of Matlab

Notes: The specific NARX neural network in the above figure is set to have 78 exogenous inputs series $\{x_t, x_{t-1},…\}$, 1 yearly frequency target GCI series $\{y_t, y_{t-1},…\}$, 2 hidden neurons in the hidden layer, and 2-year time delays for the dynamic forecasting of the target GCI series (2-year past values of GCI are considered sufficient to capture the underlying level of a particular country's GCI and its recent growth rate).

Figure 5. Example of a specific NARX neural network for yearly frequency forecasting of Global Competitiveness Index (GCI)

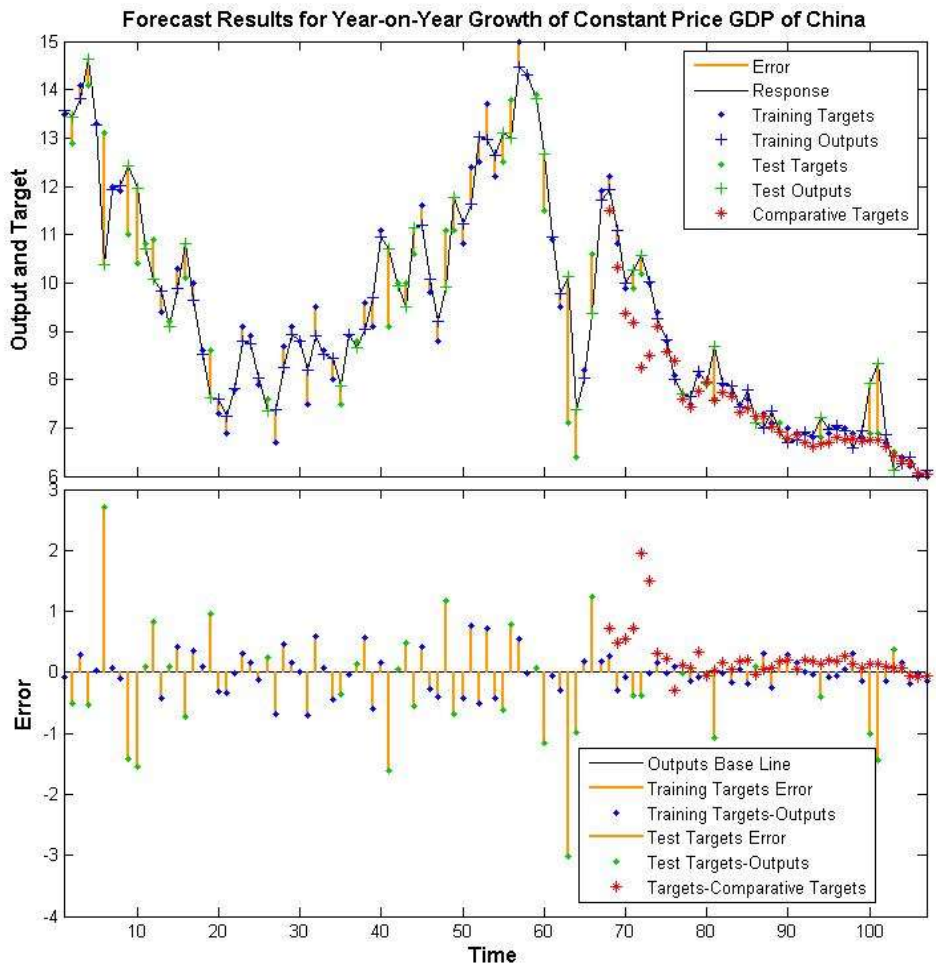

Source: National Bureau of Statistics of China; People's Bank of China; China's most famous financial database WIND; author's calculation

Notes: The upper subgraph displays the specific NARX neural network's outputs, targets, comparative targets and errors versus time, while the lower subgraph uses this neural network's outputs as the benchmark and shows the gaps between targets/comparative targets and outputs versus time. The time on the horizontal axis corresponds to each quarter from the second quarter of 1993 to the fourth quarter of 2019, which is also the time span of the exogenous inputs series or target series minus the initial four quarters as the initial time delays for forecast. It is also indicated that which time points were selected for training and testing in both subgraphs.

Figure 6. Specific NARX neural network's forecast result for year-on-year growth of constant price GDP of China

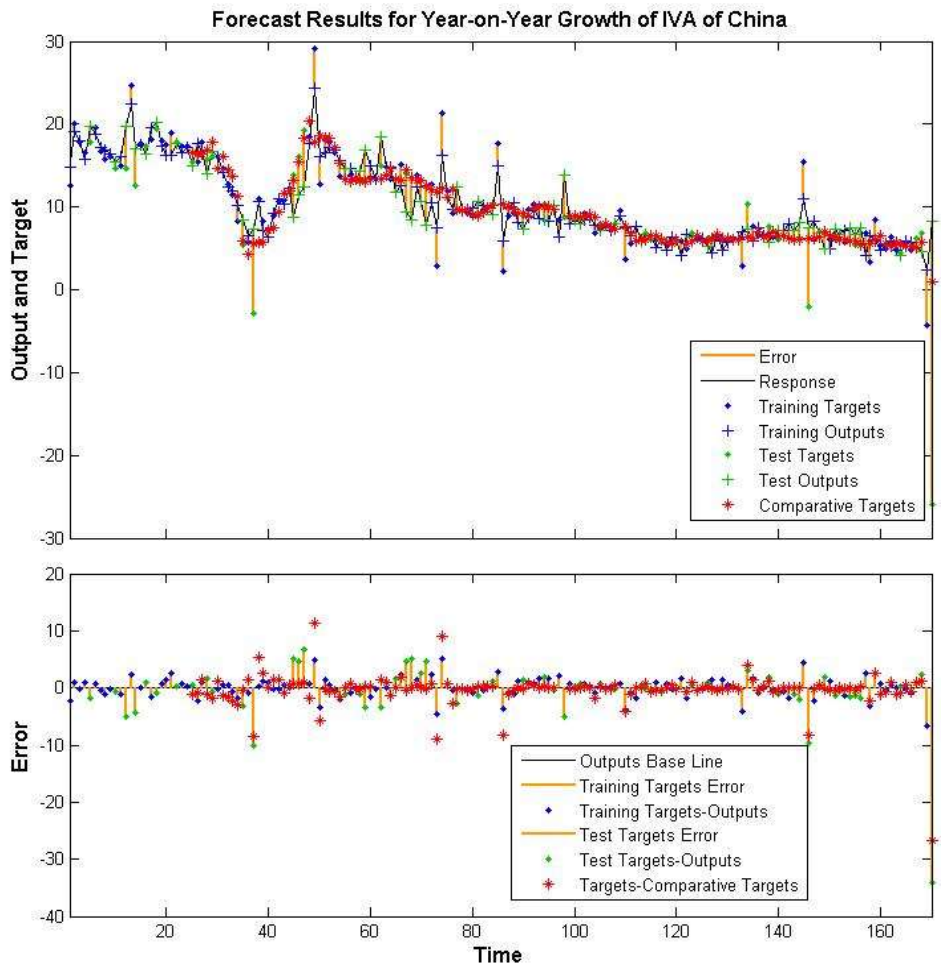

Source: National Bureau of Statistics of China; China's most famous financial database WIND; author's calculation

Notes: The upper subgraph displays the specific NARX neural network's outputs, targets, comparative targets and errors versus time, while the lower subgraph uses this neural network's outputs as the benchmark and shows the gaps between targets/comparative targets and outputs versus time. The time on the horizontal axis corresponds to each month from January 2006 to February 2020, which is also the time span of the exogenous inputs series or target series minus the initial twelve months as the initial time delays for forecast. It is also indicated that which time points were selected for training and testing in both subgraphs.

Figure 7. Specific NARX neural network's forecast result for year-on-year growth of IVA of China

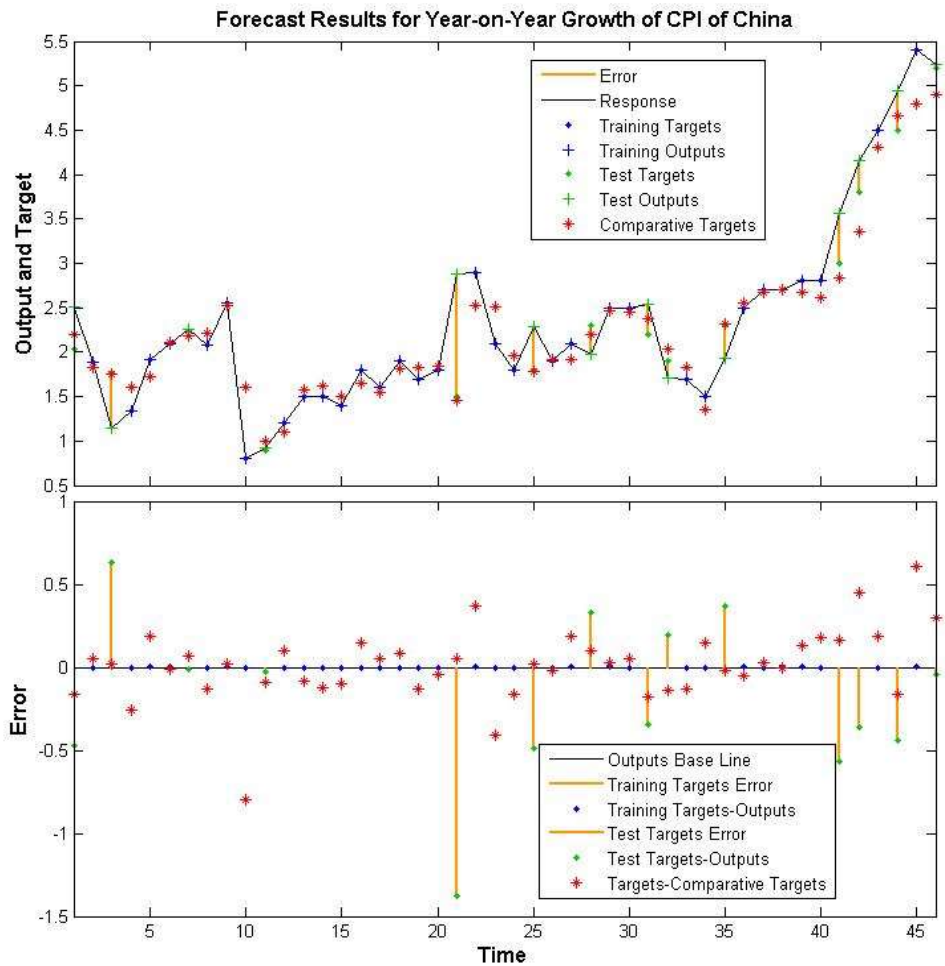

Source: National Bureau of Statistics of China; People's Bank of China; China International Capital Corporation (CICC); Nine Martingale Investment Management LP of China; China's most famous financial database WIND; author's calculation

Notes: The upper subgraph displays the specific NARX neural network's outputs, targets, comparative targets and errors versus time, while the lower subgraph uses this neural network's outputs as the benchmark and shows the gaps between targets/comparative targets and outputs versus time. The time on the horizontal axis corresponds to each month from May 2016 to February 2020, which is also the time span of the exogenous inputs series or target series minus the initial twelve months as the initial time delays for forecast. It is also indicated that which time points were selected for training and testing in both subgraphs.

Figure 8. Specific NARX neural network's forecast result for year-on-year growth of CPI of China

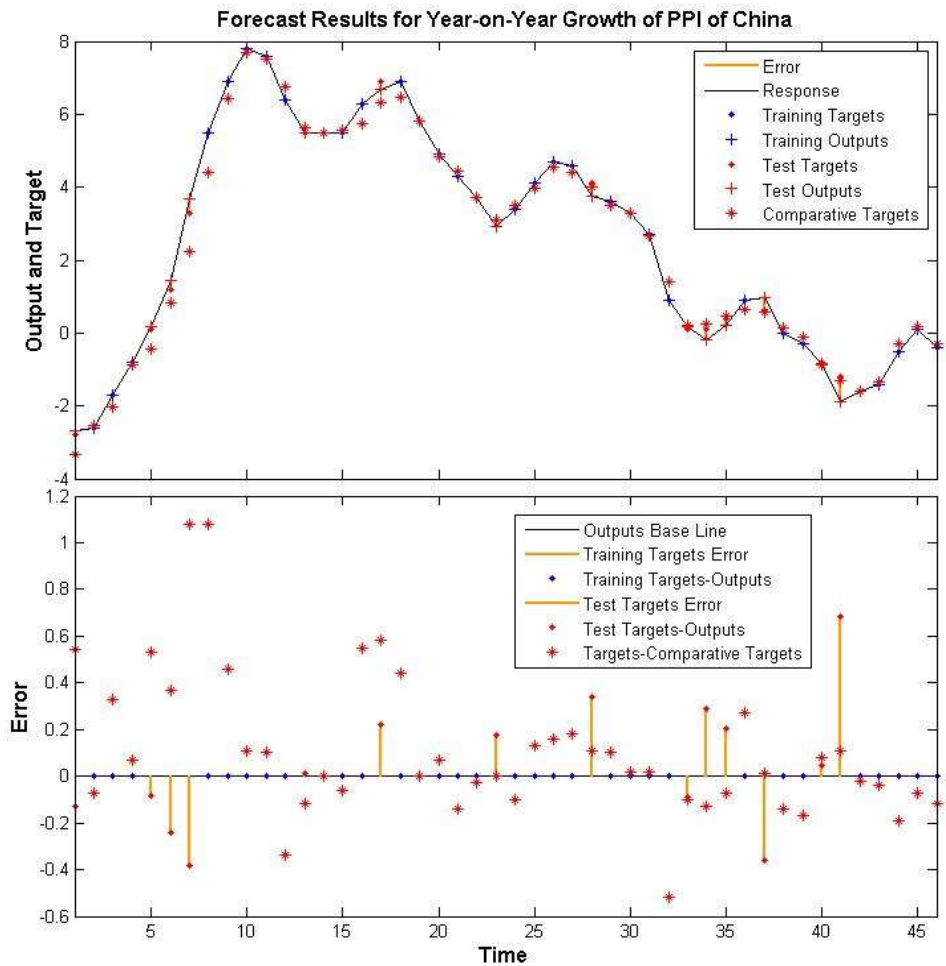

Source: National Bureau of Statistics of China; People's Bank of China; CICC; China Federation of Logistics & Purchasing (CFLP); LanGe Steel Platform of China; Postal Savings Bank of China; Cheung Kong Graduate School of Business (CKGSB); Nine Martingale Investment Management LP of China; China's most famous financial database WIND; author's calculation

Notes: The upper subgraph displays the specific NARX neural network's outputs, targets, comparative targets and errors versus time, while the lower subgraph uses this neural network's outputs as the benchmark and shows the gaps between targets/comparative targets and outputs versus time. The time on the horizontal axis corresponds to each month from May 2016 to February 2020, which is also the time span of the exogenous inputs series or target series minus the initial twelve months as the initial time delays for forecast. It is also indicated that which time points were selected for training and testing in both subgraphs.

Figure 9. Specific NARX neural network's forecast result for year-on-year growth of PPI of China

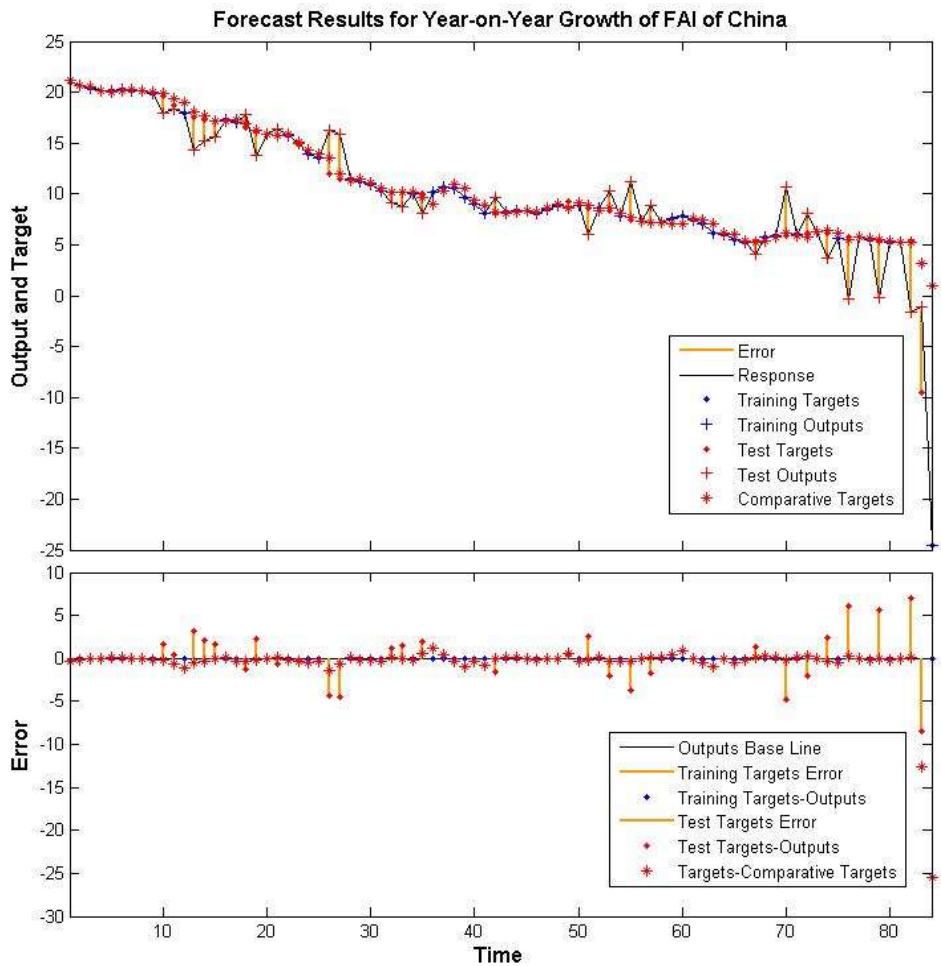

Source: National Bureau of Statistics of China; CFLP; LanGe Steel Platform of China; CKGSB; China's most famous financial database WIND; author's calculation

Notes: The upper subgraph displays the specific NARX neural network's outputs, targets, comparative targets and errors versus time, while the lower subgraph uses this neural network's outputs as the benchmark and shows the gaps between targets/comparative targets and outputs versus time. The time on the horizontal axis corresponds to each month from March 2013 to February 2020, which is also the time span of the exogenous inputs series or target series minus the initial twelve months as the initial time delays for forecast. It is also indicated that which time points were selected for training and testing in both subgraphs.

Figure 10. Specific NARX neural network's forecast result for year-on-year growth of FAI of China

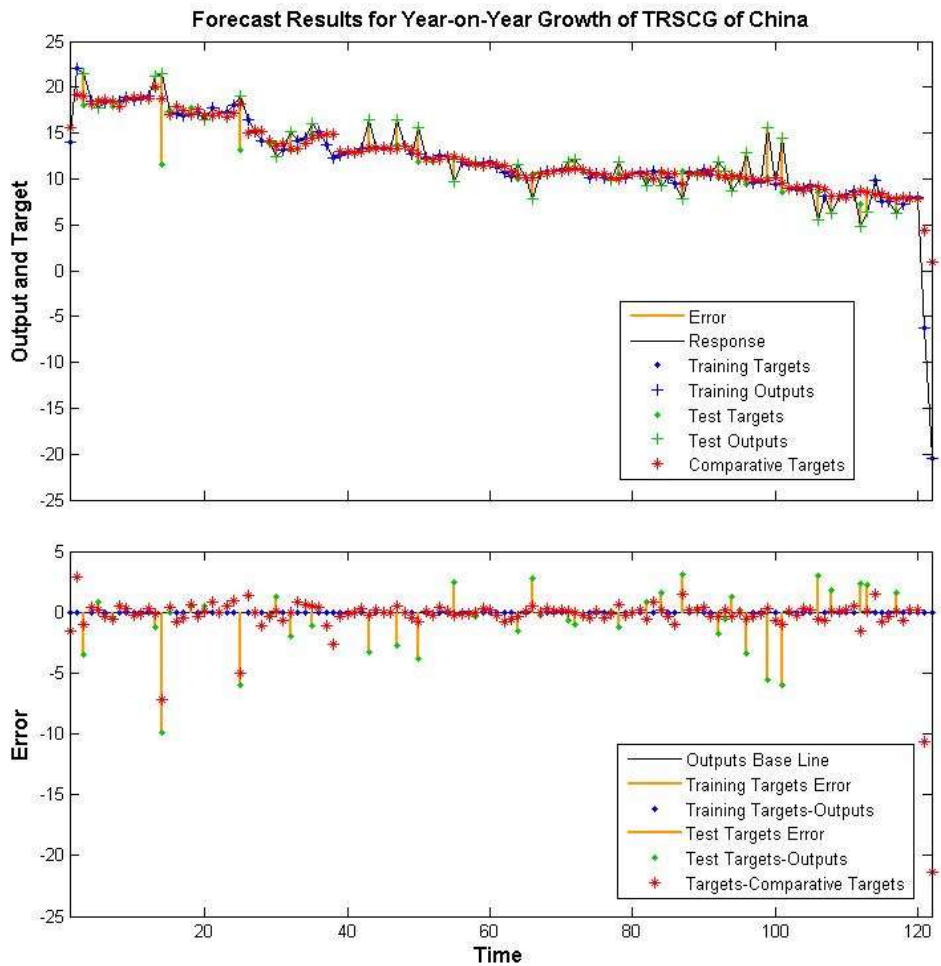

Source: National Bureau of Statistics of China; CICC; China Caixin Media & IHS Markit; CKGSB; China's most famous financial database WIND; author's calculation

Notes: The upper subgraph displays the specific NARX neural network's outputs, targets, comparative targets and errors versus time, while the lower subgraph uses this neural network's outputs as the benchmark and shows the gaps between targets/comparative targets and outputs versus time. The time on the horizontal axis corresponds to each month from January 2010 to February 2020, which is also the time span of the exogenous inputs series or target series minus the initial twelve months as the initial time delays for forecast. It is also indicated that which time points were selected for training and testing in both subgraphs.

Figure 11. Specific NARX neural network's forecast result for year-on-year growth of TRSCG of China

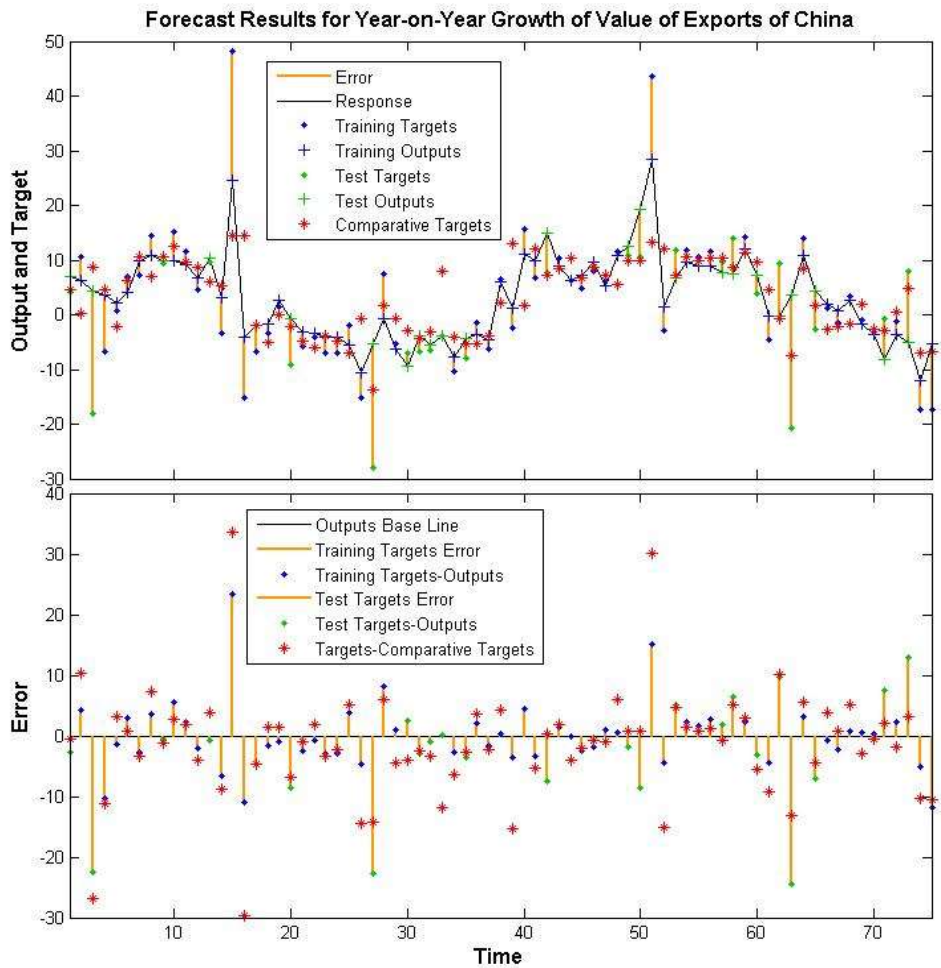

Source: National Bureau of Statistics of China; CICC; LanGe Steel Platform of China; China Customs; Ningbo Shipping Exchange of China; People's Bank of China; China's most famous financial database WIND; author's calculation

Notes: The upper subgraph displays the specific NARX neural network's outputs, targets, comparative targets and errors versus time, while the lower subgraph uses this neural network's outputs as the benchmark and shows the gaps between targets/comparative targets and outputs versus time. The time on the horizontal axis corresponds to each month from December 2013 to February 2020, which is also the time span of the exogenous inputs series or target series minus the initial twelve months as the initial time delays for forecast. It is also indicated that which time points were selected for training and testing in both subgraphs.

Figure 12. Specific NARX neural network's forecast result for year-on-year growth of value of exports of China

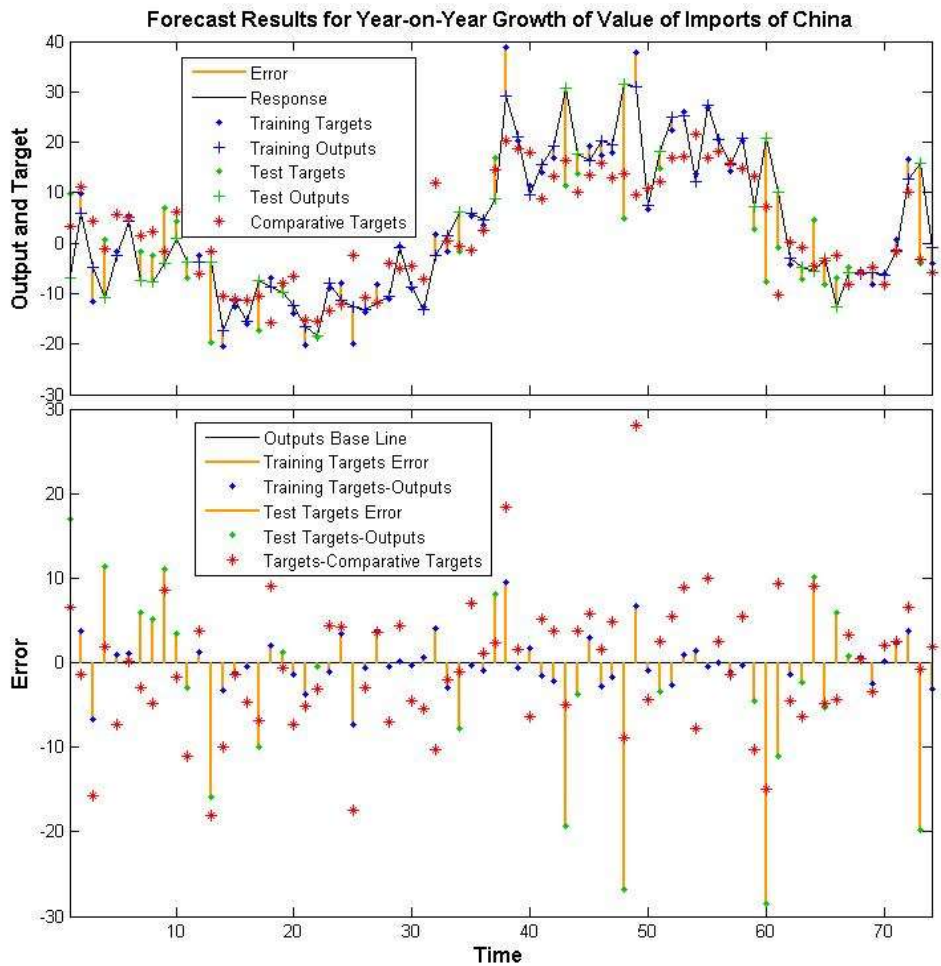

Source: National Bureau of Statistics of China; Ningbo Shipping Exchange of China; China's most famous financial database WIND; author's calculation

Notes: The upper subgraph displays the specific NARX neural network's outputs, targets, comparative targets and errors versus time, while the lower subgraph uses this neural network's outputs as the benchmark and shows the gaps between targets/comparative targets and outputs versus time. The time on the horizontal axis corresponds to each month from January 2014 to February 2020, which is also the time span of the exogenous inputs series or target series minus the initial twelve months as the initial time delays for forecast. It is also indicated that which time points were selected for training and testing in both subgraphs.

Figure 13. Specific NARX neural network's forecast result for year-on-year growth of value of imports of China

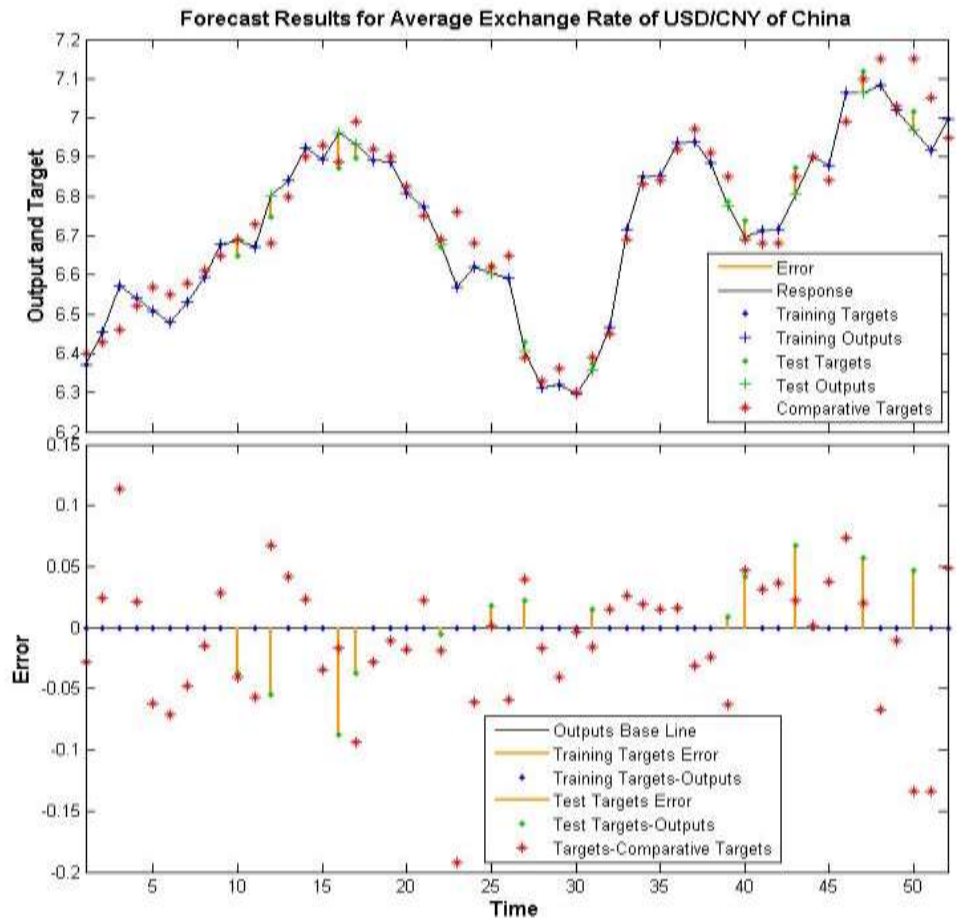

Source: National Bureau of Statistics of China; UK-based money broker ICAP; China Foreign Exchange Trade System (CFETS); China's most famous financial database WIND; author's calculation

Notes: The upper subgraph displays the specific NARX neural network's outputs, targets, comparative targets and errors versus time, while the lower subgraph uses this neural network's outputs as the benchmark and shows the gaps between targets/comparative targets and outputs versus time. The time on the horizontal axis corresponds to each month from November 2015 to February 2020, which is also the time span of the exogenous inputs series or target series minus the initial twelve months as the initial time delays for forecast. It is also indicated that which time points were selected for training and testing in both subgraphs.

Figure 14. Specific NARX neural network's forecast result for average exchange rate of

USD/CNY of China

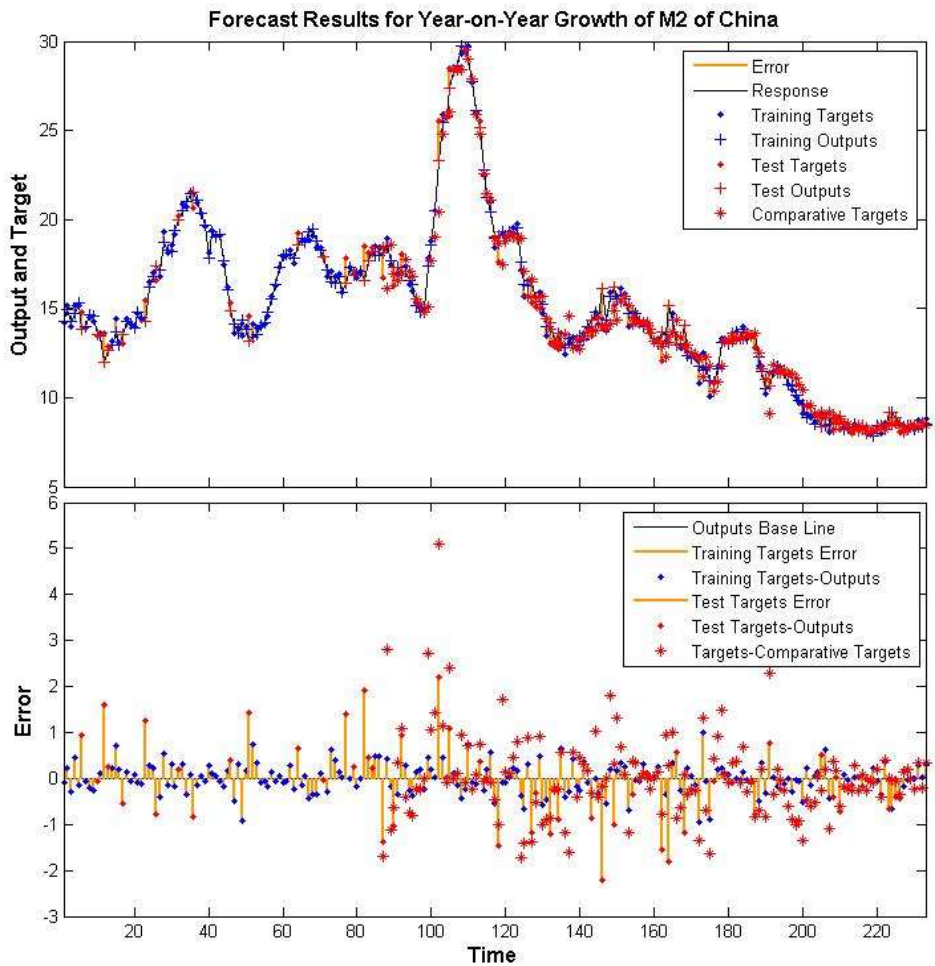

Source: National Bureau of Statistics of China; Nine Martingale Investment Management LP of China; National Development and Reform Commission of China; China's most famous financial database WIND; author's calculation

Notes: The upper subgraph displays the specific NARX neural network's outputs, targets, comparative targets and errors versus time, while the lower subgraph uses this neural network's outputs as the benchmark and shows the gaps between targets/comparative targets and outputs versus time. The time on the horizontal axis corresponds to each month from October 2000 to February 2020, which is also the time span of the exogenous inputs series or target series minus the initial twelve months as the initial time delays for forecast. It is also indicated that which time points were selected for training and testing in both subgraphs.

Figure 15. Specific NARX neural network's forecast result for year-on-year growth of M2 of China

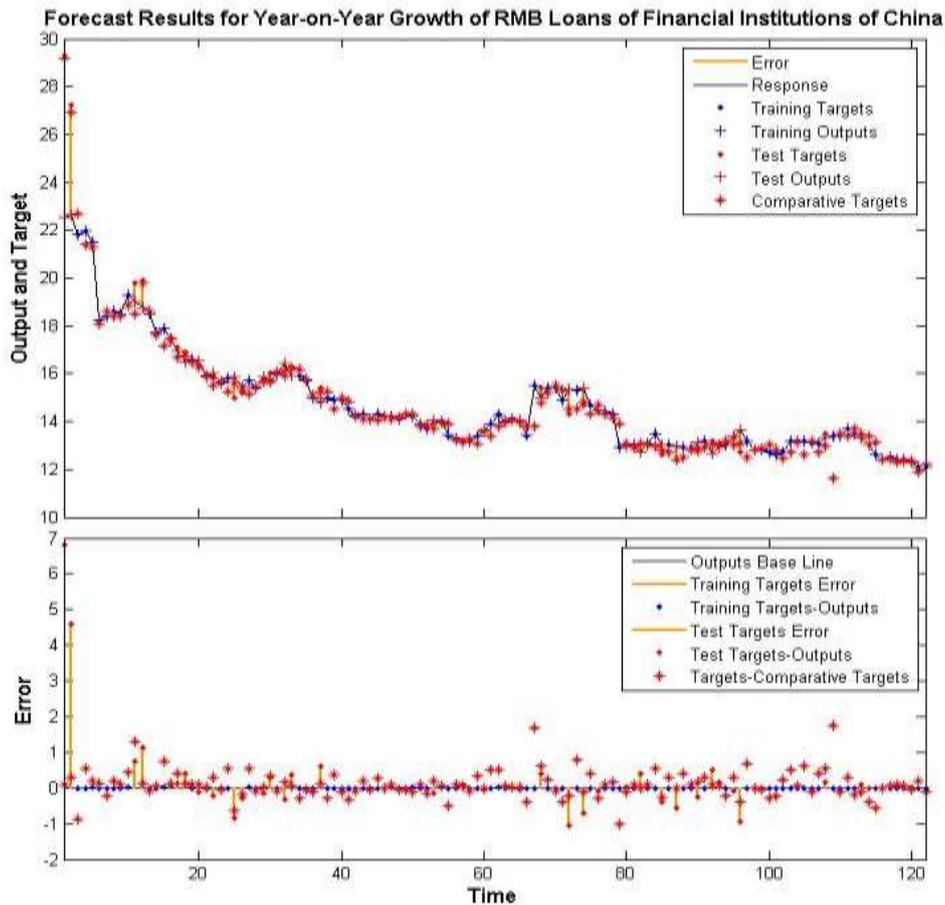

Source: National Bureau of Statistics of China; People's Bank of China; Postal Savings Bank of China; CKGSB; Nine Martingale Investment Management LP of China; China's most famous financial database WIND; author's calculation

Notes: The upper subgraph displays the specific NARX neural network's outputs, targets, comparative targets and errors versus time, while the lower subgraph uses this neural network's outputs as the benchmark and shows the gaps between targets/comparative targets and outputs versus time. The time on the horizontal axis corresponds to each month from January 2010 to February 2020, which is also the time span of the exogenous inputs series or target series minus the initial twelve months as the initial time delays for forecast. It is also indicated that which time points were selected for training and testing in both subgraphs.

Figure 16. Specific NARX neural network's forecast result for year-on-year growth of RMB loans of financial institutions of China

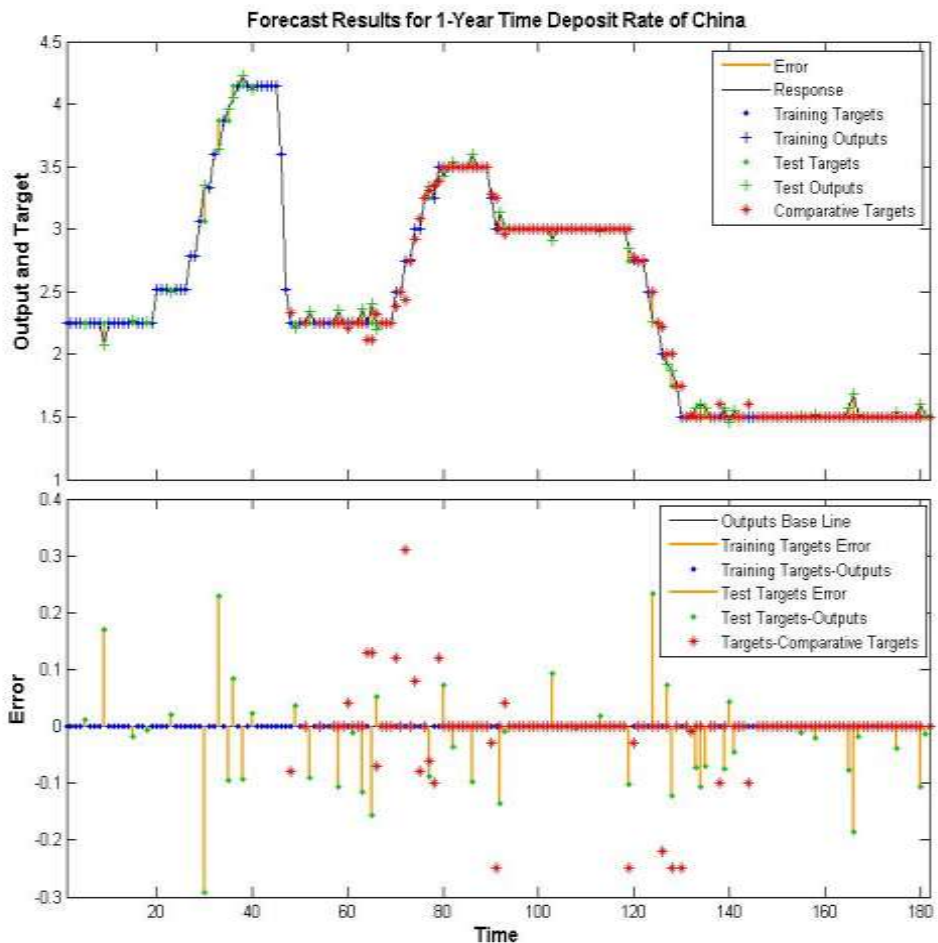

Source: National Bureau of Statistics of China; People's Bank of China; big QQ Groups from each industry; CFETS; Nine Martingale Investment Management LP of China; China's most famous financial database WIND; author's calculation

Notes: The upper subgraph displays the specific NARX neural network's outputs, targets, comparative targets and errors versus time, while the lower subgraph uses this neural network's outputs as the benchmark and shows the gaps between targets/comparative targets and outputs versus time. The time on the horizontal axis corresponds to each month from January 2005 to February 2020, which is also the time span of the exogenous inputs series or target series minus the initial twelve months as the initial time delays for forecast. It is also indicated that which time points were selected for training and testing in both subgraphs.

Figure 17. Specific NARX neural network's forecast result for 1-year time deposit rate of China

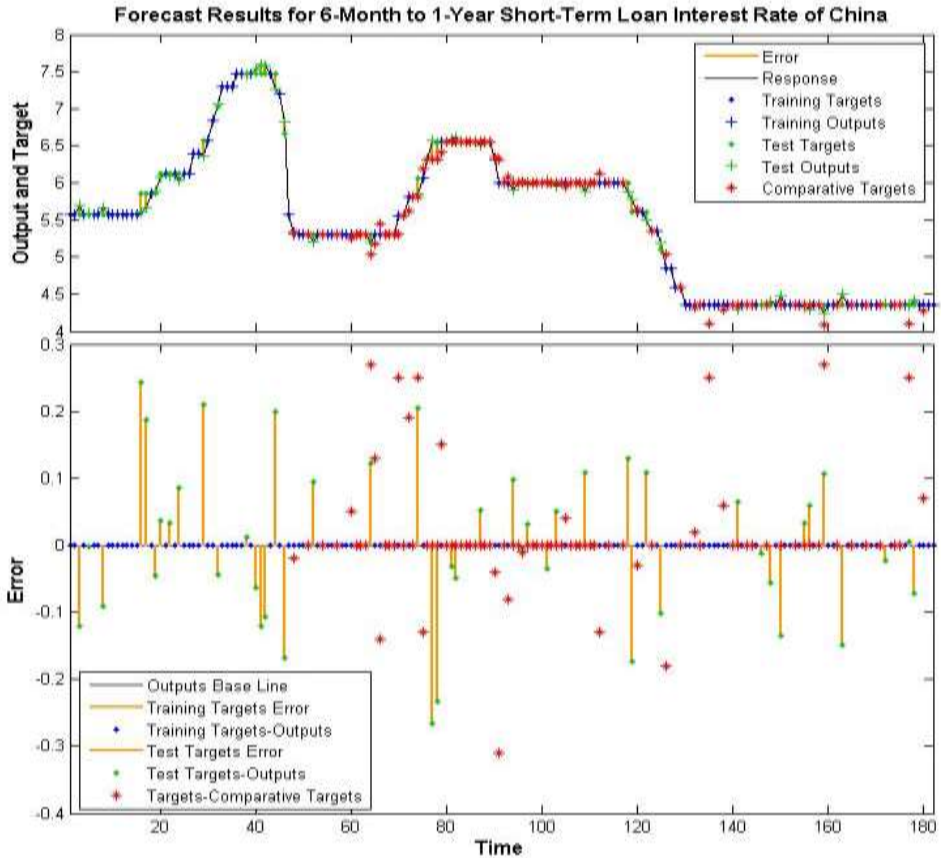

Source: National Bureau of Statistics of China; big QQ Groups from each industry; National Interbank Funding Center of China; Bank of Communications; Shanghai Pudong Development Bank; China Everbright Bank; China Minsheng Banking; Agricultural Bank of China; Bank of China; Nine Martingale Investment Management LP of China; China's most famous financial database WIND; author's calculation

Notes: The upper subgraph displays the specific NARX neural network's outputs, targets, comparative targets and errors versus time, while the lower subgraph uses this neural network's outputs as the benchmark and shows the gaps between targets/comparative targets and outputs versus time. The time on the horizontal axis corresponds to each month from January 2005 to February 2020, which is also the time span of the exogenous inputs series or target series minus the initial twelve months as the initial time delays for forecast. It is also indicated that which time points were selected for training and testing in both subgraphs.

Figure 18. Specific NARX neural network's forecast result for 6-month to 1-year short-term loan interest rate of China

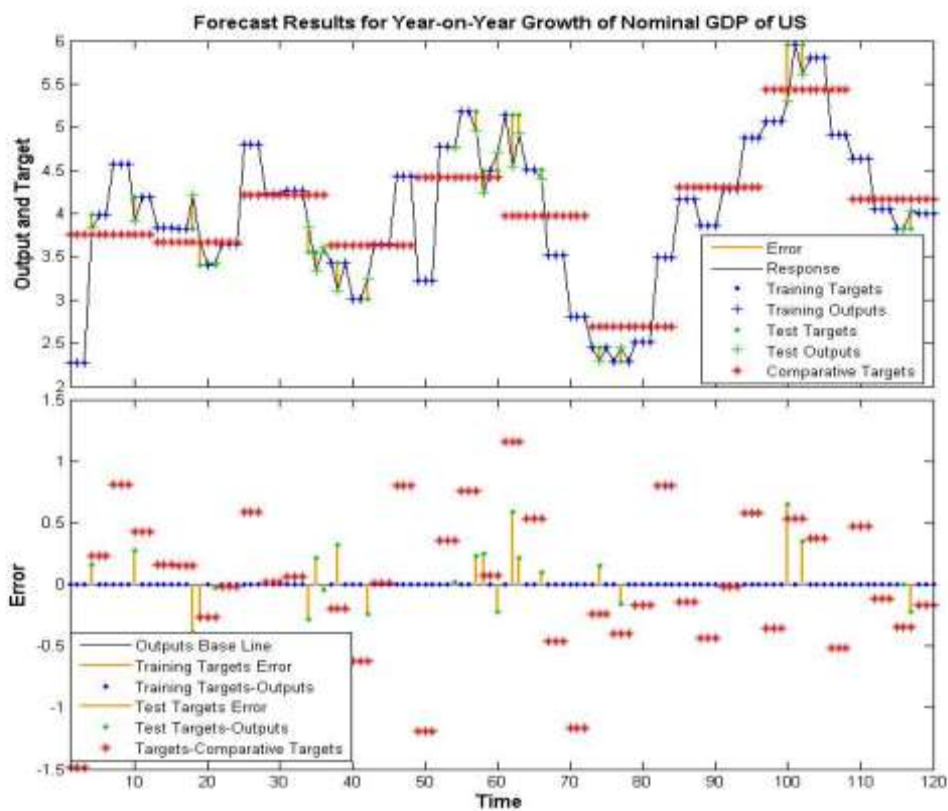

Source: Institute for Supply Management (ISM); Sentix-project; Investor's Business Daily (IBD) & Techno Metrica Market Intelligence (TIPP); US Federal Reserve Bank of Chicago; US Federal Reserve Bank of Philadelphia; US National Federation of Independent Business (NFIB) Research Foundation; US Business Roundtable; US Federal Reserve Bank of New York; US Federal Reserve Bank of St. Louis; US Federal Reserve Board; the world famous CEIC Database; World Bank (WB); International Monetary Fund (IMF); author's calculation

Notes: The upper subgraph shows the specific NARX neural network's outputs, targets, comparative targets and errors versus time, while the lower subgraph adopts this neural network's outputs as the benchmark and shows the gaps between targets/comparative targets and outputs versus time. The time on the horizontal axis corresponds to each month from January 2010 to December 2019, which is also the time span of the exogenous inputs series or target series minus the initial twelve months as the initial time delays for forecast. Both subgraphs also indicate which time points were selected for neural network training and network performance testing.

Figure 19. Specific NARX neural network's forecast result for year-on-year growth of nominal GDP of US

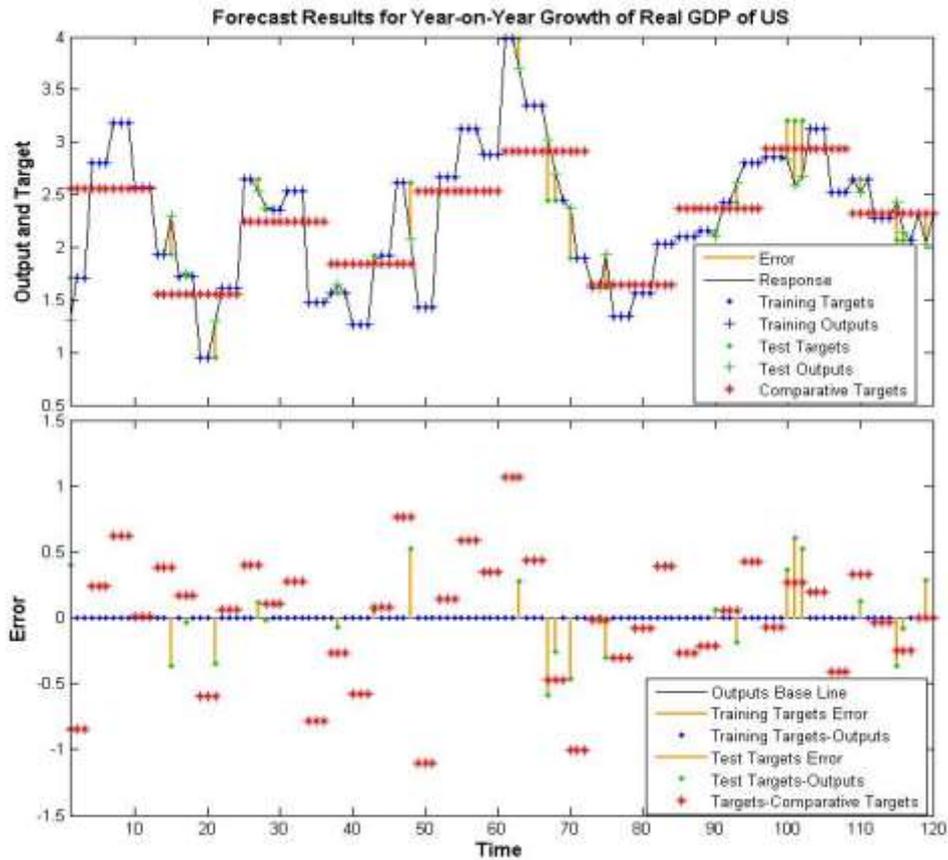

Source: ISM; Sentix-project; IBD & TIPP; US Federal Reserve Bank of Chicago; US Federal Reserve Bank of Philadelphia; US NFIB Research Foundation; US Business Roundtable; US Federal Reserve Bank of New York; US Federal Reserve Bank of St. Louis; US Federal Reserve Board; CEIC Database; WB; IMF; author's calculation

Notes: The upper subgraph shows the specific NARX neural network's outputs, targets, comparative targets and errors versus time, while the lower subgraph adopts this neural network's outputs as the benchmark and shows the gaps between targets/comparative targets and outputs versus time. The time on the horizontal axis corresponds to each month from January 2010 to December 2019, which is also the time span of the exogenous inputs series or target series minus the initial twelve months as the initial time delays for forecast. Both subgraphs also indicate which time points were selected for neural network training and network performance testing.

Figure 20. Specific NARX neural network's forecast result for year-on-year growth of real GDP of US

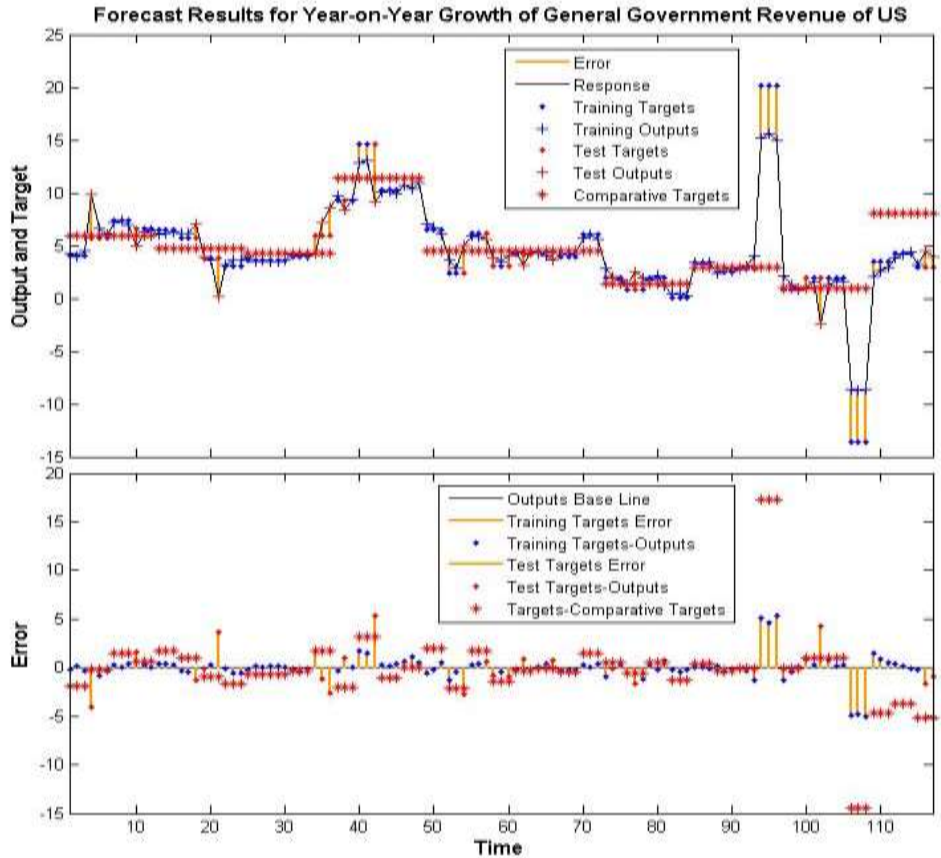

Source: ISM; Sentix-project; IBD & TIPP; US Federal Reserve Bank of Chicago; US Federal Reserve Bank of Philadelphia; US NFIB Research Foundation; US Business Roundtable; US Federal Reserve Bank of New York; US Federal Reserve Bank of St. Louis; US Federal Reserve Board; CEIC Database; WB; IMF; author's calculation

Notes: The upper subgraph shows the specific NARX neural network's outputs, targets, comparative targets and errors versus time, while the lower subgraph adopts this neural network's outputs as the benchmark and shows the gaps between targets/comparative targets and outputs versus time. The time on the horizontal axis corresponds to each month from January 2010 to September 2019, which is also the time span of the exogenous inputs series or target series minus the initial twelve months as the initial time delays for forecast. Both subgraphs also indicate which time points were selected for neural network training and network performance testing.

Figure 21. Specific NARX neural network's forecast result for year-on-year growth of general government revenue of US

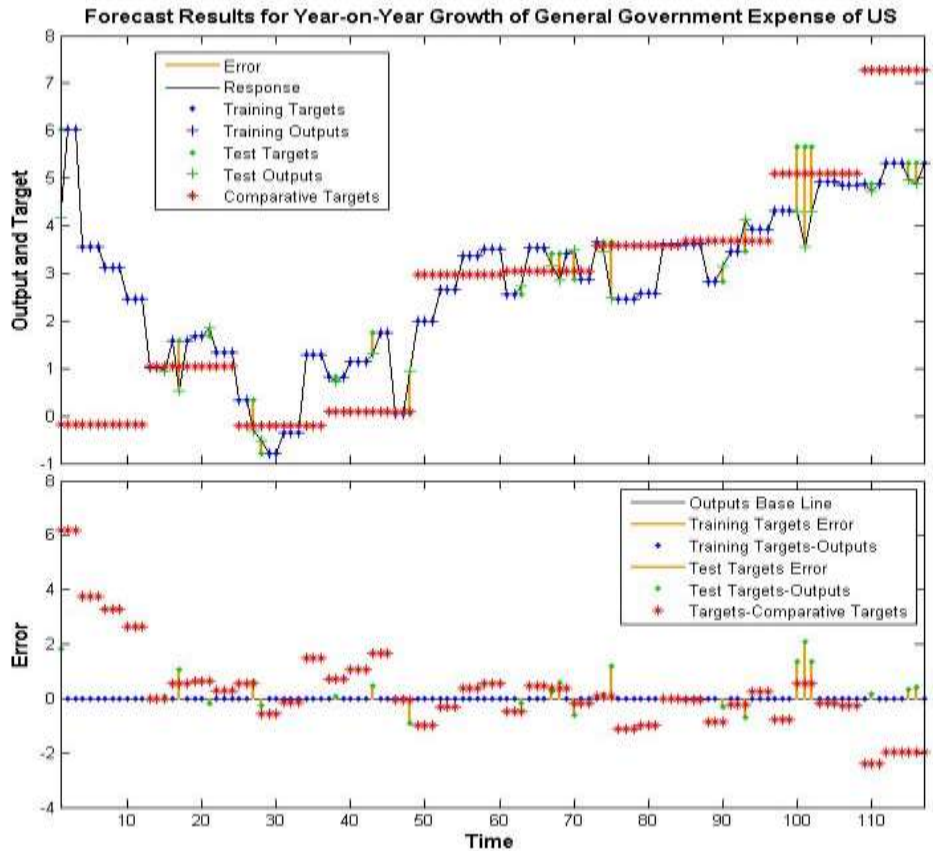

Source: ISM; Sentix-project; IBD & TIPP; US Federal Reserve Bank of Chicago; US Federal Reserve Bank of Philadelphia; US NFIB Research Foundation; US Business Roundtable; US Federal Reserve Bank of New York; US Federal Reserve Bank of St. Louis; US Federal Reserve Board; CEIC Database; WB; IMF; author's calculation

Notes: The upper subgraph shows the specific NARX neural network's outputs, targets, comparative targets and errors versus time, while the lower subgraph adopts this neural network's outputs as the benchmark and shows the gaps between targets/comparative targets and outputs versus time. The time on the horizontal axis corresponds to each month from January 2010 to September 2019, which is also the time span of the exogenous inputs series or target series minus the initial twelve months as the initial time delays for forecast. Both subgraphs also indicate which time points were selected for neural network training and network performance testing.

Figure 22. Specific NARX neural network's forecast result for year-on-year growth of general government expense of US

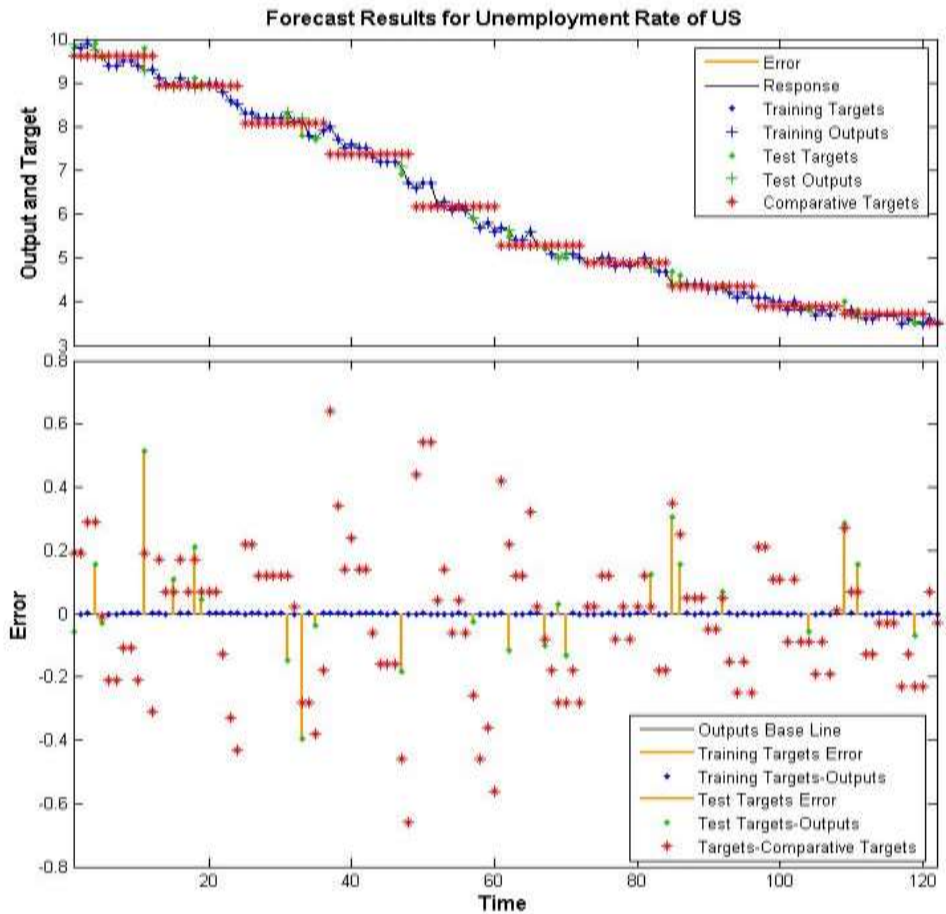

Source: ISM; Sentix-project; IBD & TIPP; US Federal Reserve Bank of Chicago; US Federal Reserve Bank of Philadelphia; US NFIB Research Foundation; US Business Roundtable; US Federal Reserve Bank of New York; US Federal Reserve Bank of St. Louis; US Federal Reserve Board; CEIC Database; WB; IMF; author's calculation

Notes: The upper subgraph shows the specific NARX neural network's outputs, targets, comparative targets and errors versus time, while the lower subgraph adopts this neural network's outputs as the benchmark and shows the gaps between targets/comparative targets and outputs versus time. The time on the horizontal axis corresponds to each month from January 2010 to February 2020, which is also the time span of the exogenous inputs series or target series minus the initial twelve months as the initial time delays for forecast. Both subgraphs also indicate which time points were selected for neural network training and network performance testing.

Figure 23. Specific NARX neural network's forecast result for unemployment rate of US

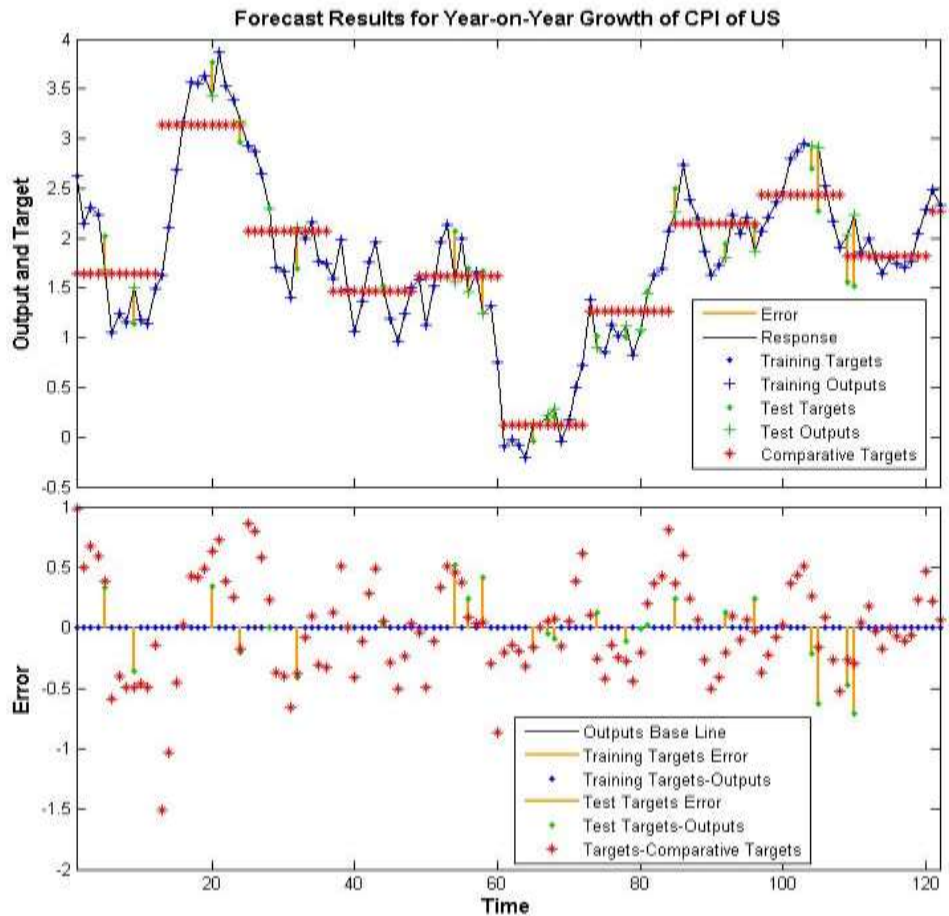

Source: ISM; Sentix-project; IBD & TIPP; US Federal Reserve Bank of Chicago; US Federal Reserve Bank of Philadelphia; US NFIB Research Foundation; US Business Roundtable; US Federal Reserve Bank of New York; US Federal Reserve Bank of St. Louis; US Federal Reserve Board; CEIC Database; WB; IMF; author's calculation

Notes: The upper subgraph shows the specific NARX neural network's outputs, targets, comparative targets and errors versus time, while the lower subgraph adopts this neural network's outputs as the benchmark and shows the gaps between targets/comparative targets and outputs versus time. The time on the horizontal axis corresponds to each month from January 2010 to February 2020, which is also the time span of the exogenous inputs series or target series minus the initial twelve months as the initial time delays for forecast. Both subgraphs also indicate which time points were selected for neural network training and network performance testing.

Figure 24. Specific NARX neural network's forecast result for year-on-year growth of CPI of US

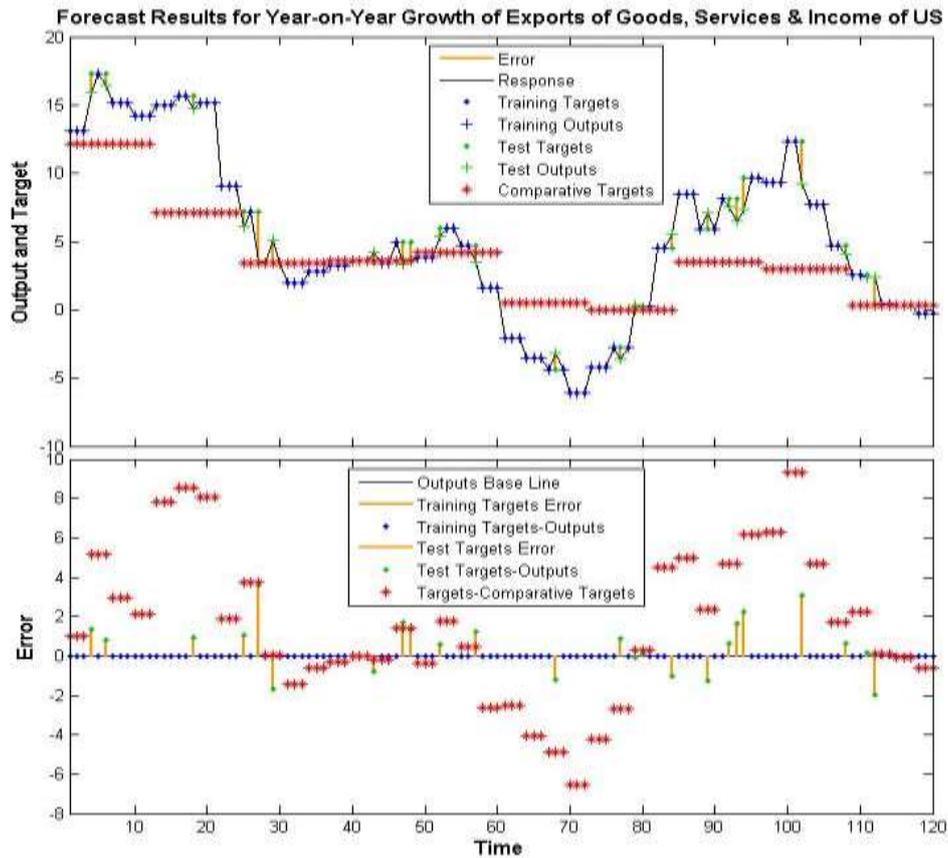

Source: ISM; Sentix-project; IBD & TIPP; US Federal Reserve Bank of Chicago; US Federal Reserve Bank of Philadelphia; US NFIB Research Foundation; US Business Roundtable; US Federal Reserve Bank of New York; US Federal Reserve Bank of St. Louis; US Federal Reserve Board; CEIC Database; WB; IMF; author's calculation

Notes: The upper subgraph shows the specific NARX neural network's outputs, targets, comparative targets and errors versus time, while the lower subgraph adopts this neural network's outputs as the benchmark and shows the gaps between targets/comparative targets and outputs versus time. The time on the horizontal axis corresponds to each month from January 2010 to December 2019, which is also the time span of the exogenous inputs series or target series minus the initial twelve months as the initial time delays for forecast. Both subgraphs also indicate which time points were selected for neural network training and network performance testing.

Figure 25. Specific NARX neural network's forecast result for year-on-year growth of exports of goods, services & income of US

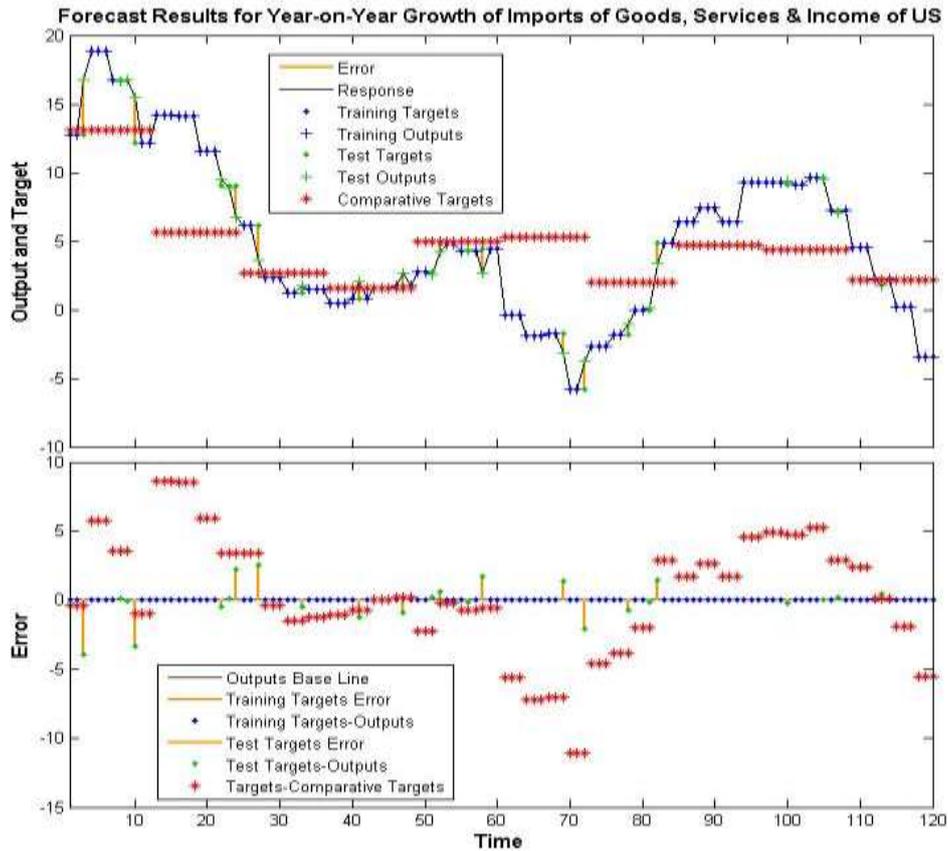

Source: ISM; Sentix-project; IBD & TIPP; US Federal Reserve Bank of Chicago; US Federal Reserve Bank of Philadelphia; US NFIB Research Foundation; US Business Roundtable; US Federal Reserve Bank of New York; US Federal Reserve Bank of St. Louis; US Federal Reserve Board; CEIC Database; WB; IMF; author's calculation

Notes: The upper subgraph shows the specific NARX neural network's outputs, targets, comparative targets and errors versus time, while the lower subgraph adopts this neural network's outputs as the benchmark and shows the gaps between targets/comparative targets and outputs versus time. The time on the horizontal axis corresponds to each month from January 2010 to December 2019, which is also the time span of the exogenous inputs series or target series minus the initial twelve months as the initial time delays for forecast. Both subgraphs also indicate which time points were selected for neural network training and network performance testing.

Figure 26. Specific NARX neural network's forecast result for year-on-year growth of imports of goods, services & income of US

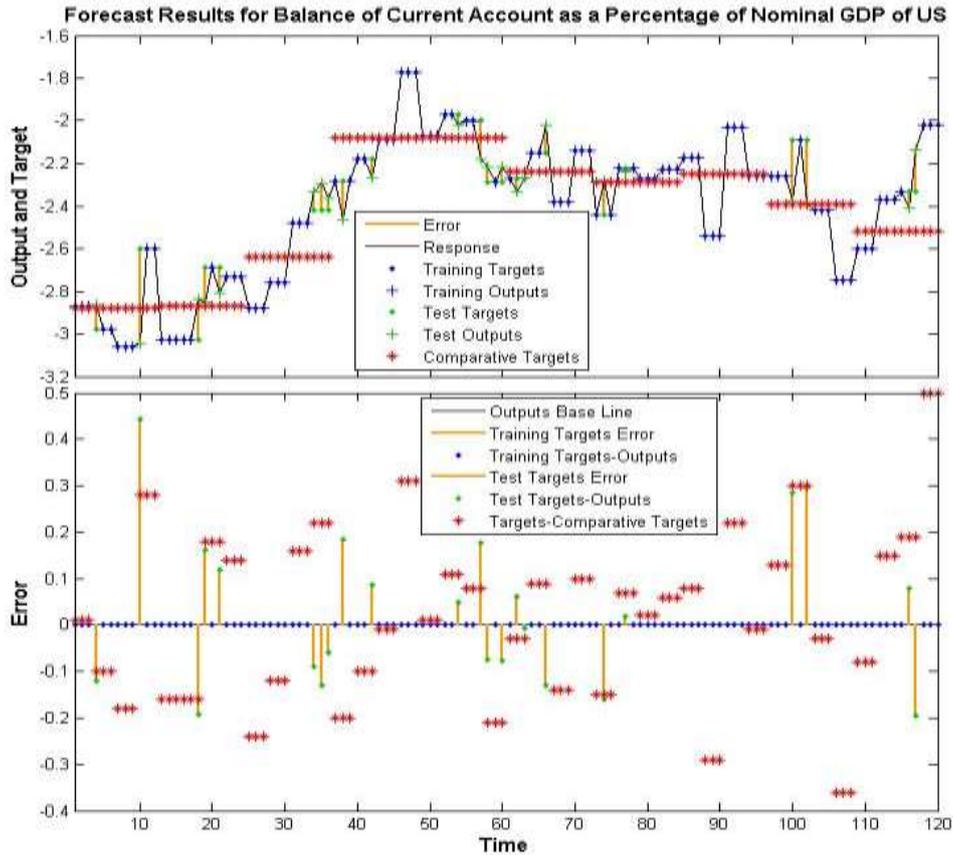

Source: ISM; Sentix-project; IBD & TIPP; US Federal Reserve Bank of Chicago; US Federal Reserve Bank of Philadelphia; US NFIB Research Foundation; US Business Roundtable; US Federal Reserve Bank of New York; US Federal Reserve Bank of St. Louis; US Federal Reserve Board; CEIC Database; WB; IMF; author's calculation

Notes: The upper subgraph shows the specific NARX neural network's outputs, targets, comparative targets and errors versus time, while the lower subgraph adopts this neural network's outputs as the benchmark and shows the gaps between targets/comparative targets and outputs versus time. The time on the horizontal axis corresponds to each month from January 2010 to December 2019, which is also the time span of the exogenous inputs series or target series minus the initial twelve months as the initial time delays for forecast. Both subgraphs also indicate which time points were selected for neural network training and network performance testing.

Figure 27. Specific NARX neural network's forecast result for balance of current account as a percentage of nominal GDP of US

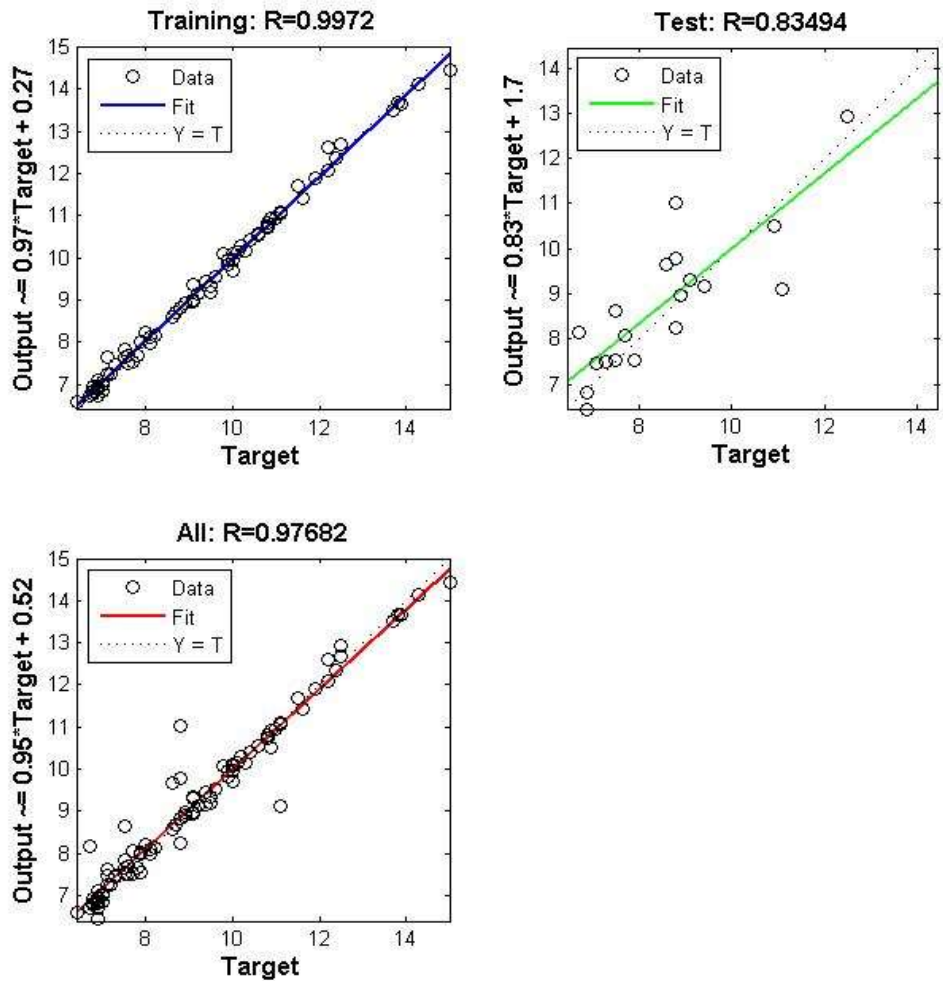

Source: National Bureau of Statistics of China; People's Bank of China; China's most famous financial database WIND; author's calculation

Notes: The upper left subgraph displays the specific NARX neural network's forecast results versus target values at time points for training, the upper right subgraph shows this neural network's forecast results versus target values at time points for testing, while the lower left subgraph displays this neural network's forecast results versus target values at all time points (time points for both training and testing). The square root of the coefficient of determination $R^2$ for evaluating the prediction performance of this specific NARX neural network is displayed at the top of each subgraph.

Figure 28. The regression graph and the coefficient of determination $R^2$ for evaluating the performance of the specific NARX neural network's forecasting for year-on-year growth of real GDP of China

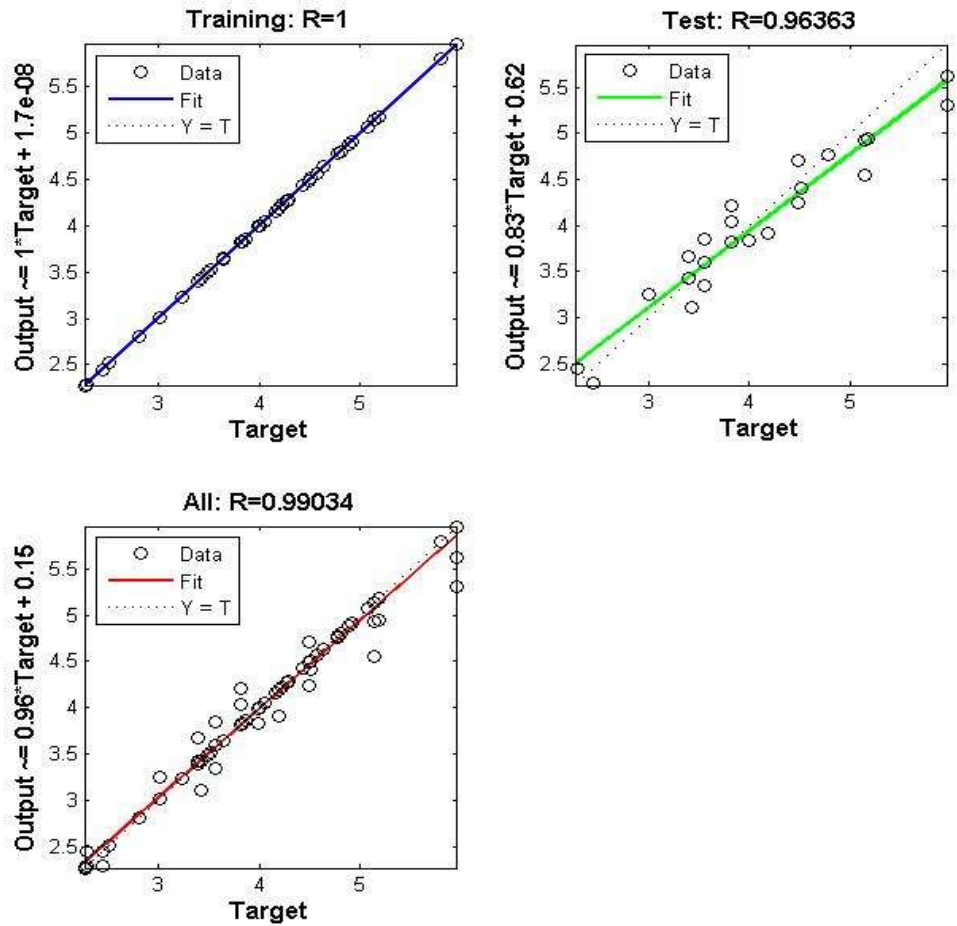

Source: ISM; Sentix-project; IBD & TIPP; US Federal Reserve Bank of Chicago; US Federal Reserve Bank of Philadelphia; US NFIB Research Foundation; US Business Roundtable; US Federal Reserve Bank of New York; US Federal Reserve Bank of St. Louis; US Federal Reserve Board; CEIC Database; WB; IMF; author's calculation

Notes: The upper left subgraph displays the specific NARX neural network's forecast results versus target values at time points for training, the upper right subgraph shows this neural network's forecast results versus target values at time points for testing, while the lower left subgraph displays this neural network's forecast results versus target values at all time points. The square root of the coefficient of determination $R^2$ for evaluating the prediction performance of this specific NARX neural network is displayed at the top of each subgraph.

Figure 29. The regression graph and the coefficient of determination $R^2$ for evaluating the performance of the specific NARX neural network's forecasting for year-on-year growth of real GDP of US

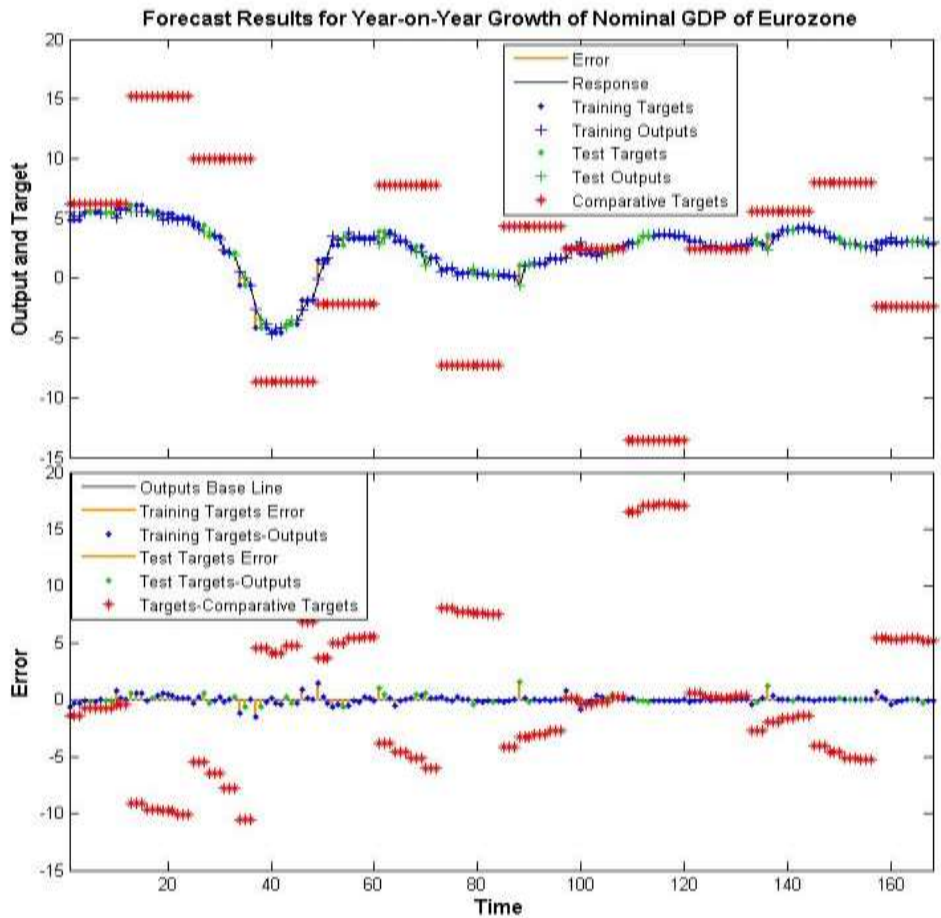

Source: European Central Bank (ECB); European Commission's Directorate-General for Economic and Financial Affairs; Sentix-project; CEIC Database; WB; IMF; author's calculation

Notes: After training the specific NARX neural network based on overall regional exogenous inputs, the upper subgraph shows this neural network's outputs, targets, comparative targets and errors versus time, while the lower subgraph adopts this neural network's outputs as the benchmark and shows the gaps between targets/comparative targets and outputs versus time. The time on the horizontal axis corresponds to each month from January 2006 to December 2019, which is also the time span of the exogenous inputs series or target series minus the initial twelve months as the initial time delays for forecast. Both subgraphs also indicate which time points were selected for neural network training and network performance testing.

Figure 30. Specific NARX neural network's forecast result for year-on-year growth of nominal GDP of Eurozone based on whole area related exogenous inputs

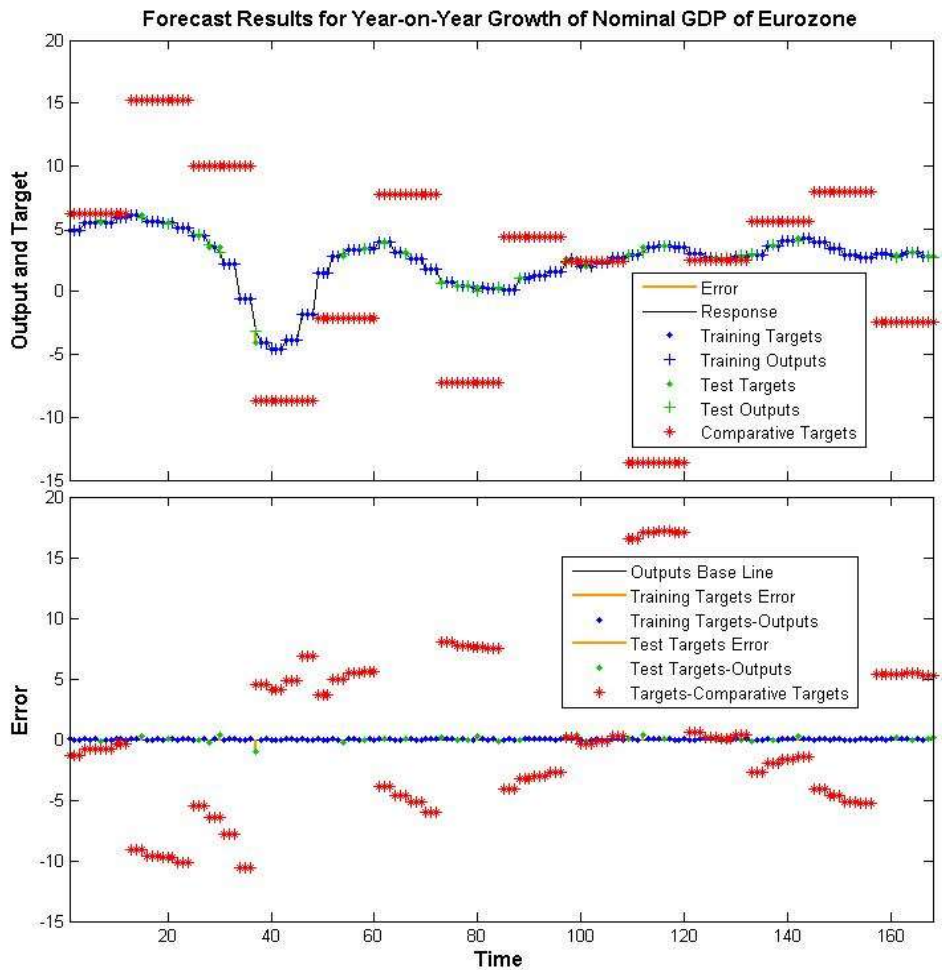

Source: ECB; European Commission's Directorate-General for Economic and Financial Affairs; Sentix-project; CEIC Database; WB; IMF; author's calculation

Notes: After training the specific NARX neural network based on both overall regional & country-specific exogenous inputs, the upper subgraph shows this neural network's outputs, targets, comparative targets and errors versus time, while the lower subgraph adopts this neural network's outputs as the benchmark and shows the gaps between targets/comparative targets and outputs versus time. The time on the horizontal axis corresponds to each month from January 2006 to December 2019, which is also the time span of the exogenous inputs series or target series minus the initial twelve months as the initial time delays for forecast. Both subgraphs also indicate which time points were selected for neural network training and network performance testing.

Figure 31. Specific NARX neural network's forecast result for year-on-year growth of nominal GDP of Eurozone based on both whole area and subdivision area related exogenous inputs

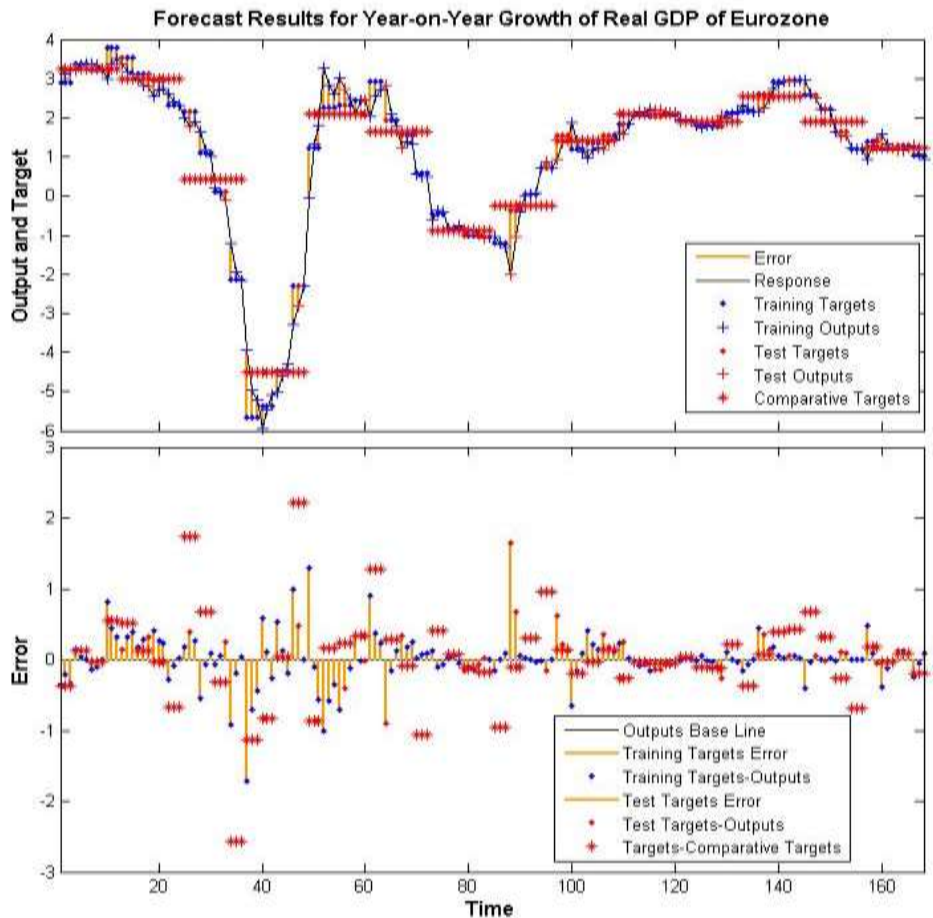

Source: ECB; European Commission's Directorate-General for Economic and Financial Affairs; Sentix-project; CEIC Database; WB; IMF; author's calculation

Notes: After training the specific NARX neural network based on overall regional exogenous inputs, the upper subgraph shows this neural network's outputs, targets, comparative targets and errors versus time, while the lower subgraph adopts this neural network's outputs as the benchmark and shows the gaps between targets/comparative targets and outputs versus time. The time on the horizontal axis corresponds to each month from January 2006 to December 2019, which is also the time span of the exogenous inputs series or target series minus the initial twelve months as the initial time delays for forecast. Both subgraphs also indicate which time points were selected for neural network training and network performance testing.

Figure 32. Specific NARX neural network's forecast result for year-on-year growth of real GDP of Eurozone based on whole area related exogenous inputs

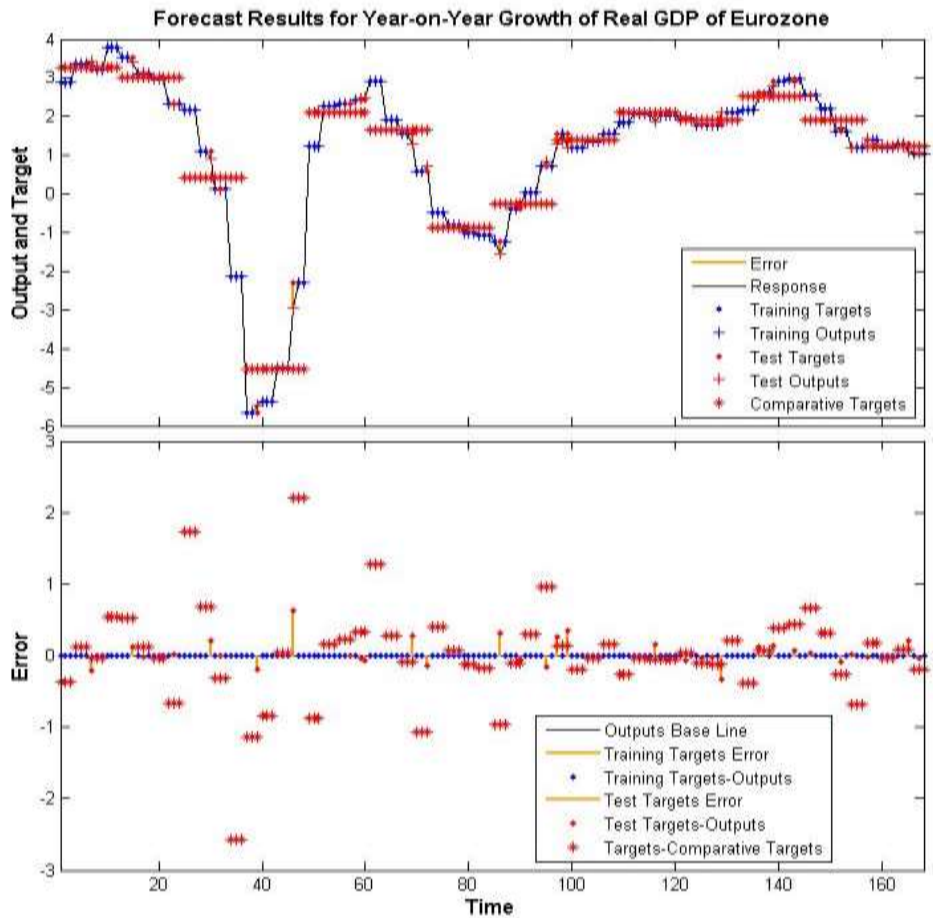

Source: ECB; European Commission's Directorate-General for Economic and Financial Affairs; Sentix-project; CEIC Database; WB; IMF; author's calculation

Notes: After training the specific NARX neural network based on both overall regional & country-specific exogenous inputs, the upper subgraph shows this neural network's outputs, targets, comparative targets and errors versus time, while the lower subgraph adopts this neural network's outputs as the benchmark and shows the gaps between targets/comparative targets and outputs versus time. The time on the horizontal axis corresponds to each month from January 2006 to December 2019, which is also the time span of the exogenous inputs series or target series minus the initial twelve months as the initial time delays for forecast. Both subgraphs also indicate which time points were selected for neural network training and network performance testing.

Figure 33. Specific NARX neural network's forecast result for year-on-year growth of real GDP of Eurozone based on both whole area and subdivision area related exogenous inputs

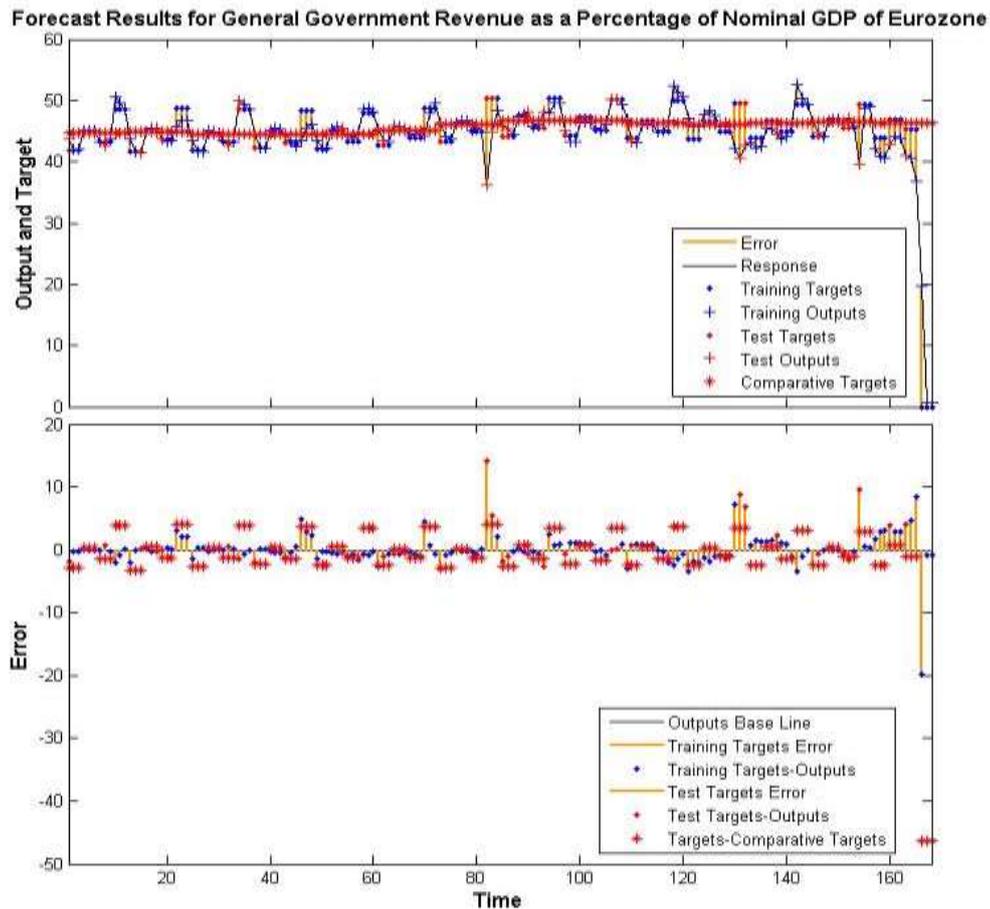

Source: ECB; European Commission's Directorate-General for Economic and Financial Affairs; Sentix-project; CEIC Database; WB; IMF; author's calculation

Notes: After training the specific NARX neural network based on overall regional exogenous inputs, the upper subgraph shows this neural network's outputs, targets, comparative targets and errors versus time, while the lower subgraph adopts this neural network's outputs as the benchmark and shows the gaps between targets/comparative targets and outputs versus time. The time on the horizontal axis corresponds to each month from January 2006 to December 2019, which is also the time span of the exogenous inputs series or target series minus the initial twelve months as the initial time delays for forecast. Both subgraphs also indicate which time points were selected for neural network training and network performance testing.

Figure 34. Specific NARX neural network's forecast result for general government revenue as a percentage of nominal GDP of Eurozone based on whole area related exogenous inputs

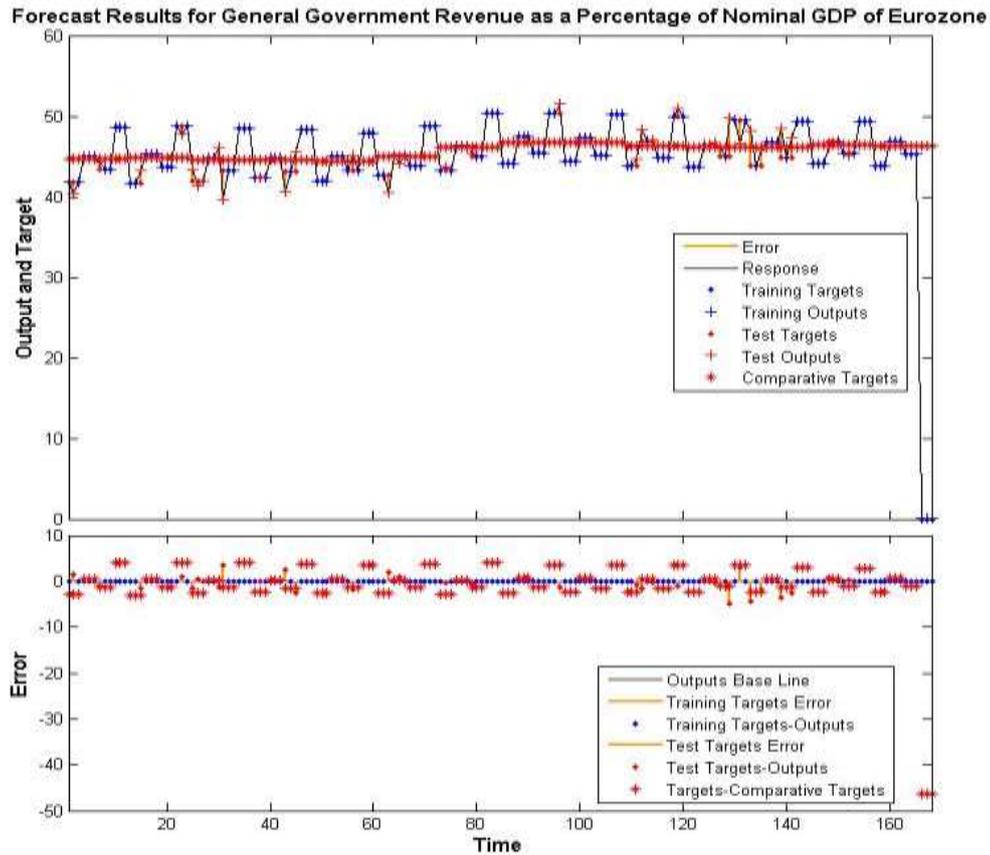

Source: ECB; European Commission's Directorate-General for Economic and Financial Affairs; Sentix-project; CEIC Database; WB; IMF; author's calculation

Notes: After training the specific NARX neural network based on both overall regional & country-specific exogenous inputs, the upper subgraph shows this neural network's outputs, targets, comparative targets and errors versus time, while the lower subgraph adopts this neural network's outputs as the benchmark and shows the gaps between targets/comparative targets and outputs versus time. The time on the horizontal axis corresponds to each month from January 2006 to December 2019, which is also the time span of the exogenous inputs series or target series minus the initial twelve months as the initial time delays for forecast. Both subgraphs also indicate which time points were selected for neural network training and network performance testing.

Figure 35. Specific NARX neural network's forecast result for general government revenue as a percentage of nominal GDP of Eurozone based on both whole area and subdivision area related exogenous inputs

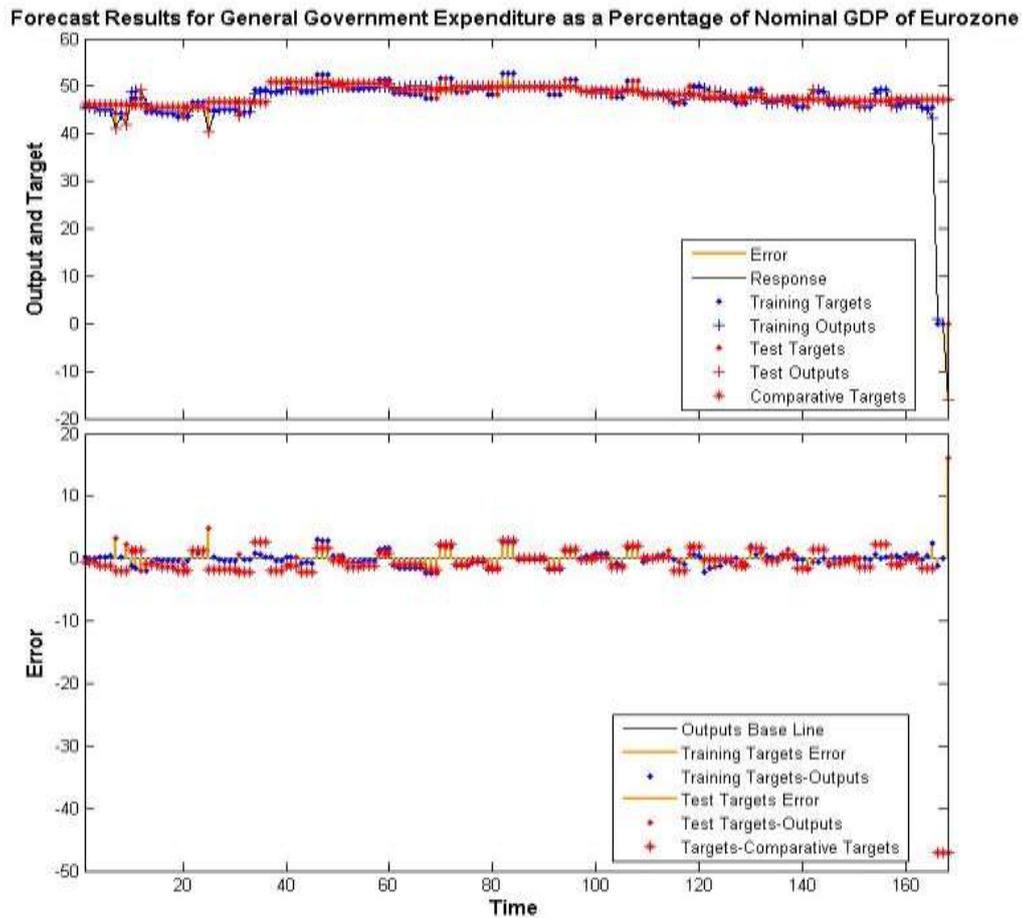

Source: ECB; European Commission's Directorate-General for Economic and Financial Affairs; Sentix-project; CEIC Database; WB; IMF; author's calculation

Notes: After training the specific NARX neural network based on overall regional exogenous inputs, the upper subgraph shows this neural network's outputs, targets, comparative targets and errors versus time, while the lower subgraph adopts this neural network's outputs as the benchmark and shows the gaps between targets/comparative targets and outputs versus time. The time on the horizontal axis corresponds to each month from January 2006 to December 2019, which is also the time span of the exogenous inputs series or target series minus the initial twelve months as the initial time delays for forecast. Both subgraphs also indicate which time points were selected for neural network training and network performance testing.

Figure 36. Specific NARX neural network's forecast result for general government expenditure as a percentage of nominal GDP of Eurozone based on whole area related exogenous inputs

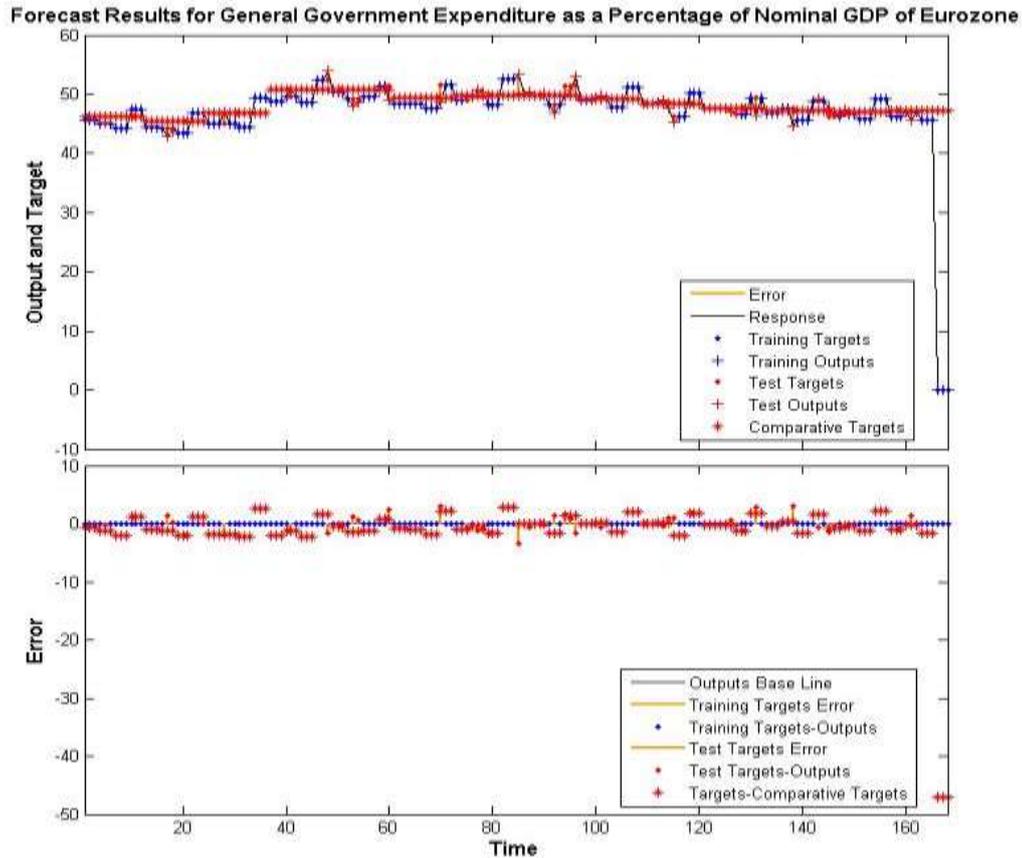

Source: ECB; European Commission's Directorate-General for Economic and Financial Affairs; Sentix-project; CEIC Database; WB; IMF; author's calculation

Notes: After training the specific NARX neural network based on both overall regional & country-specific exogenous inputs, the upper subgraph shows this neural network's outputs, targets, comparative targets and errors versus time, while the lower subgraph adopts this neural network's outputs as the benchmark and shows the gaps between targets/comparative targets and outputs versus time. The time on the horizontal axis corresponds to each month from January 2006 to December 2019, which is also the time span of the exogenous inputs series or target series minus the initial twelve months as the initial time delays for forecast. Both subgraphs also indicate which time points were selected for neural network training and network performance testing.

Figure 37. Specific NARX neural network's forecast result for general government expenditure as a percentage of nominal GDP of Eurozone based on both whole area and subdivision area related exogenous inputs

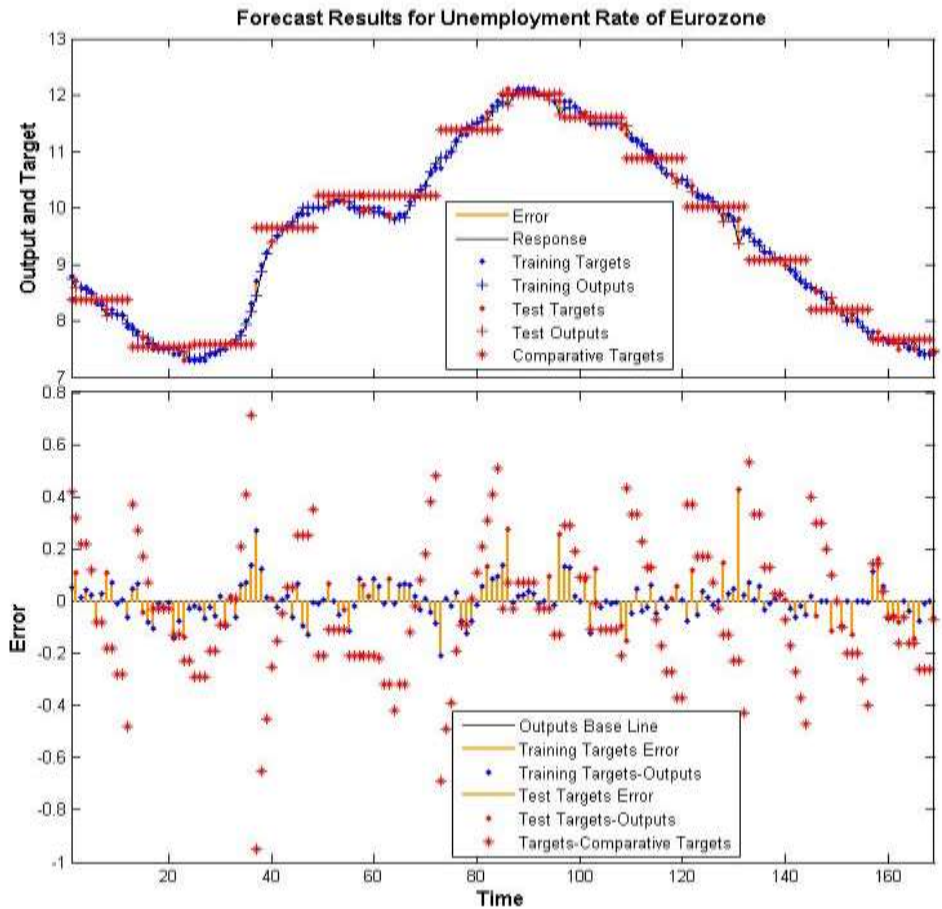

Source: ECB; European Commission's Directorate-General for Economic and Financial Affairs; Sentix-project; CEIC Database; WB; IMF; author's calculation

Notes: After training the specific NARX neural network based on overall regional exogenous inputs, the upper subgraph shows this neural network's outputs, targets, comparative targets and errors versus time, while the lower subgraph adopts this neural network's outputs as the benchmark and shows the gaps between targets/comparative targets and outputs versus time. The time on the horizontal axis corresponds to each month from January 2006 to January 2020, which is also the time span of the exogenous inputs series or target series minus the initial twelve months as the initial time delays for forecast. Both subgraphs also indicate which time points were selected for neural network training and network performance testing.

Figure 38. Specific NARX neural network's forecast result for unemployment rate of Eurozone based on whole area related exogenous inputs

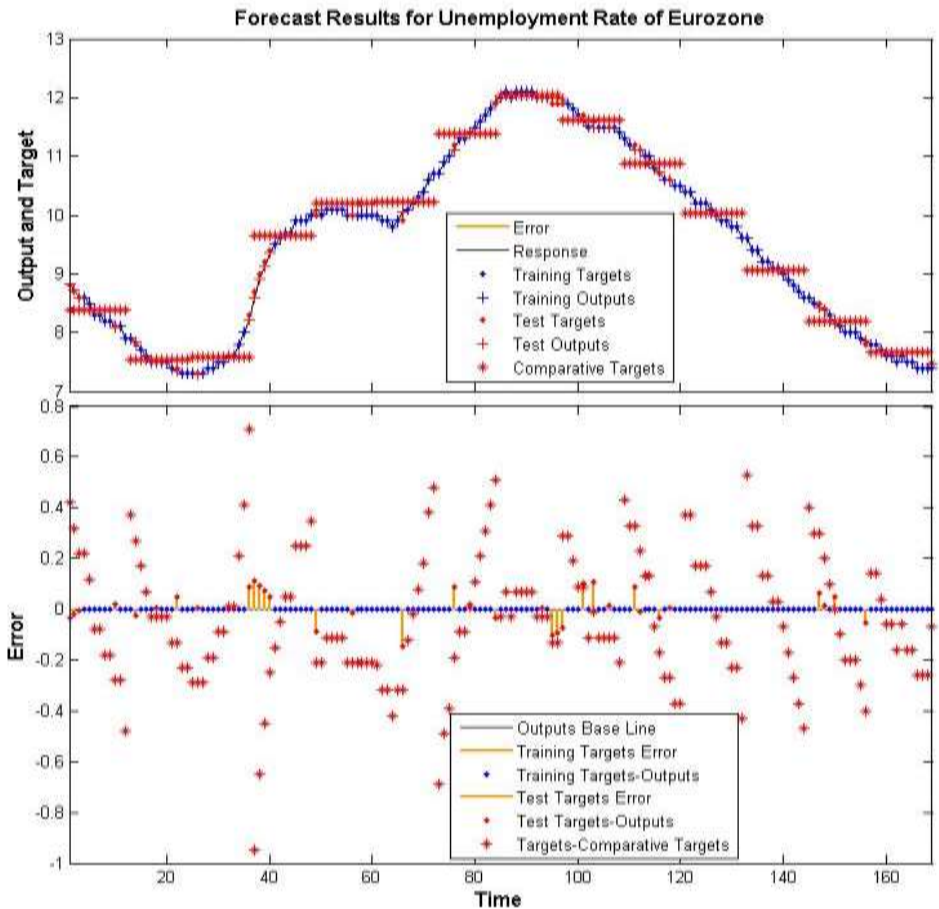

Source: ECB; European Commission's Directorate-General for Economic and Financial Affairs; Sentix-project; CEIC Database; WB; IMF; author's calculation

Notes: After training the specific NARX neural network based on both overall regional & country-specific exogenous inputs, the upper subgraph shows this neural network's outputs, targets, comparative targets and errors versus time, while the lower subgraph adopts this neural network's outputs as the benchmark and shows the gaps between targets/comparative targets and outputs versus time. The time on the horizontal axis corresponds to each month from January 2006 to January 2020, which is also the time span of the exogenous inputs series or target series minus the initial twelve months as the initial time delays for forecast. Both subgraphs also indicate which time points were selected for neural network training and network performance testing.

Figure 39. Specific NARX neural network's forecast result for unemployment rate of Eurozone based on both whole area and subdivision area related exogenous inputs

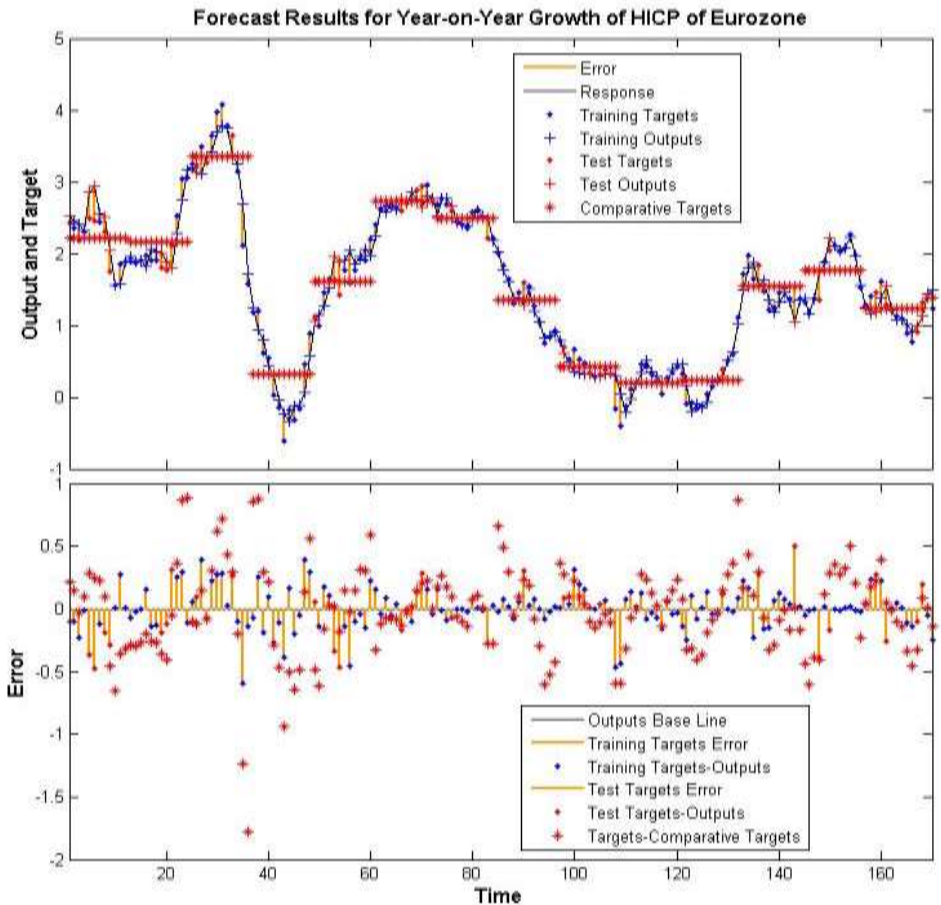

Source: ECB; European Commission's Directorate-General for Economic and Financial Affairs; Sentix-project; CEIC Database; WB; IMF; author's calculation

Notes: After training the specific NARX neural network based on overall regional exogenous inputs, the upper subgraph shows this neural network's outputs, targets, comparative targets and errors versus time, while the lower subgraph adopts this neural network's outputs as the benchmark and shows the gaps between targets/comparative targets and outputs versus time. The time on the horizontal axis corresponds to each month from January 2006 to February 2020, which is also the time span of the exogenous inputs series or target series minus the initial twelve months as the initial time delays for forecast. Both subgraphs also indicate which time points were selected for neural network training and network performance testing.

Figure 40. Specific NARX neural network's forecast result for year-on-year growth of HICP of Eurozone based on whole area related exogenous inputs

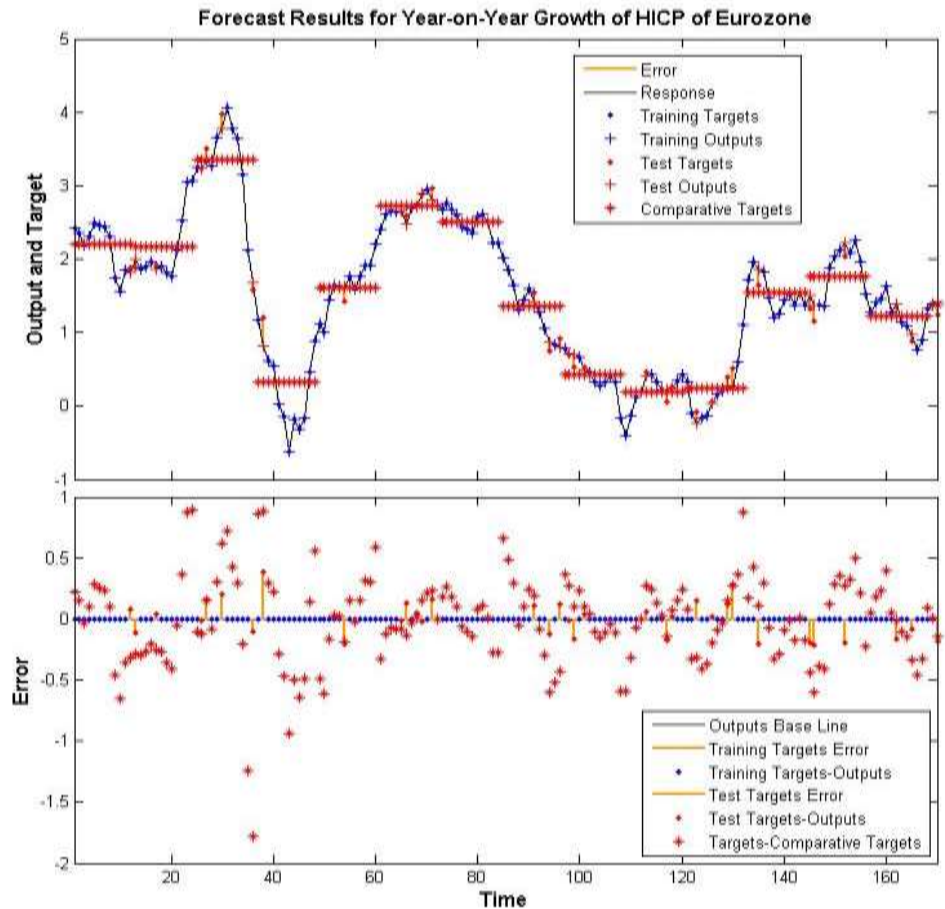

Source: ECB; European Commission's Directorate-General for Economic and Financial Affairs; Sentix-project; CEIC Database; WB; IMF; author's calculation

Notes: After training the specific NARX neural network based on both overall regional & country-specific exogenous inputs, the upper subgraph shows this neural network's outputs, targets, comparative targets and errors versus time, while the lower subgraph adopts this neural network's outputs as the benchmark and shows the gaps between targets/comparative targets and outputs versus time. The time on the horizontal axis corresponds to each month from January 2006 to February 2020, which is also the time span of the exogenous inputs series or target series minus the initial twelve months as the initial time delays for forecast. Both subgraphs also indicate which time points were selected for neural network training and network performance testing.

Figure 41. Specific NARX neural network's forecast result for year-on-year growth of HICP of Eurozone based on both whole area and subdivision area related exogenous inputs

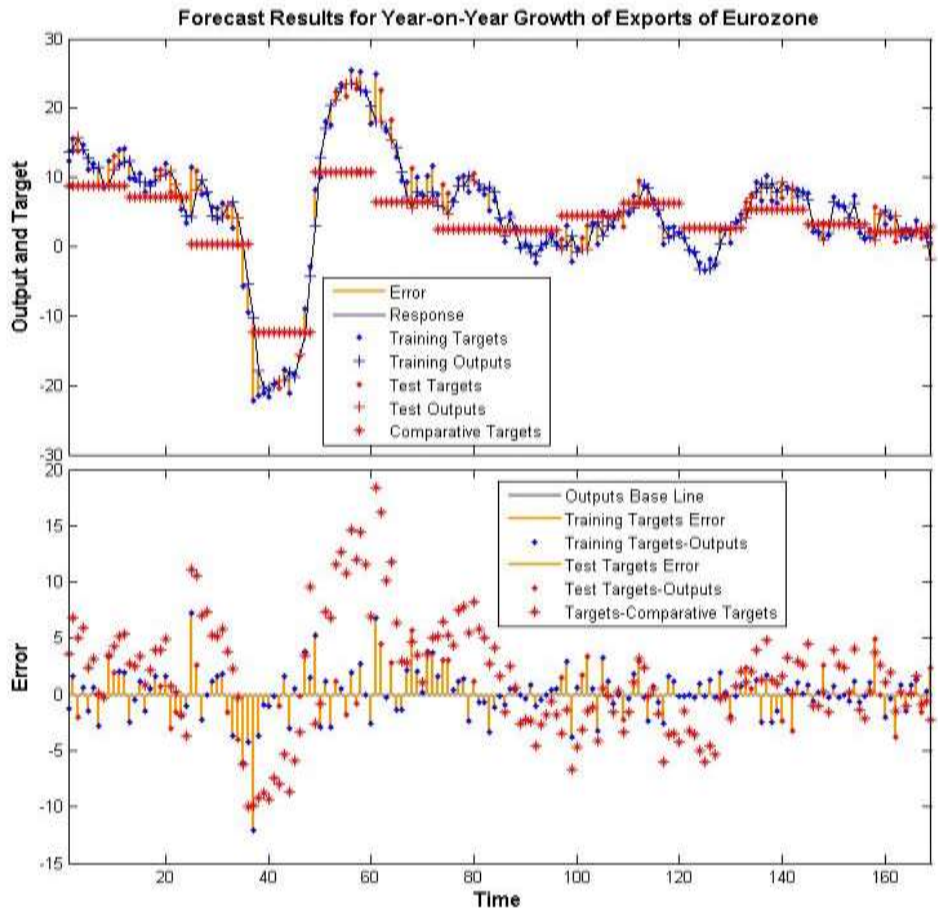

Source: ECB; European Commission's Directorate-General for Economic and Financial Affairs; Sentix-project; CEIC Database; WB; IMF; author's calculation

Notes: After training the specific NARX neural network based on overall regional exogenous inputs, the upper subgraph shows this neural network's outputs, targets, comparative targets and errors versus time, while the lower subgraph adopts this neural network's outputs as the benchmark and shows the gaps between targets/comparative targets and outputs versus time. The time on the horizontal axis corresponds to each month from January 2006 to January 2020, which is also the time span of the exogenous inputs series or target series minus the initial twelve months as the initial time delays for forecast. Both subgraphs also indicate which time points were selected for neural network training and network performance testing.

Figure 42. Specific NARX neural network's forecast result for year-on-year growth of exports of Eurozone based on whole area related exogenous inputs

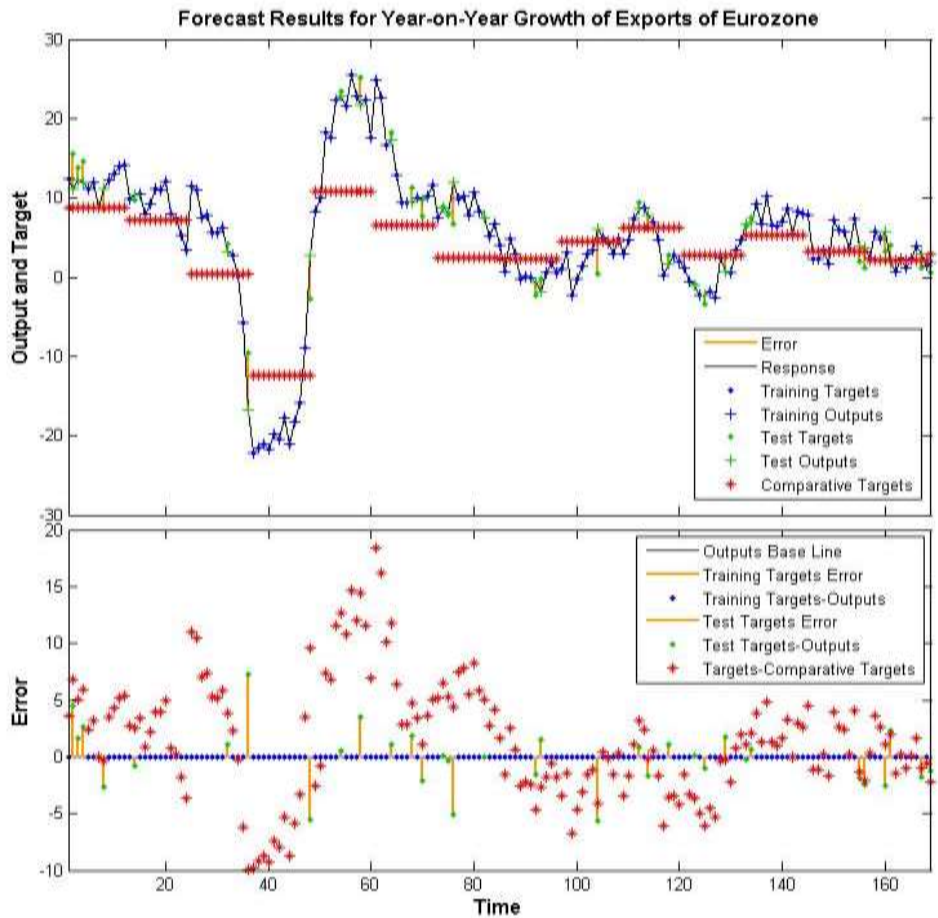

Source: ECB; European Commission's Directorate-General for Economic and Financial Affairs; Sentix-project; CEIC Database; WB; IMF; author's calculation

Notes: After training the specific NARX neural network based on both overall regional & country-specific exogenous inputs, the upper subgraph shows this neural network's outputs, targets, comparative targets and errors versus time, while the lower subgraph adopts this neural network's outputs as the benchmark and shows the gaps between targets/comparative targets and outputs versus time. The time on the horizontal axis corresponds to each month from January 2006 to January 2020, which is also the time span of the exogenous inputs series or target series minus the initial twelve months as the initial time delays for forecast. Both subgraphs also indicate which time points were selected for neural network training and network performance testing.

Figure 43. Specific NARX neural network's forecast result for year-on-year growth of exports of Eurozone based on both whole area and subdivision area related exogenous inputs

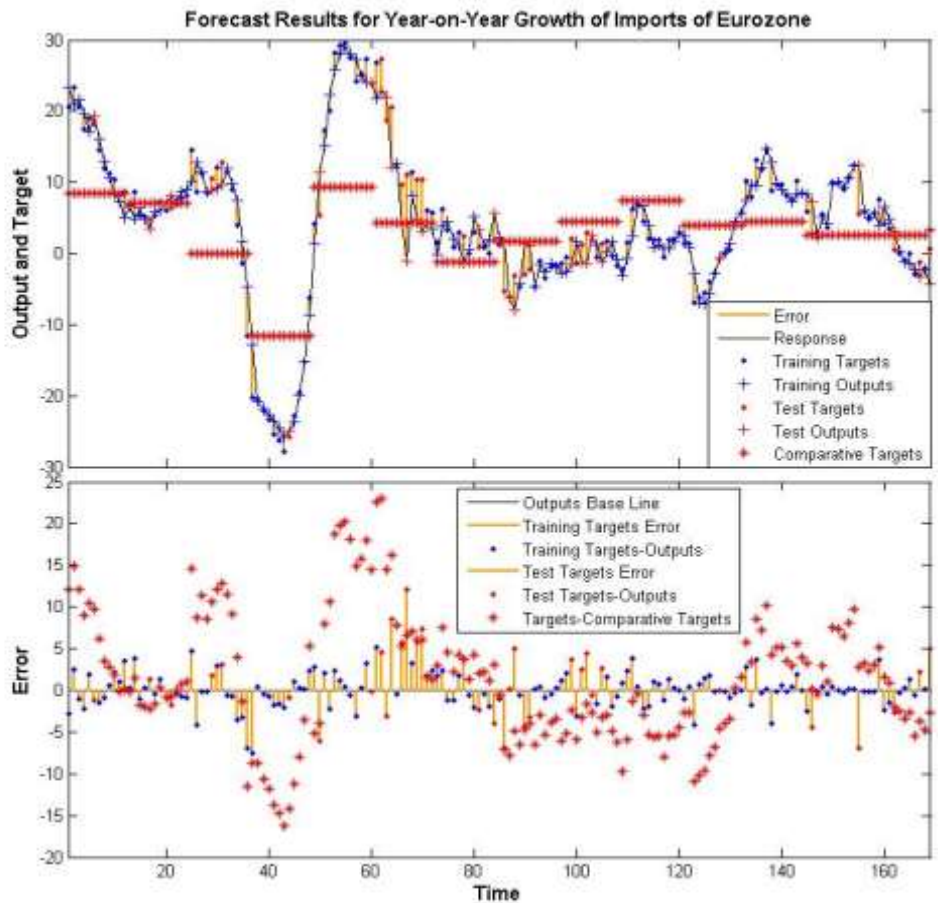

Source: ECB; European Commission's Directorate-General for Economic and Financial Affairs; Sentix-project; CEIC Database; WB; IMF; author's calculation

Notes: After training the specific NARX neural network based on overall regional exogenous inputs, the upper subgraph shows this neural network's outputs, targets, comparative targets and errors versus time, while the lower subgraph adopts this neural network's outputs as the benchmark and shows the gaps between targets/comparative targets and outputs versus time. The time on the horizontal axis corresponds to each month from January 2006 to January 2020, which is also the time span of the exogenous inputs series or target series minus the initial twelve months as the initial time delays for forecast. Both subgraphs also indicate which time points were selected for neural network training and network performance testing.

Figure 44. Specific NARX neural network's forecast result for year-on-year growth of imports of Eurozone based on whole area related exogenous inputs

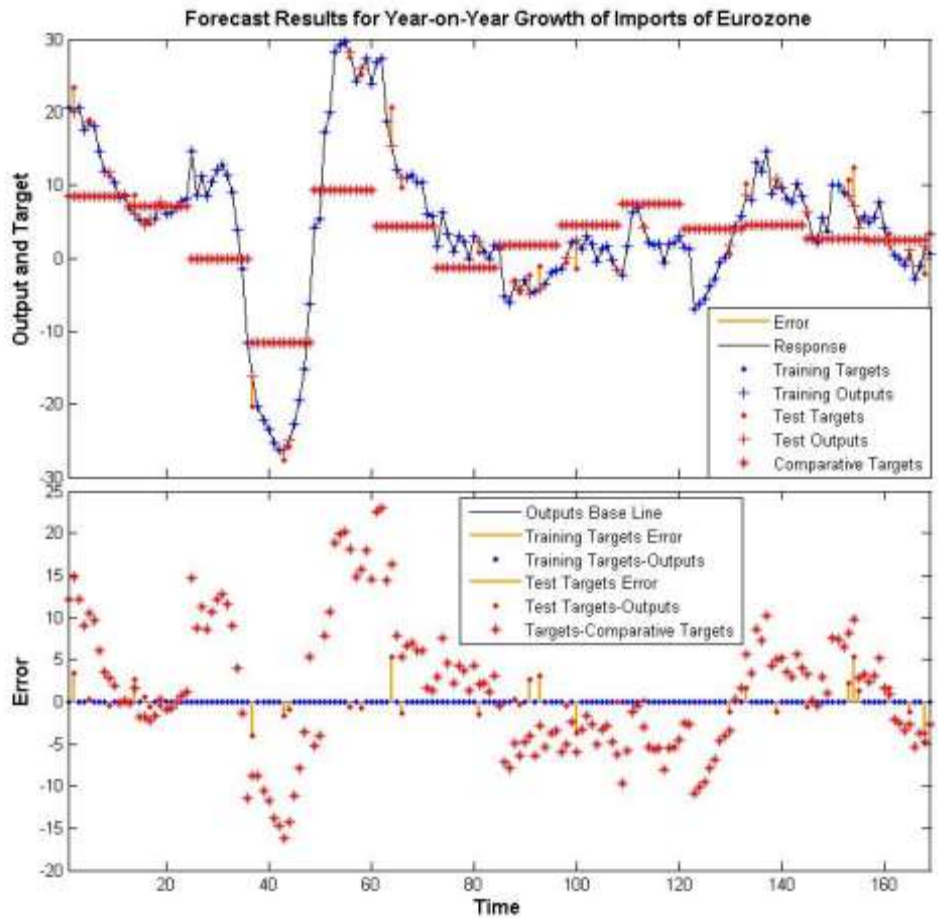

Source: ECB; European Commission's Directorate-General for Economic and Financial Affairs; Sentix-project; CEIC Database; WB; IMF; author's calculation

Notes: After training the specific NARX neural network based on both overall regional & country-specific exogenous inputs, the upper subgraph shows this neural network's outputs, targets, comparative targets and errors versus time, while the lower subgraph adopts this neural network's outputs as the benchmark and shows the gaps between targets/comparative targets and outputs versus time. The time on the horizontal axis corresponds to each month from January 2006 to January 2020, which is also the time span of the exogenous inputs series or target series minus the initial twelve months as the initial time delays for forecast. Both subgraphs also indicate which time points were selected for neural network training and network performance testing.

Figure 45. Specific NARX neural network's forecast result for year-on-year growth of imports of Eurozone based on both whole area and subdivision area related exogenous inputs

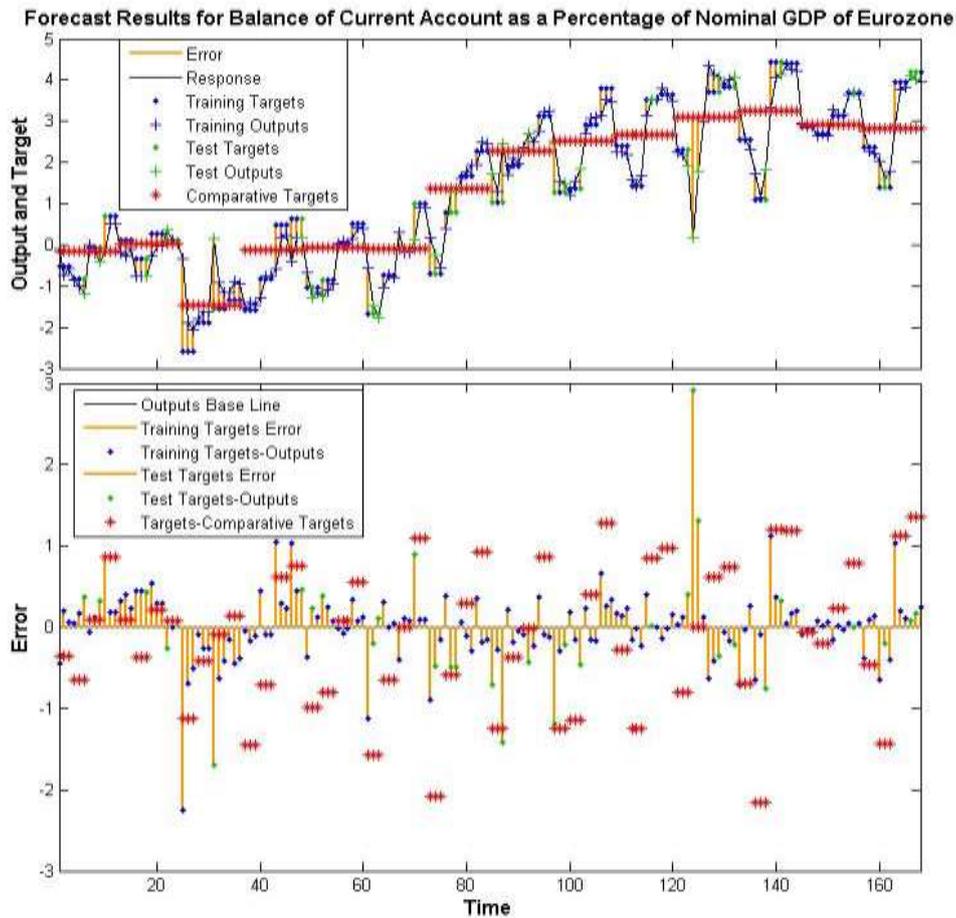

Source: ECB; European Commission's Directorate-General for Economic and Financial Affairs; Sentix-project; CEIC Database; WB; IMF; author's calculation

Notes: After training the specific NARX neural network based on overall regional exogenous inputs, the upper subgraph shows this neural network's outputs, targets, comparative targets and errors versus time, while the lower subgraph adopts this neural network's outputs as the benchmark and shows the gaps between targets/comparative targets and outputs versus time. The time on the horizontal axis corresponds to each month from January 2006 to December 2019, which is also the time span of the exogenous inputs series or target series minus the initial twelve months as the initial time delays for forecast. Both subgraphs also indicate which time points were selected for neural network training and network performance testing.

Figure 46. Specific NARX neural network's forecast result for balance of current account as a percentage of nominal GDP of Eurozone based on whole area related exogenous inputs

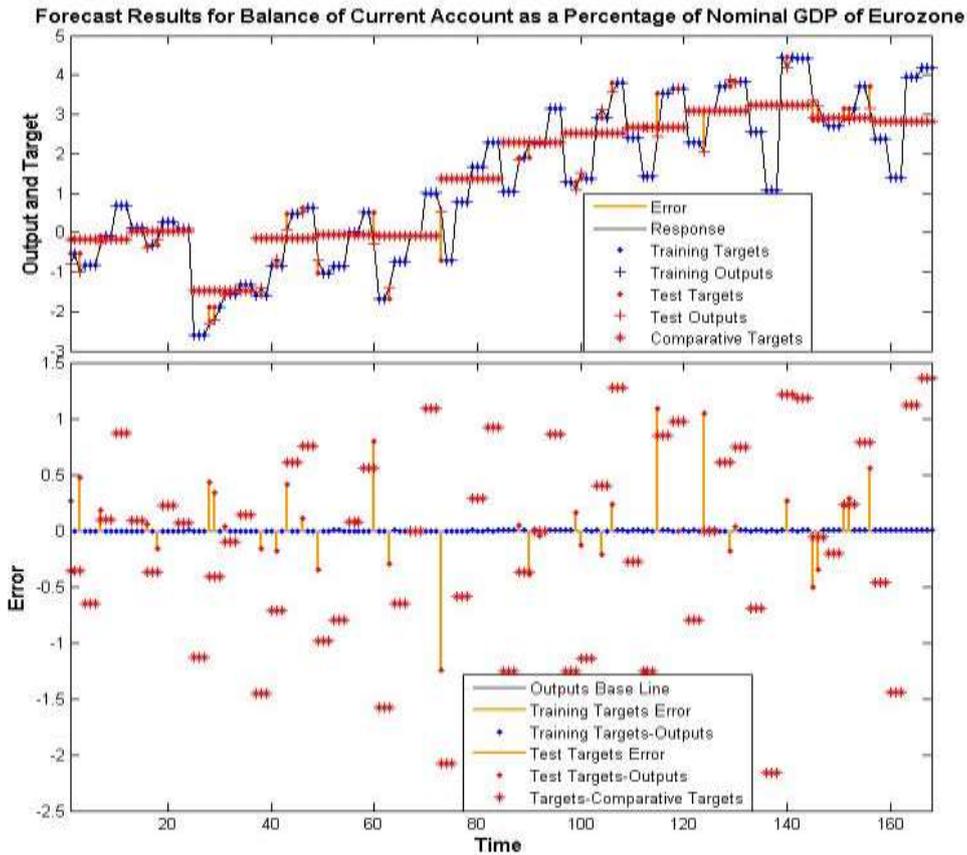

Source: ECB; European Commission's Directorate-General for Economic and Financial Affairs; Sentix-project; CEIC Database; WB; IMF; author's calculation

Notes: After training the specific NARX neural network based on both overall regional & country-specific exogenous inputs, the upper subgraph shows this neural network's outputs, targets, comparative targets and errors versus time, while the lower subgraph adopts this neural network's outputs as the benchmark and shows the gaps between targets/comparative targets and outputs versus time. The time on the horizontal axis corresponds to each month from January 2006 to December 2019, which is also the time span of the exogenous inputs series or target series minus the initial twelve months as the initial time delays for forecast. Both subgraphs also indicate which time points were selected for neural network training and network performance testing.

Figure 47. Specific NARX neural network's forecast result for balance of current account as a percentage of nominal GDP of Eurozone based on both whole area and subdivision area related exogenous inputs

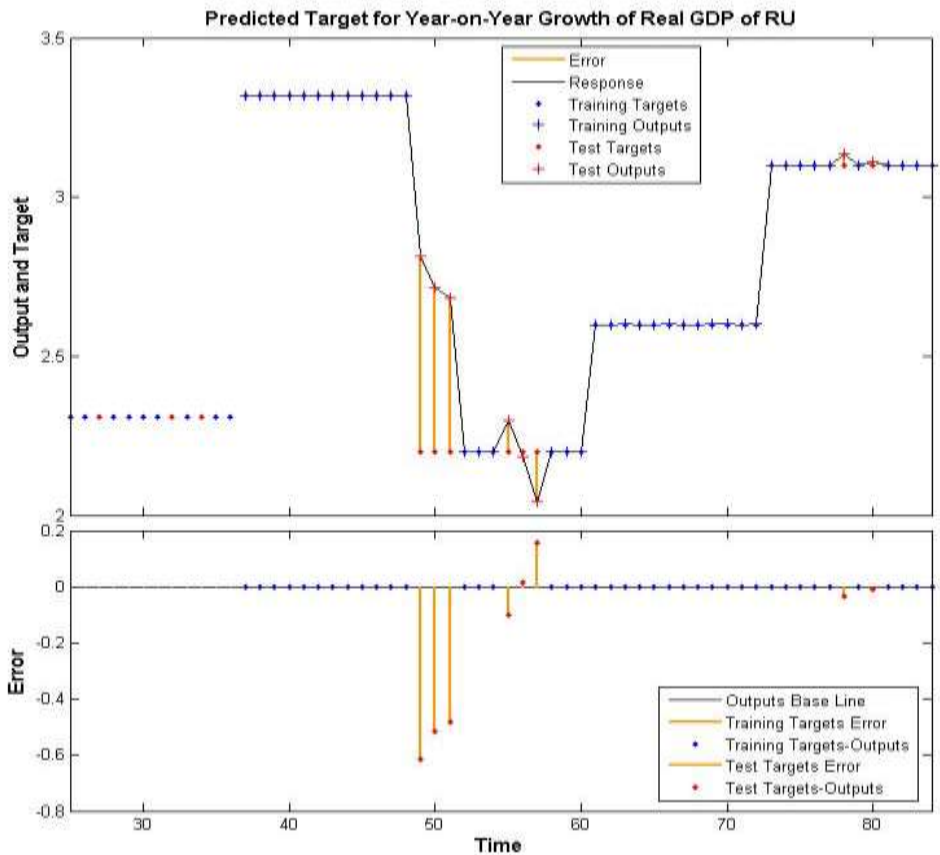

Source: Russia Federal State Statistics Service (RFSSS); author's calculation

Notes: After training the specific NARX neural network based on limited & most relevant exogenous inputs, the upper subgraph displays this neural network's predicted targets, actual official targets, and errors versus time, while the lower subgraph adopts this neural network's predicted targets as the benchmark and shows the gaps between actual official targets and predicted targets versus time. The time on the horizontal axis corresponds to each month from January 2014 to December 2020, which is also the time span of the exogenous inputs series (from January 2013 to December 2020) minus the initial twelve months as the initial time delays for target prediction, while there are no predicted targets in the initial several periods, due to the training target missing caused by the data missing in the actual official targets series (only from January 2016 to December 2020). Both subgraphs indicate which time points were selected for neural network training and network performance testing.

Figure 48. Specific NARX neural network's prediction of target for year-on-year growth of real GDP of RU based on limited & most relevant exogenous inputs

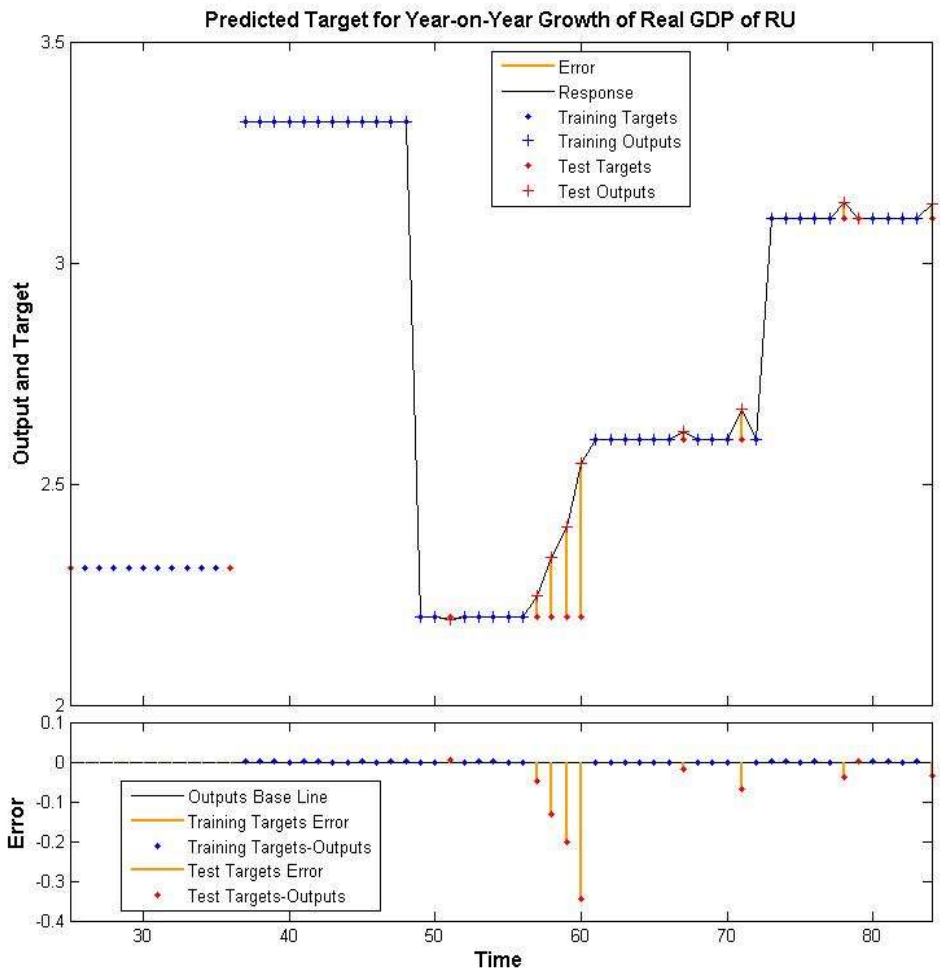

Source: RFSSS; Central Bank of the Russian Federation (CBRF); Russia Federal Treasury (RFT); Federal Tax Service of Russia (FTSR); Ministry of Finance of the Russian Federation (MFRF); Ministry of Labour and Social Security of the Russian Federation (MLSSRF); Russia Federal Customs Service (RFCS); Ministry of Economic Development of the Russian Federation (MEDRF); Moscow Exchange; Institute of World Economy and International Relations, Russian Academy of Sciences (IMEMO RAS); Vnesheconombank of Russia (VEB); Higher School Of Economics Russia (HSER); Bank for International Settlements (BIS); IMF; WB; CEIC Database; author's calculation

Notes: After training the specific NARX neural network based on abundant & comprehensive exogenous inputs, the upper subgraph displays this neural network's predicted targets, actual official targets, and errors versus time, while the lower subgraph adopts this neural network's predicted targets as the benchmark and shows the gaps between actual official targets and predicted targets versus time. The time on the horizontal axis corresponds to each month from January 2014 to December 2020, which is also the time span of the exogenous inputs series (from January 2013 to December 2020) minus the initial twelve months as the initial time delays for target prediction, while there are no predicted targets in the initial several periods, due to the training target missing caused by the data missing in the actual official targets series (only from January 2016 to December 2020). Both subgraphs indicate which time points were selected for neural network training and network performance testing.

Figure 49. Specific NARX neural network's prediction of target for year-on-year growth of real GDP of RU based on abundant & comprehensive exogenous inputs

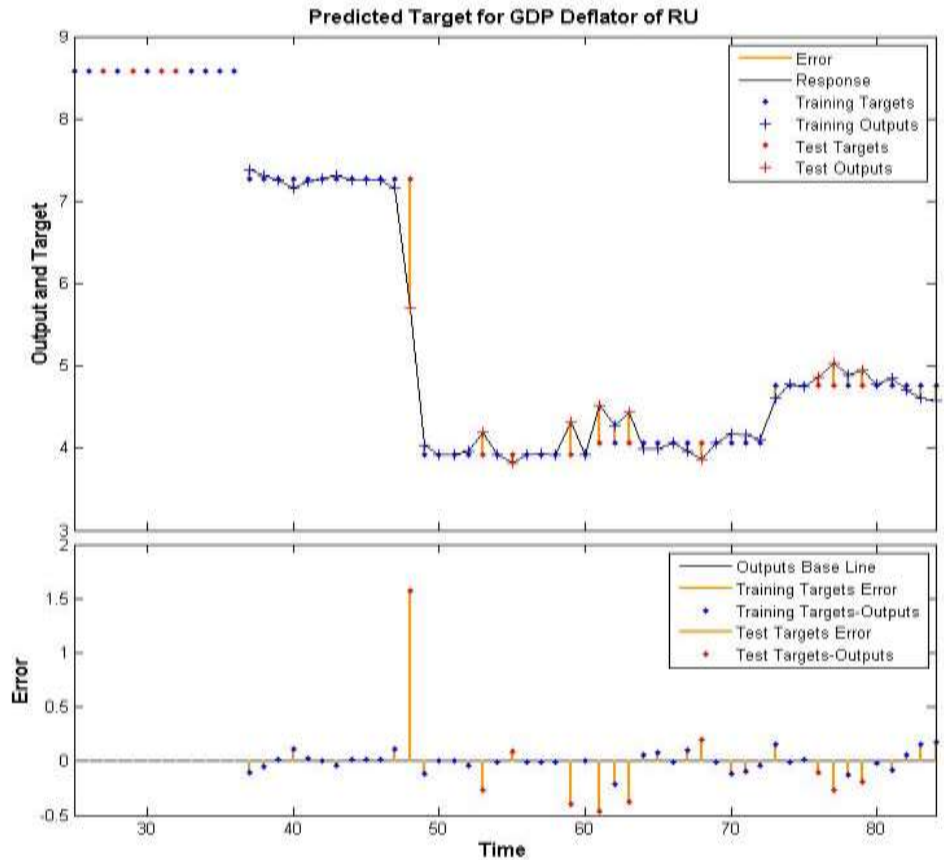

Source: RFSSS; author's calculation

Notes: After training the specific NARX neural network based on limited & most relevant exogenous inputs, the upper subgraph displays this neural network's predicted targets, actual official targets, and errors versus time, while the lower subgraph adopts this neural network's predicted targets as the benchmark and shows the gaps between actual official targets and predicted targets versus time. The time on the horizontal axis corresponds to each month from January 2014 to December 2020, which is also the time span of the exogenous inputs series (from January 2013 to December 2020) minus the initial twelve months as the initial time delays for target prediction, while there are no predicted targets in the initial several periods, due to the training target missing caused by the data missing in the actual official targets series (only from January 2016 to December 2020). Both subgraphs indicate which time points were selected for neural network training and network performance testing.

Figure 50. Specific NARX neural network's prediction of target for GDP deflator of RU based on limited & most relevant exogenous inputs

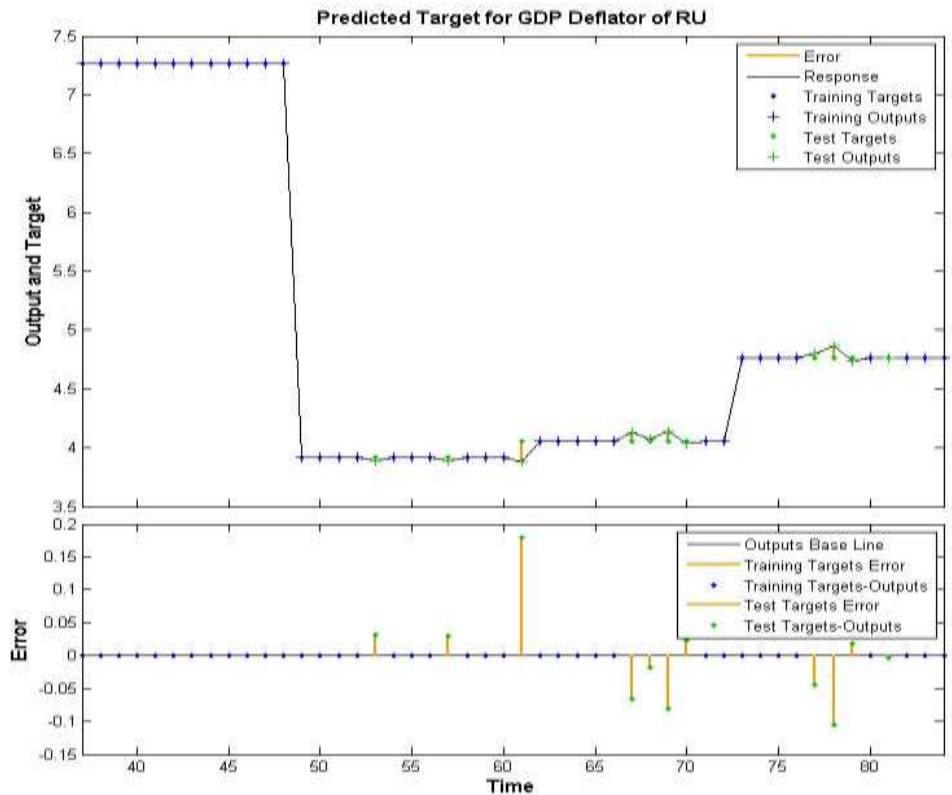

Source: RFSSS; CBRF; RFT; FTSR; MFRF; MLSSRF; RFCS; MEDRF; Moscow Exchange; IMEMO RAS; VEB; HSER; BIS; IMF; WB; CEIC Database; author's calculation

Notes: After training the specific NARX neural network based on abundant & comprehensive exogenous inputs, the upper subgraph displays this neural network's predicted targets, actual official targets, and errors versus time, while the lower subgraph adopts this neural network's predicted targets as the benchmark and shows the gaps between actual official targets and predicted targets versus time. The time on the horizontal axis corresponds to each month from January 2014 to December 2020, which is also the time span of the exogenous inputs series (from January 2013 to December 2020) minus the initial twelve months as the initial time delays for target prediction, while there are no predicted targets in the initial several periods, due to the training target missing caused by the data missing in the actual official targets series (only from January 2016 to December 2020). Both subgraphs indicate which time points were selected for neural network training and network performance testing.

Figure 51. Specific NARX neural network's prediction of target for GDP deflator of RU based on abundant & comprehensive exogenous inputs

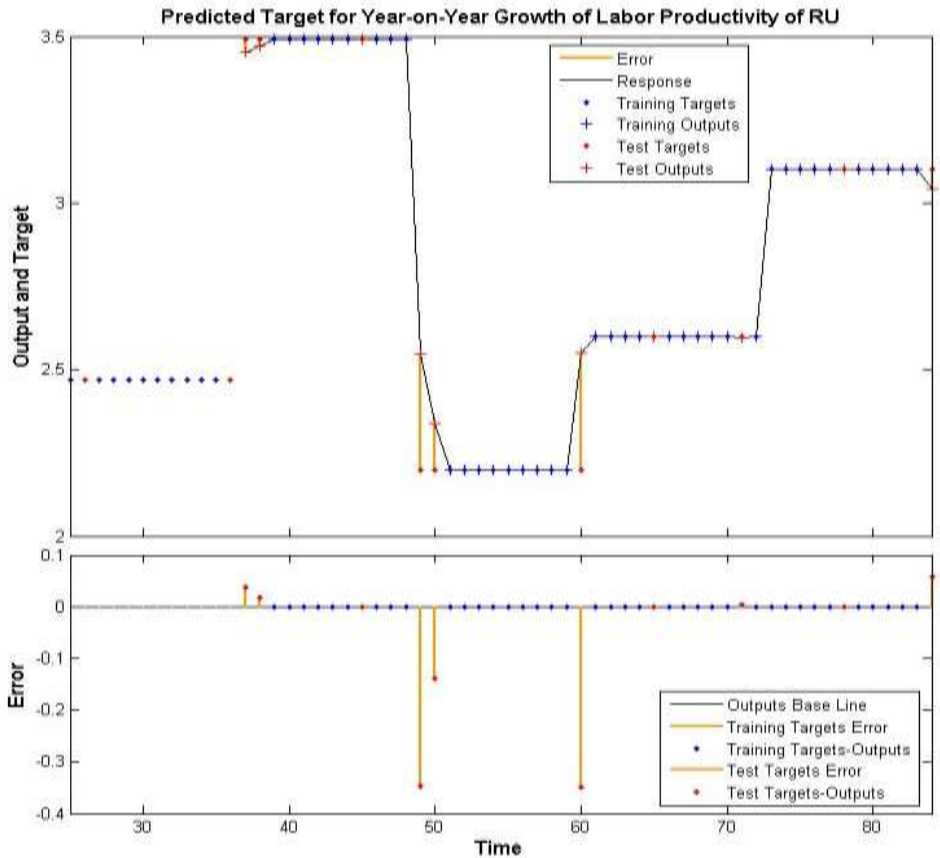

Source: RFSSS; author's calculation

Notes: After training the specific NARX neural network based on limited & most relevant exogenous inputs, the upper subgraph displays this neural network's predicted targets, actual official targets, and errors versus time, while the lower subgraph adopts this neural network's predicted targets as the benchmark and shows the gaps between actual official targets and predicted targets versus time. The time on the horizontal axis corresponds to each month from January 2014 to December 2020, which is also the time span of the exogenous inputs series (from January 2013 to December 2020) minus the initial twelve months as the initial time delays for target prediction, while there are no predicted targets in the initial several periods, due to the training target missing caused by the data missing in the actual official targets series (only from January 2016 to December 2020). Both subgraphs indicate which time points were selected for neural network training and network performance testing.

Figure 52. Specific NARX neural network's prediction of target for year-on-year growth of labor productivity of RU based on limited & most relevant exogenous inputs

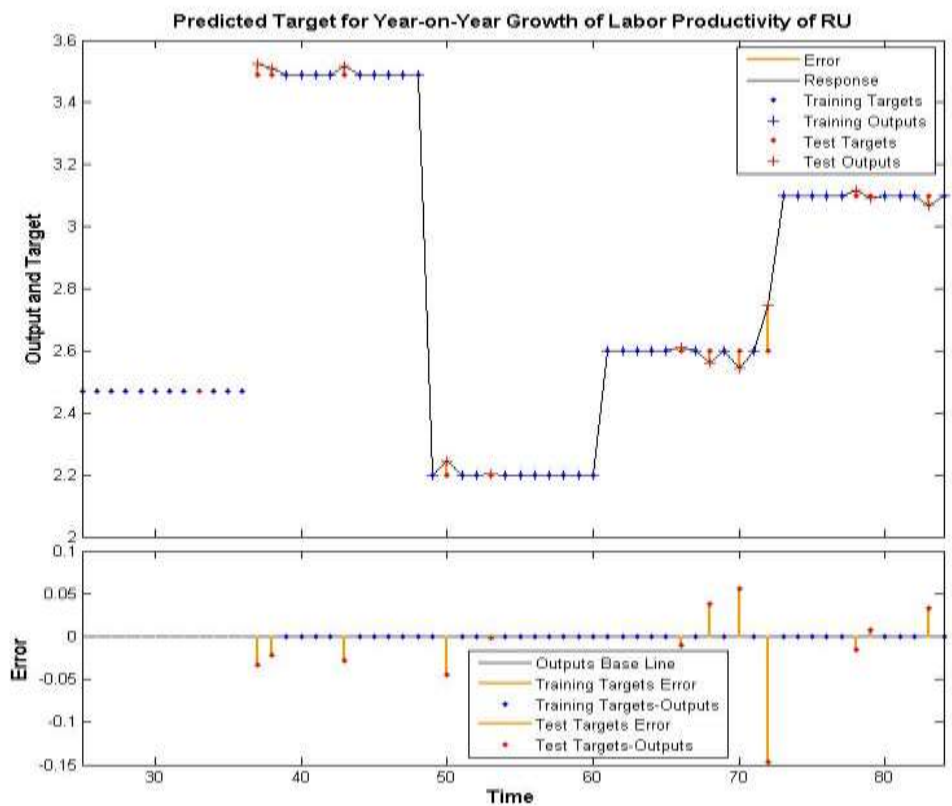

Source: RFSSS; CBRF; RFT; FTSR; MFRF; MLSSRF; RFCS; MEDRF; Moscow Exchange; IMEMO RAS; VEB; HSER; BIS; IMF; WB; CEIC Database; author's calculation

Notes: After training the specific NARX neural network based on abundant & comprehensive exogenous inputs, the upper subgraph displays this neural network's predicted targets, actual official targets, and errors versus time, while the lower subgraph adopts this neural network's predicted targets as the benchmark and shows the gaps between actual official targets and predicted targets versus time. The time on the horizontal axis corresponds to each month from January 2014 to December 2020, which is also the time span of the exogenous inputs series (from January 2013 to December 2020) minus the initial twelve months as the initial time delays for target prediction, while there are no predicted targets in the initial several periods, due to the training target missing caused by the data missing in the actual official targets series (only from January 2016 to December 2020). Both subgraphs indicate which time points were selected for neural network training and network performance testing.

Figure 53. Specific NARX neural network's prediction of target for year-on-year growth of labor productivity of RU based on abundant & comprehensive exogenous inputs

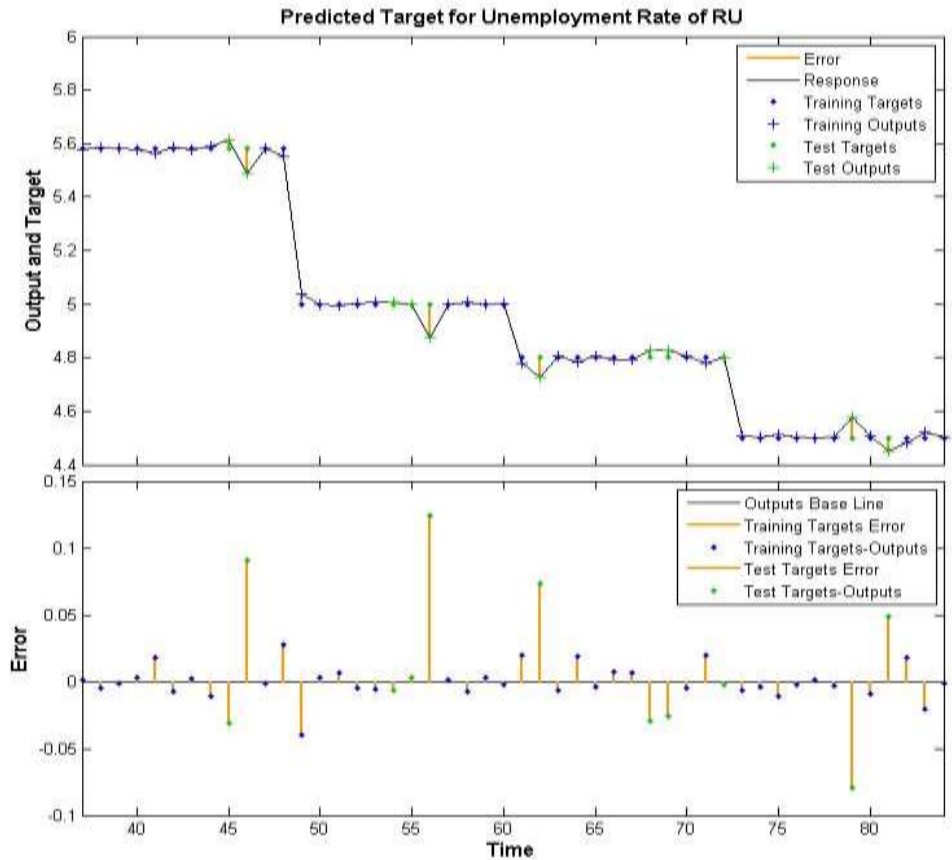

Source: RFSSS; CEIC Database; author's calculation

Notes: After training the specific NARX neural network based on limited & most relevant exogenous inputs, the upper subgraph displays this neural network's predicted targets, actual official targets, and errors versus time, while the lower subgraph adopts this neural network's predicted targets as the benchmark and shows the gaps between actual official targets and predicted targets versus time. The time on the horizontal axis corresponds to each month from January 2014 to December 2020, which is also the time span of the exogenous inputs series (from January 2013 to December 2020) minus the initial twelve months as the initial time delays for target prediction, while there are no predicted targets in the initial several periods, due to the training target missing caused by the data missing in the actual official targets series (only from January 2016 to December 2020). Both subgraphs indicate which time points were selected for neural network training and network performance testing.

Figure 54. Specific NARX neural network's prediction of target for unemployment rate of RU based on limited & most relevant exogenous inputs

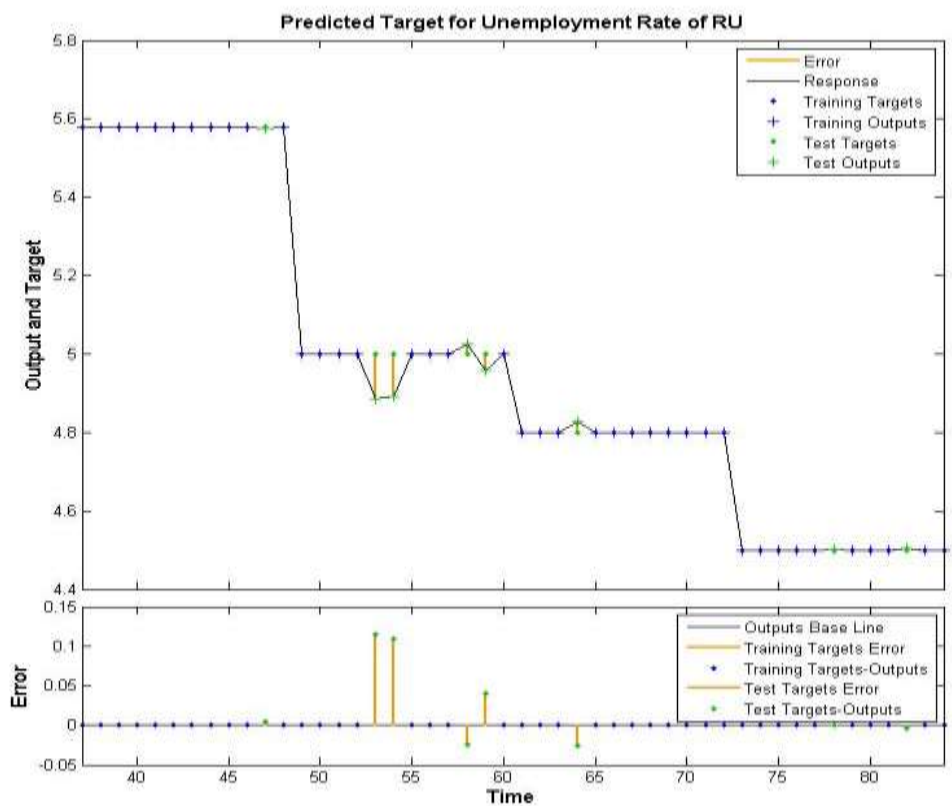

Source: RFSSS; CBRF; RFT; FTSR; MFRF; MLSSRF; RFCS; MEDRF; Moscow Exchange; IMEMO RAS; VEB; HSER; BIS; IMF; WB; CEIC Database; author's calculation

Notes: After training the specific NARX neural network based on abundant & comprehensive exogenous inputs, the upper subgraph displays this neural network's predicted targets, actual official targets, and errors versus time, while the lower subgraph adopts this neural network's predicted targets as the benchmark and shows the gaps between actual official targets and predicted targets versus time. The time on the horizontal axis corresponds to each month from January 2014 to December 2020, which is also the time span of the exogenous inputs series (from January 2013 to December 2020) minus the initial twelve months as the initial time delays for target prediction, while there are no predicted targets in the initial several periods, due to the training target missing caused by the data missing in the actual official targets series (only from January 2016 to December 2020). Both subgraphs indicate which time points were selected for neural network training and network performance testing.

Figure 55. Specific NARX neural network's prediction of target for unemployment rate of RU based on abundant & comprehensive exogenous inputs

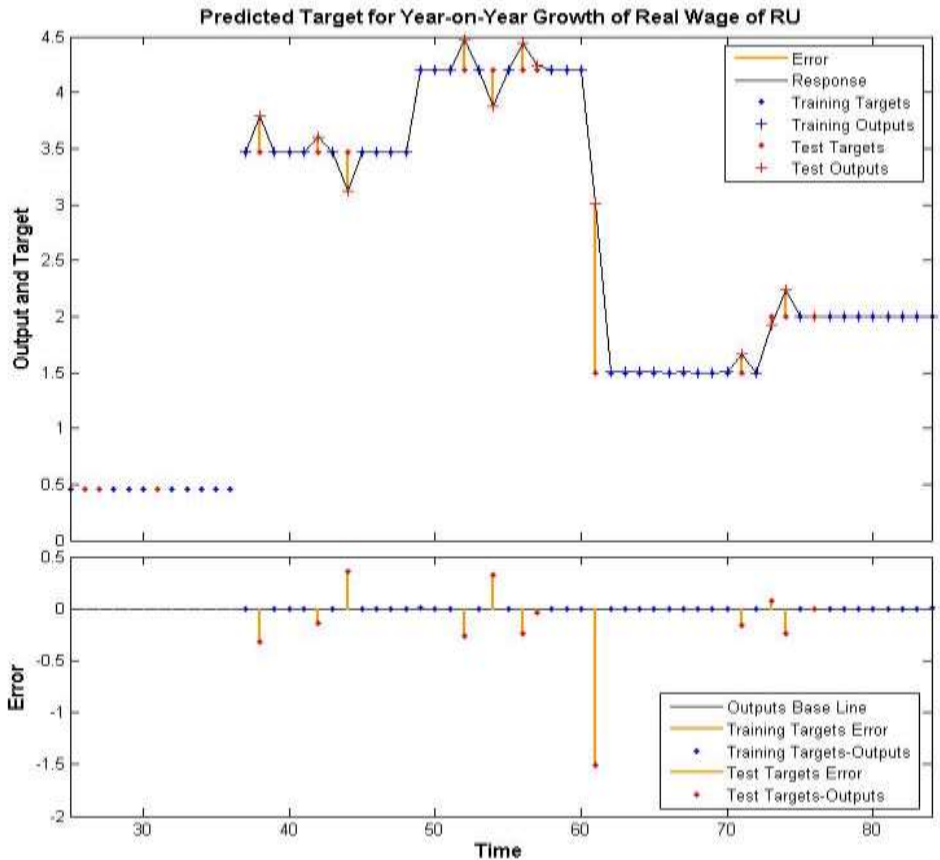

Source: RFSSS; MLSSRF; author's calculation

Notes: After training the specific NARX neural network based on limited & most relevant exogenous inputs, the upper subgraph displays this neural network's predicted targets, actual official targets, and errors versus time, while the lower subgraph adopts this neural network's predicted targets as the benchmark and shows the gaps between actual official targets and predicted targets versus time. The time on the horizontal axis corresponds to each month from January 2014 to December 2020, which is also the time span of the exogenous inputs series (from January 2013 to December 2020) minus the initial twelve months as the initial time delays for target prediction, while there are no predicted targets in the initial several periods, due to the training target missing caused by the data missing in the actual official targets series (only from January 2016 to December 2020). Both subgraphs indicate which time points were selected for neural network training and network performance testing.

Figure 56. Specific NARX neural network's prediction of target for year-on-year growth of real wage of RU based on limited & most relevant exogenous inputs

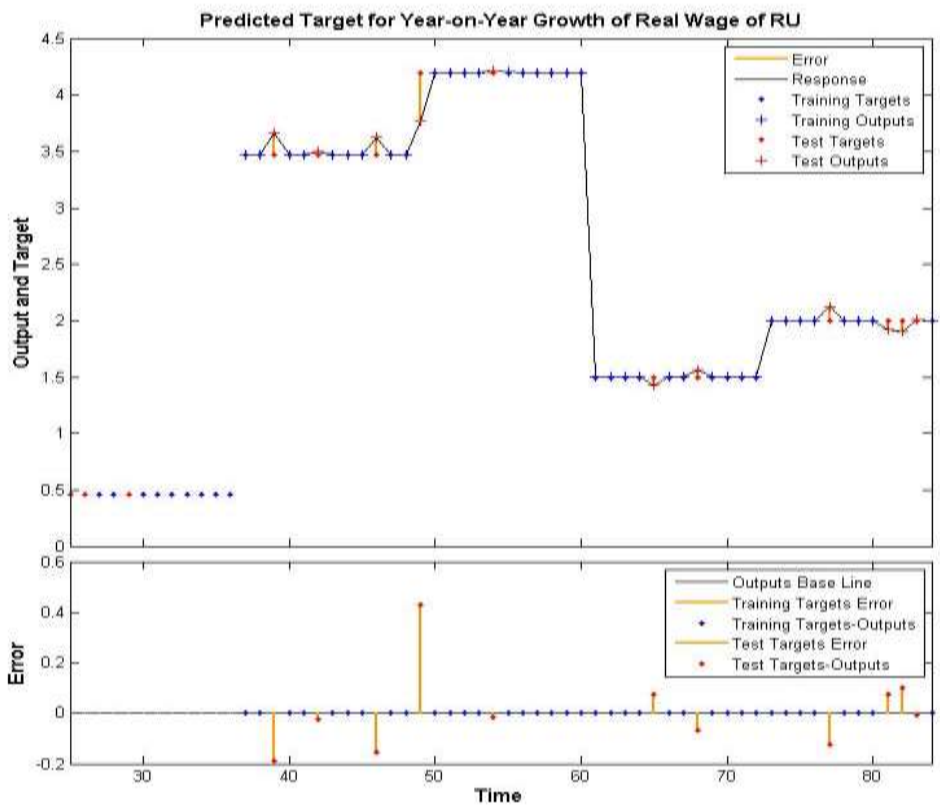

Source: RFSSS; CBRF; RFT; FTSR; MFRF; MLSSRF; RFCS; MEDRF; Moscow Exchange; IMEMO RAS; VEB; HSER; BIS; IMF; WB; CEIC Database; author's calculation

Notes: After training the specific NARX neural network based on abundant & comprehensive exogenous inputs, the upper subgraph displays this neural network's predicted targets, actual official targets, and errors versus time, while the lower subgraph adopts this neural network's predicted targets as the benchmark and shows the gaps between actual official targets and predicted targets versus time. The time on the horizontal axis corresponds to each month from January 2014 to December 2020, which is also the time span of the exogenous inputs series (from January 2013 to December 2020) minus the initial twelve months as the initial time delays for target prediction, while there are no predicted targets in the initial several periods, due to the training target missing caused by the data missing in the actual official targets series (only from January 2016 to December 2020). Both subgraphs indicate which time points were selected for neural network training and network performance testing.

Figure 57. Specific NARX neural network's prediction of target for year-on-year growth of real wage of RU based on abundant & comprehensive exogenous inputs

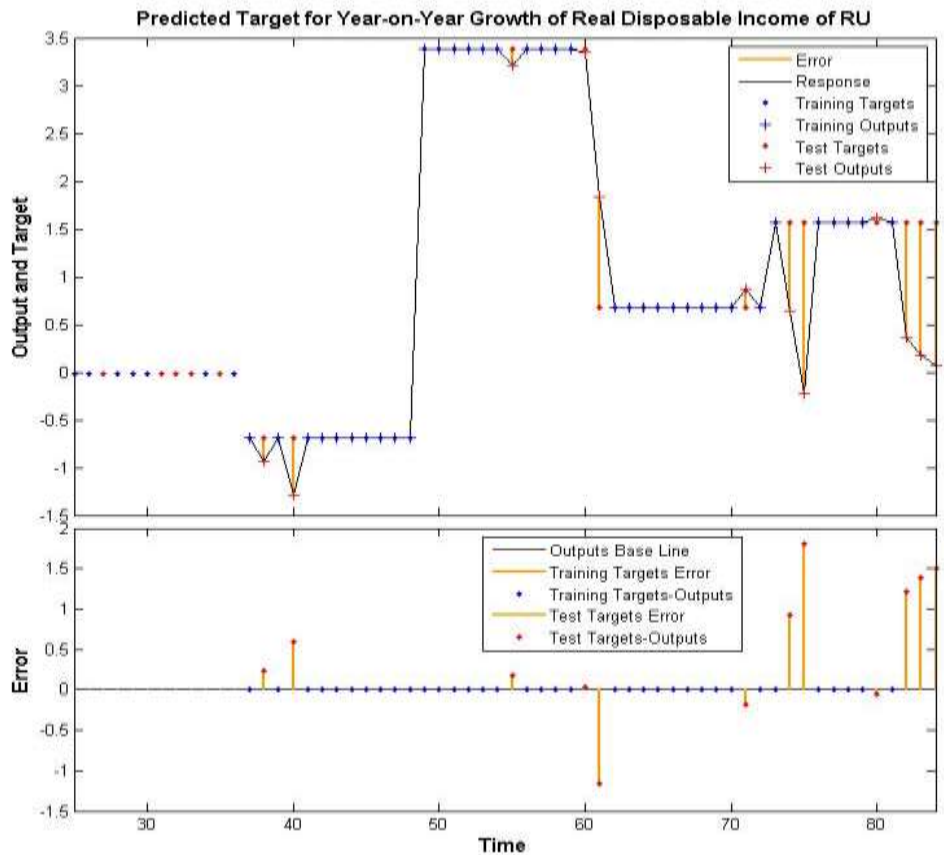

Source: RFSSS; author's calculation

Notes: After training the specific NARX neural network based on limited & most relevant exogenous inputs, the upper subgraph displays this neural network's predicted targets, actual official targets, and errors versus time, while the lower subgraph adopts this neural network's predicted targets as the benchmark and shows the gaps between actual official targets and predicted targets versus time. The time on the horizontal axis corresponds to each month from January 2014 to December 2020, which is also the time span of the exogenous inputs series (from January 2013 to December 2020) minus the initial twelve months as the initial time delays for target prediction, while there are no predicted targets in the initial several periods, due to the training target missing caused by the data missing in the actual official targets series (only from January 2016 to December 2020). Both subgraphs indicate which time points were selected for neural network training and network performance testing.

Figure 58. Specific NARX neural network's prediction of target for year-on-year growth of real disposable income of RU based on limited & most relevant exogenous inputs

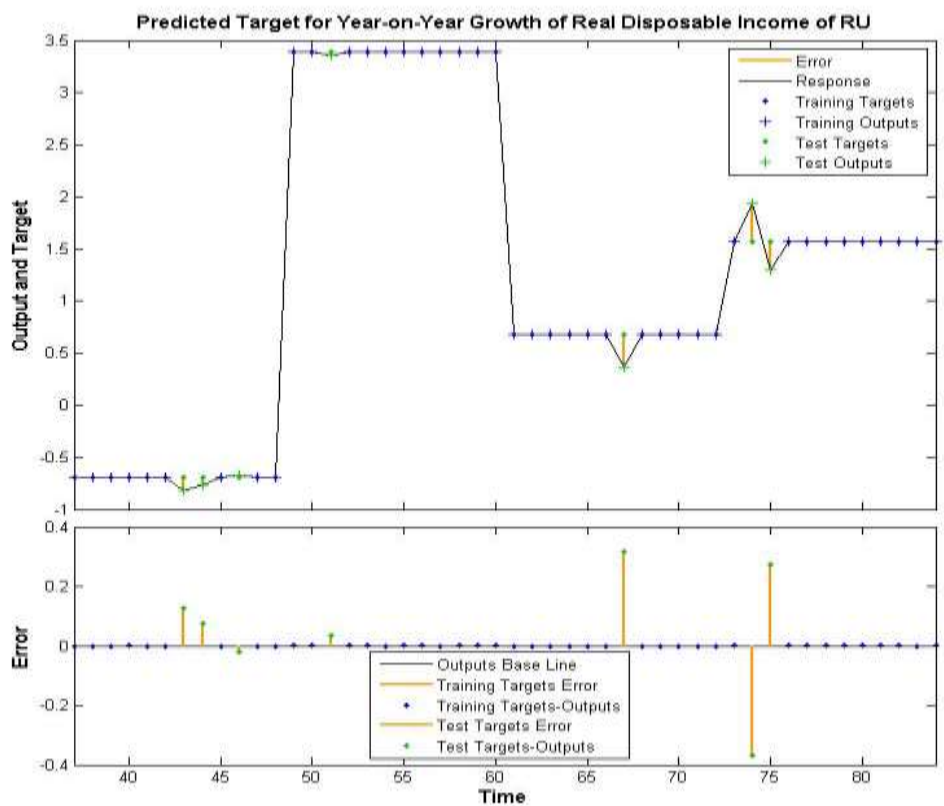

Source: RFSSS; CBRF; RFT; FTSR; MFRF; MLSSRF; RFCS; MEDRF; Moscow Exchange; IMEMO RAS; VEB; HSER; BIS; IMF; WB; CEIC Database; author's calculation

Notes: After training the specific NARX neural network based on abundant & comprehensive exogenous inputs, the upper subgraph displays this neural network's predicted targets, actual official targets, and errors versus time, while the lower subgraph adopts this neural network's predicted targets as the benchmark and shows the gaps between actual official targets and predicted targets versus time. The time on the horizontal axis corresponds to each month from January 2014 to December 2020, which is also the time span of the exogenous inputs series (from January 2013 to December 2020) minus the initial twelve months as the initial time delays for target prediction, while there are no predicted targets in the initial several periods, due to the training target missing caused by the data missing in the actual official targets series (only from January 2016 to December 2020). Both subgraphs indicate which time points were selected for neural network training and network performance testing.

Figure 59. Specific NARX neural network's prediction of target for year-on-year growth of real disposable income of RU based on abundant & comprehensive exogenous inputs

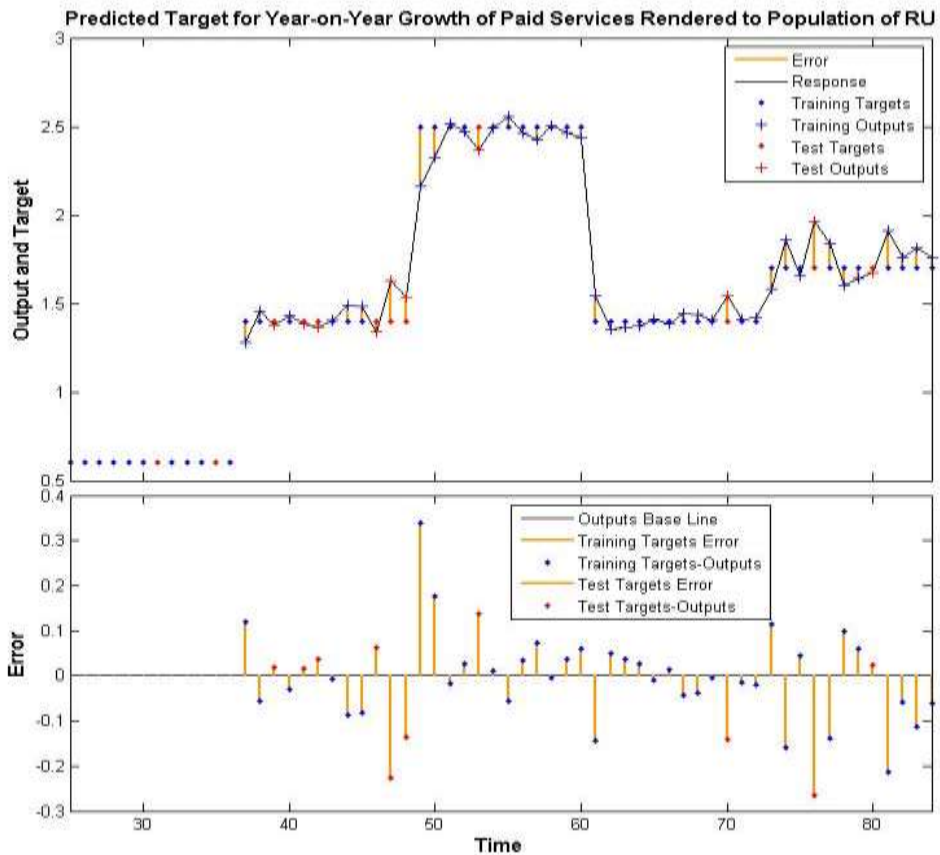

Source: RFSSS; author's calculation

Notes: After training the specific NARX neural network based on limited & most relevant exogenous inputs, the upper subgraph displays this neural network's predicted targets, actual official targets, and errors versus time, while the lower subgraph adopts this neural network's predicted targets as the benchmark and shows the gaps between actual official targets and predicted targets versus time. The time on the horizontal axis corresponds to each month from January 2014 to December 2020, which is also the time span of the exogenous inputs series (from January 2013 to December 2020) minus the initial twelve months as the initial time delays for target prediction, while there are no predicted targets in the initial several periods, due to the training target missing caused by the data missing in the actual official targets series (only from January 2016 to December 2020). Both subgraphs indicate which time points were selected for neural network training and network performance testing.

Figure 60. Specific NARX neural network's prediction of target for year-on-year growth of paid services rendered to population of RU based on limited & most relevant exogenous inputs

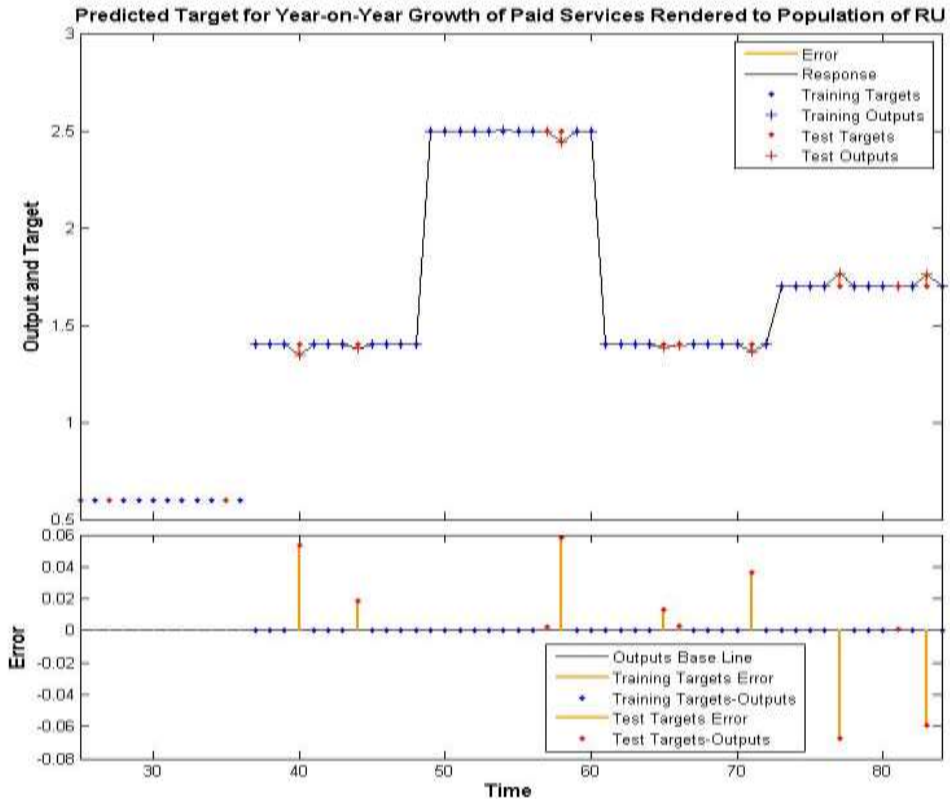

Source: RFSSS; CBRF; RFT; FTSR; MFRF; MLSSRF; RFCS; MEDRF; Moscow Exchange; IMEMO RAS; VEB; HSER; BIS; IMF; WB; CEIC Database; author's calculation

Notes: After training the specific NARX neural network based on abundant & comprehensive exogenous inputs, the upper subgraph displays this neural network's predicted targets, actual official targets, and errors versus time, while the lower subgraph adopts this neural network's predicted targets as the benchmark and shows the gaps between actual official targets and predicted targets versus time. The time on the horizontal axis corresponds to each month from January 2014 to December 2020, which is also the time span of the exogenous inputs series (from January 2013 to December 2020) minus the initial twelve months as the initial time delays for target prediction, while there are no predicted targets in the initial several periods, due to the training target missing caused by the data missing in the actual official targets series (only from January 2016 to December 2020). Both subgraphs indicate which time points were selected for neural network training and network performance testing.

Figure 61. Specific NARX neural network's prediction of target for year-on-year growth of paid services rendered to population of RU based on abundant & comprehensive exogenous inputs

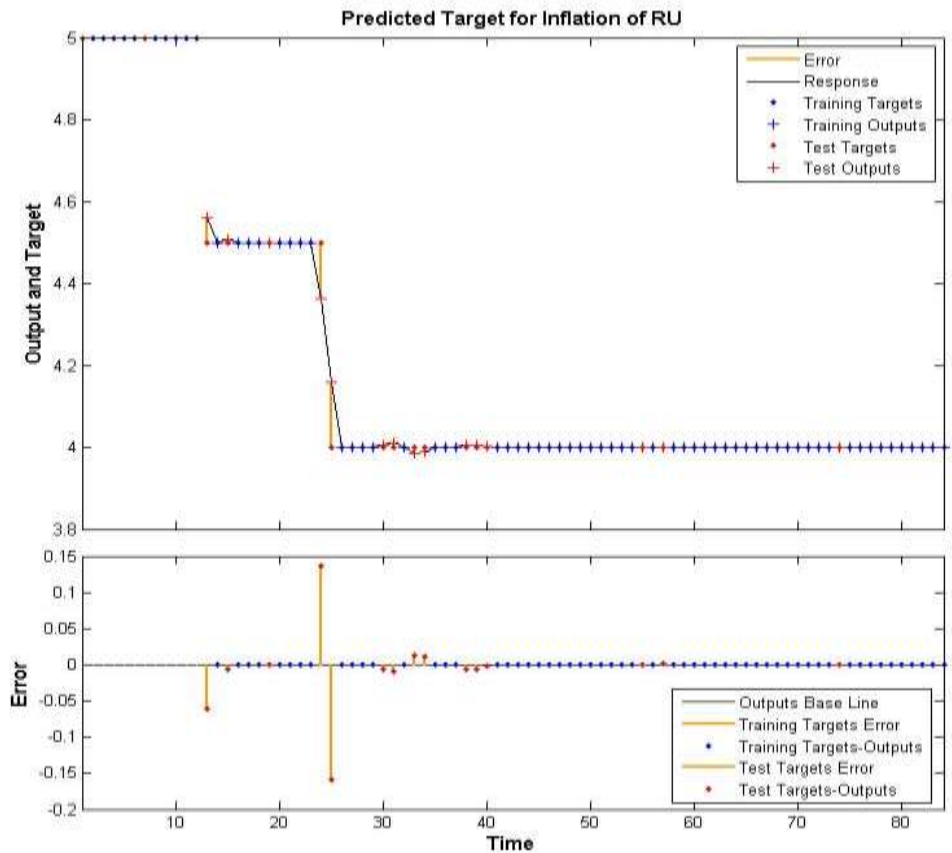

Source: CBRF; RFSSS; author's calculation

Notes: After training the specific NARX neural network based on limited & most relevant exogenous inputs, the upper subgraph displays this neural network's predicted targets, actual official targets, and errors versus time, while the lower subgraph adopts this neural network's predicted targets as the benchmark and shows the gaps between actual official targets and predicted targets versus time. The time on the horizontal axis corresponds to each month from January 2014 to December 2020, which is also the time span of the exogenous inputs series (from January 2013 to December 2020) minus the initial twelve months as the initial time delays for target prediction, while there are no predicted targets in the initial several periods, due to the training target missing caused by the data missing in the actual official targets series (only from January 2016 to December 2020). Both subgraphs indicate which time points were selected for neural network training and network performance testing.

Figure 62. Specific NARX neural network's prediction of target for inflation of RU based on limited & most relevant exogenous inputs

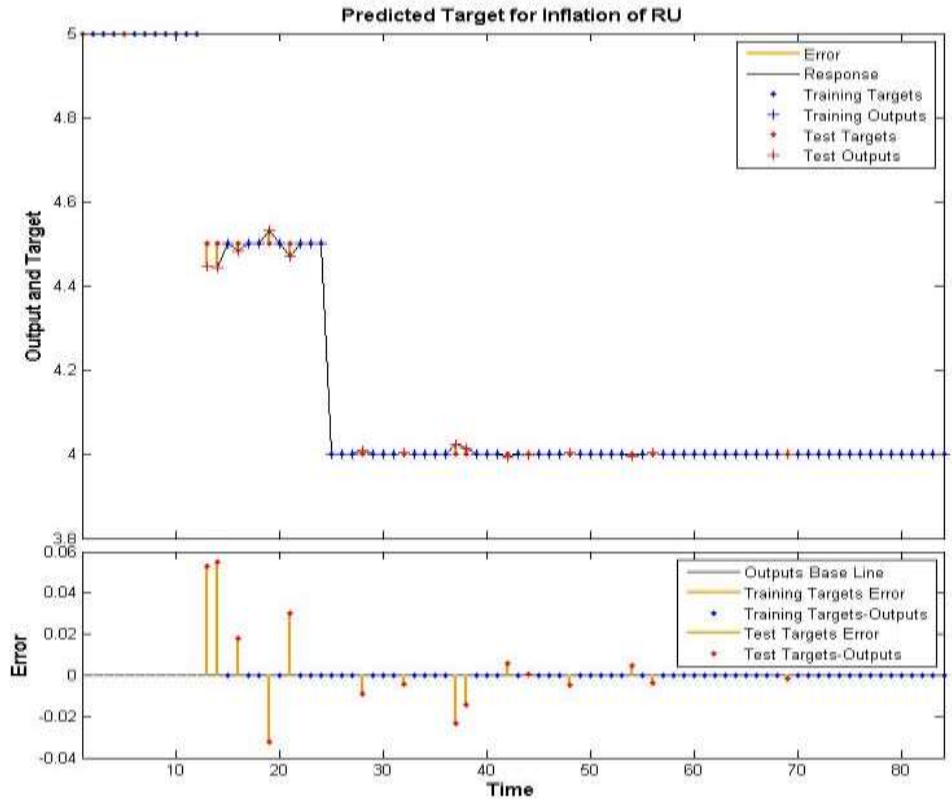

Source: RFSSS; CBRF; RFT; FTSR; MFRF; MLSSRF; RFCS; MEDRF; Moscow Exchange; IMEMO RAS; VEB; HSER; BIS; IMF; WB; CEIC Database; author's calculation

Notes: After training the specific NARX neural network based on abundant & comprehensive exogenous inputs, the upper subgraph displays this neural network's predicted targets, actual official targets, and errors versus time, while the lower subgraph adopts this neural network's predicted targets as the benchmark and shows the gaps between actual official targets and predicted targets versus time. The time on the horizontal axis corresponds to each month from January 2014 to December 2020, which is also the time span of the exogenous inputs series (from January 2013 to December 2020) minus the initial twelve months as the initial time delays for target prediction, while there are no predicted targets in the initial several periods, due to the training target missing caused by the data missing in the actual official targets series (only from January 2016 to December 2020). Both subgraphs indicate which time points were selected for neural network training and network performance testing.

Figure 63. Specific NARX neural network's prediction of target for inflation of RU based on abundant & comprehensive exogenous inputs

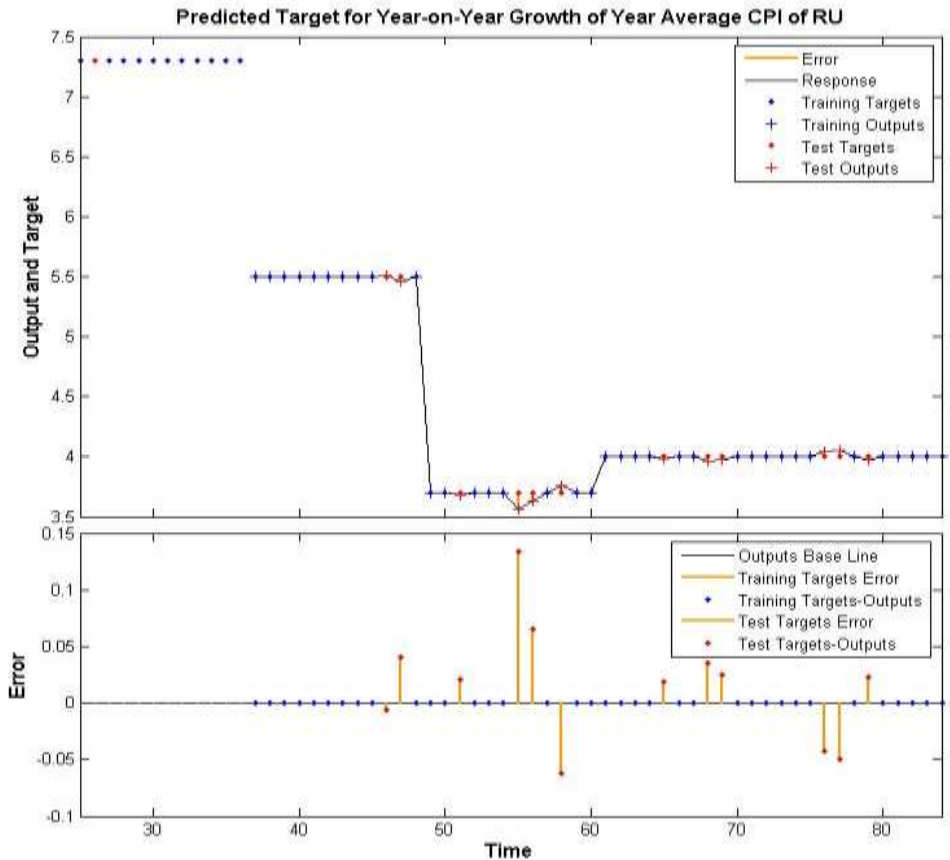

Source: RFSSS; author's calculation

Notes: After training the specific NARX neural network based on limited & most relevant exogenous inputs, the upper subgraph displays this neural network's predicted targets, actual official targets, and errors versus time, while the lower subgraph adopts this neural network's predicted targets as the benchmark and shows the gaps between actual official targets and predicted targets versus time. The time on the horizontal axis corresponds to each month from January 2014 to December 2020, which is also the time span of the exogenous inputs series (from January 2013 to December 2020) minus the initial twelve months as the initial time delays for target prediction, while there are no predicted targets in the initial several periods, due to the training target missing caused by the data missing in the actual official targets series (only from January 2016 to December 2020). Both subgraphs indicate which time points were selected for neural network training and network performance testing.

Figure 64. Specific NARX neural network's prediction of target for year-on-year growth of year average CPI of RU based on limited & most relevant exogenous inputs

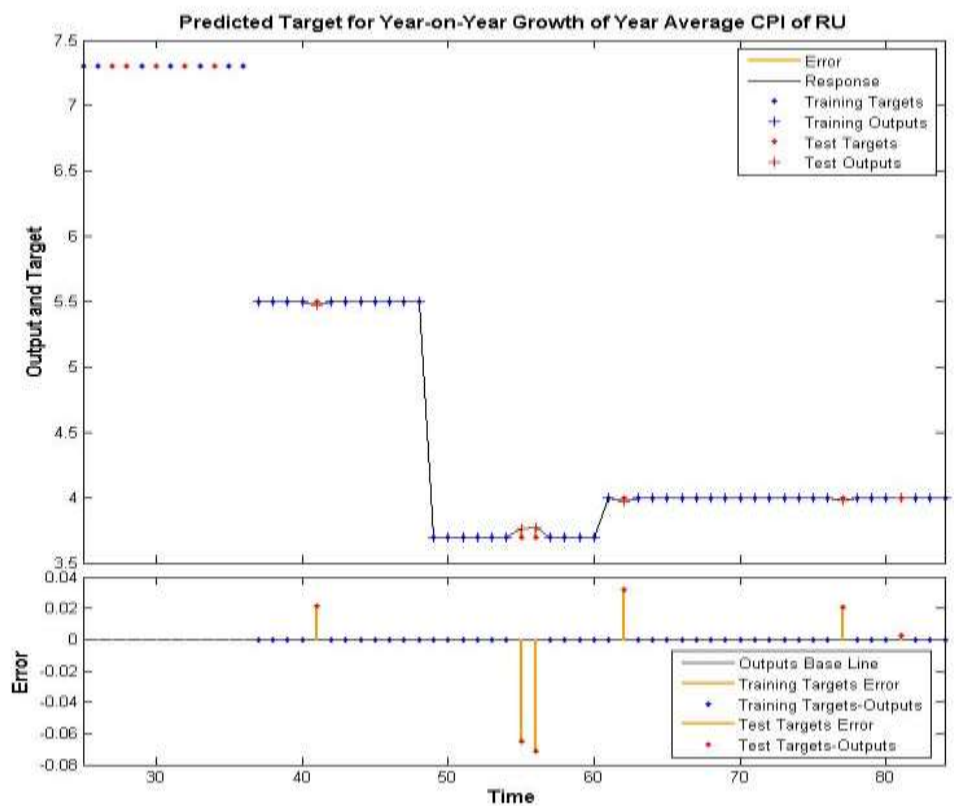

Source: RFSSS; CBRF; RFT; FTSR; MFRF; MLSSRF; RFCS; MEDRF; Moscow Exchange; IMEMO RAS; VEB; HSER; BIS; IMF; WB; CEIC Database; author's calculation

Notes: After training the specific NARX neural network based on abundant & comprehensive exogenous inputs, the upper subgraph displays this neural network's predicted targets, actual official targets, and errors versus time, while the lower subgraph adopts this neural network's predicted targets as the benchmark and shows the gaps between actual official targets and predicted targets versus time. The time on the horizontal axis corresponds to each month from January 2014 to December 2020, which is also the time span of the exogenous inputs series (from January 2013 to December 2020) minus the initial twelve months as the initial time delays for target prediction, while there are no predicted targets in the initial several periods, due to the training target missing caused by the data missing in the actual official targets series (only from January 2016 to December 2020). Both subgraphs indicate which time points were selected for neural network training and network performance testing.

Figure 65. Specific NARX neural network's prediction of target for year-on-year growth of year average CPI of RU based on abundant & comprehensive exogenous inputs

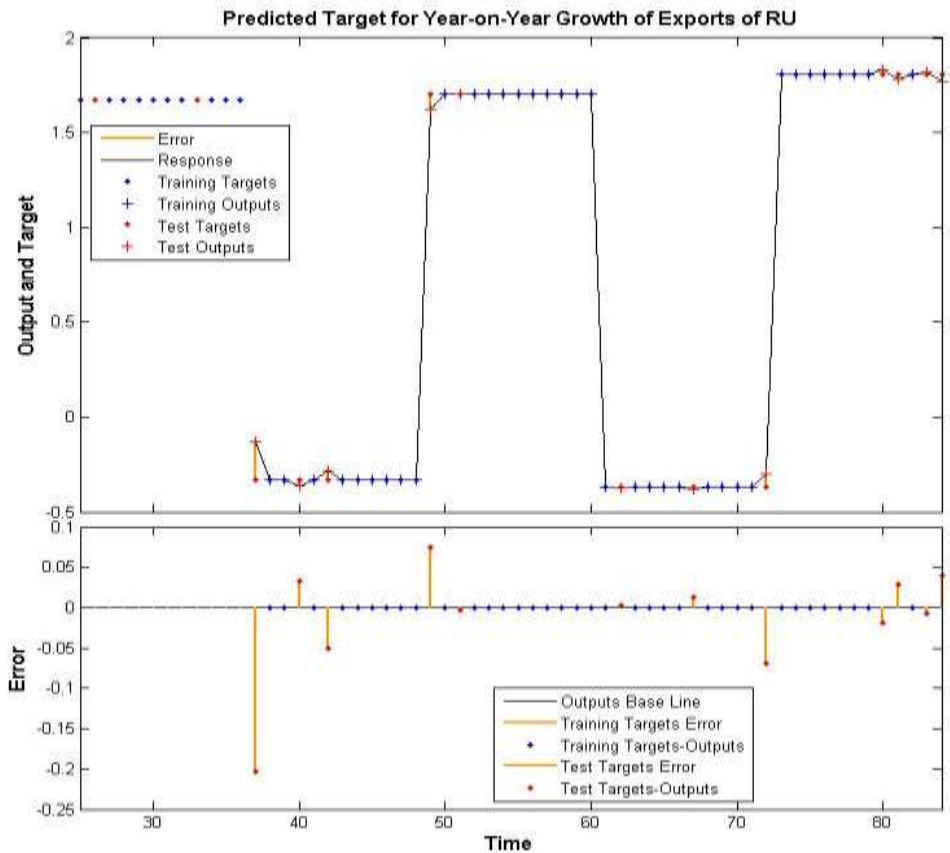

Source: RFCS; MEDRF; author's calculation

Notes: After training the specific NARX neural network based on limited & most relevant exogenous inputs, the upper subgraph displays this neural network's predicted targets, actual official targets, and errors versus time, while the lower subgraph adopts this neural network's predicted targets as the benchmark and shows the gaps between actual official targets and predicted targets versus time. The time on the horizontal axis corresponds to each month from January 2014 to December 2020, which is also the time span of the exogenous inputs series (from January 2013 to December 2020) minus the initial twelve months as the initial time delays for target prediction, while there are no predicted targets in the initial several periods, due to the training target missing caused by the data missing in the actual official targets series (only from January 2016 to December 2020). Both subgraphs indicate which time points were selected for neural network training and network performance testing.

Figure 66. Specific NARX neural network's prediction of target for year-on-year growth of exports of RU based on limited & most relevant exogenous inputs

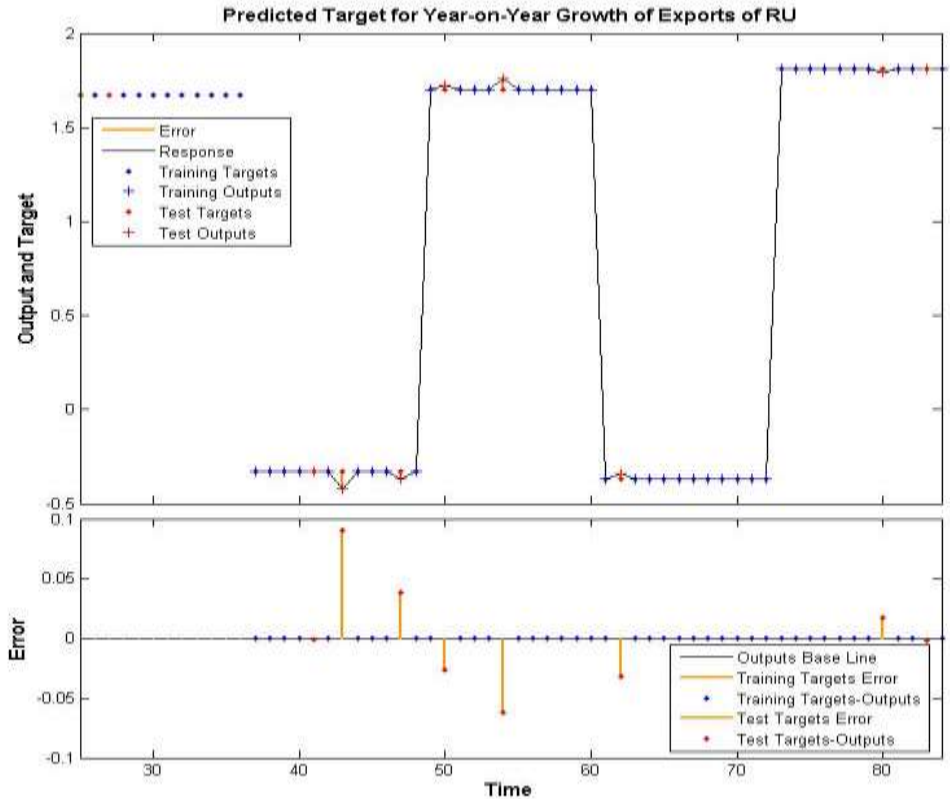

Source: RFSSS; CBRF; RFT; FTSR; MFRF; MLSSRF; RFCS; MEDRF; Moscow Exchange; IMEMO RAS; VEB; HSER; BIS; IMF; WB; CEIC Database; author's calculation

Notes: After training the specific NARX neural network based on abundant & comprehensive exogenous inputs, the upper subgraph displays this neural network's predicted targets, actual official targets, and errors versus time, while the lower subgraph adopts this neural network's predicted targets as the benchmark and shows the gaps between actual official targets and predicted targets versus time. The time on the horizontal axis corresponds to each month from January 2014 to December 2020, which is also the time span of the exogenous inputs series (from January 2013 to December 2020) minus the initial twelve months as the initial time delays for target prediction, while there are no predicted targets in the initial several periods, due to the training target missing caused by the data missing in the actual official targets series (only from January 2016 to December 2020). Both subgraphs indicate which time points were selected for neural network training and network performance testing.

Figure 67. Specific NARX neural network's prediction of target for year-on-year growth of exports of RU based on abundant & comprehensive exogenous inputs

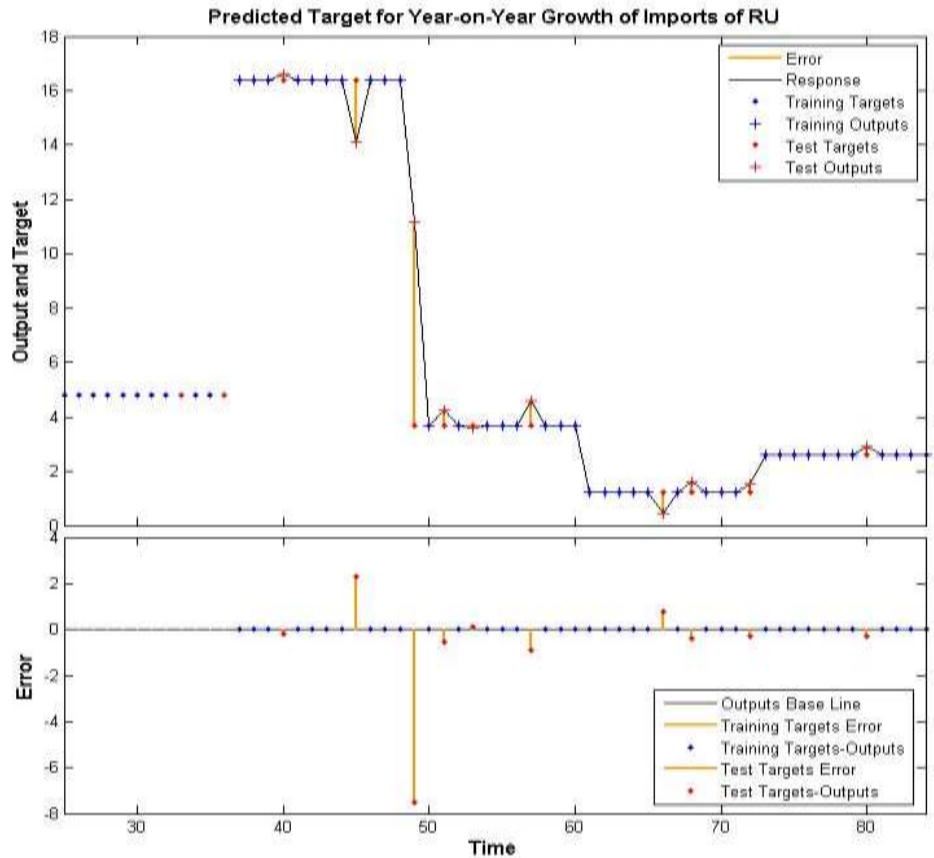

Source: RFCS; MEDRF; author's calculation

Notes: After training the specific NARX neural network based on limited & most relevant exogenous inputs, the upper subgraph displays this neural network's predicted targets, actual official targets, and errors versus time, while the lower subgraph adopts this neural network's predicted targets as the benchmark and shows the gaps between actual official targets and predicted targets versus time. The time on the horizontal axis corresponds to each month from January 2014 to December 2020, which is also the time span of the exogenous inputs series (from January 2013 to December 2020) minus the initial twelve months as the initial time delays for target prediction, while there are no predicted targets in the initial several periods, due to the training target missing caused by the data missing in the actual official targets series (only from January 2016 to December 2020). Both subgraphs indicate which time points were selected for neural network training and network performance testing.

Figure 68. Specific NARX neural network's prediction of target for year-on-year growth of imports of RU based on limited & most relevant exogenous inputs

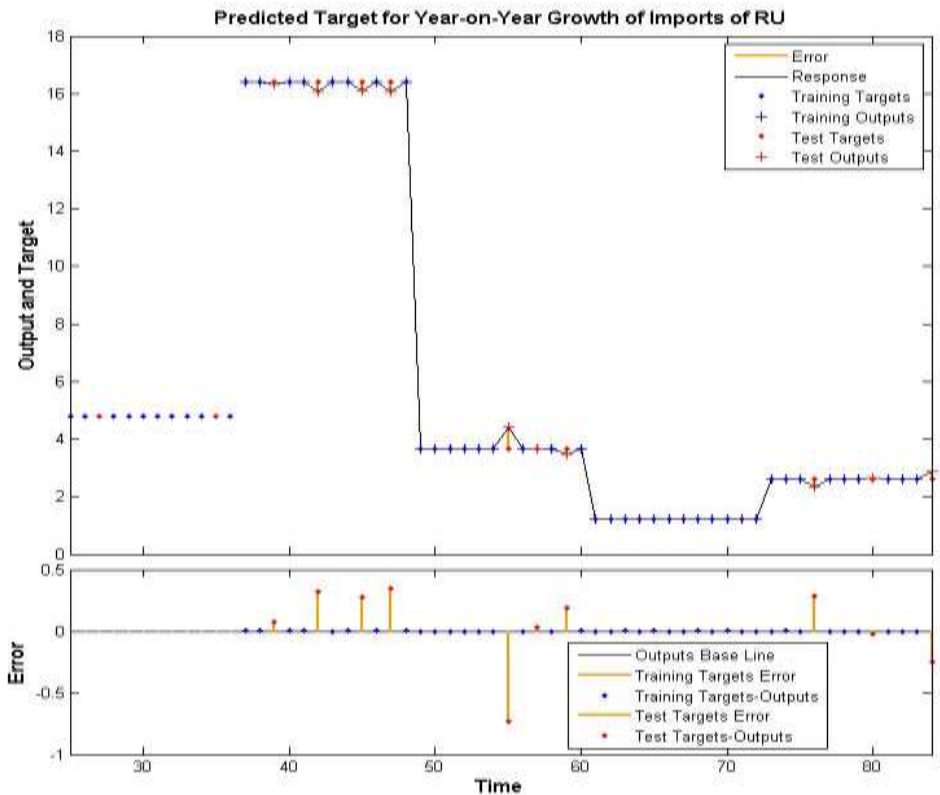

Source: RFSSS; CBRF; RFT; FTSR; MFRF; MLSSRF; RFCS; MEDRF; Moscow Exchange; IMEMO RAS; VEB; HSER; BIS; IMF; WB; CEIC Database; author's calculation

Notes: After training the specific NARX neural network based on abundant & comprehensive exogenous inputs, the upper subgraph displays this neural network's predicted targets, actual official targets, and errors versus time, while the lower subgraph adopts this neural network's predicted targets as the benchmark and shows the gaps between actual official targets and predicted targets versus time. The time on the horizontal axis corresponds to each month from January 2014 to December 2020, which is also the time span of the exogenous inputs series (from January 2013 to December 2020) minus the initial twelve months as the initial time delays for target prediction, while there are no predicted targets in the initial several periods, due to the training target missing caused by the data missing in the actual official targets series (only from January 2016 to December 2020). Both subgraphs indicate which time points were selected for neural network training and network performance testing.

Figure 69. Specific NARX neural network's prediction of target for year-on-year growth of imports of RU based on abundant & comprehensive exogenous inputs

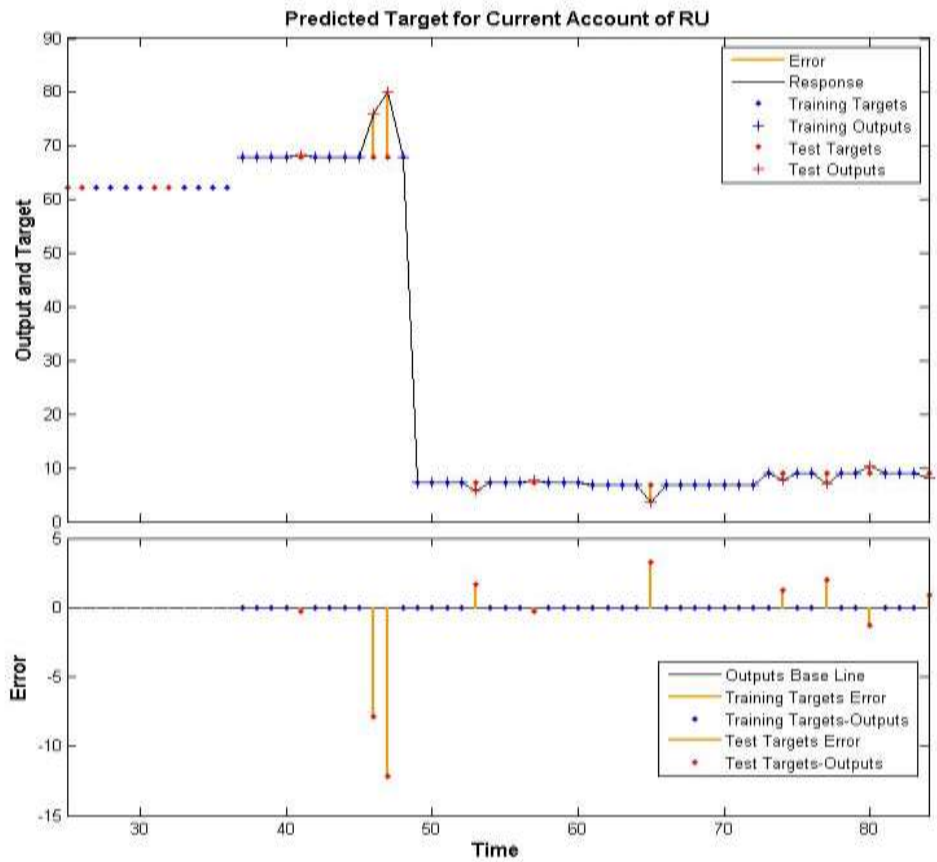

Source: CBRF; author's calculation

Notes: After training the specific NARX neural network based on limited & most relevant exogenous inputs, the upper subgraph displays this neural network's predicted targets, actual official targets, and errors versus time, while the lower subgraph adopts this neural network's predicted targets as the benchmark and shows the gaps between actual official targets and predicted targets versus time. The time on the horizontal axis corresponds to each month from January 2014 to December 2020, which is also the time span of the exogenous inputs series (from January 2013 to December 2020) minus the initial twelve months as the initial time delays for target prediction, while there are no predicted targets in the initial several periods, due to the training target missing caused by the data missing in the actual official targets series (only from January 2016 to December 2020). Both subgraphs indicate which time points were selected for neural network training and network performance testing.

Figure 70. Specific NARX neural network's prediction of target for current account of RU based on limited & most relevant exogenous inputs

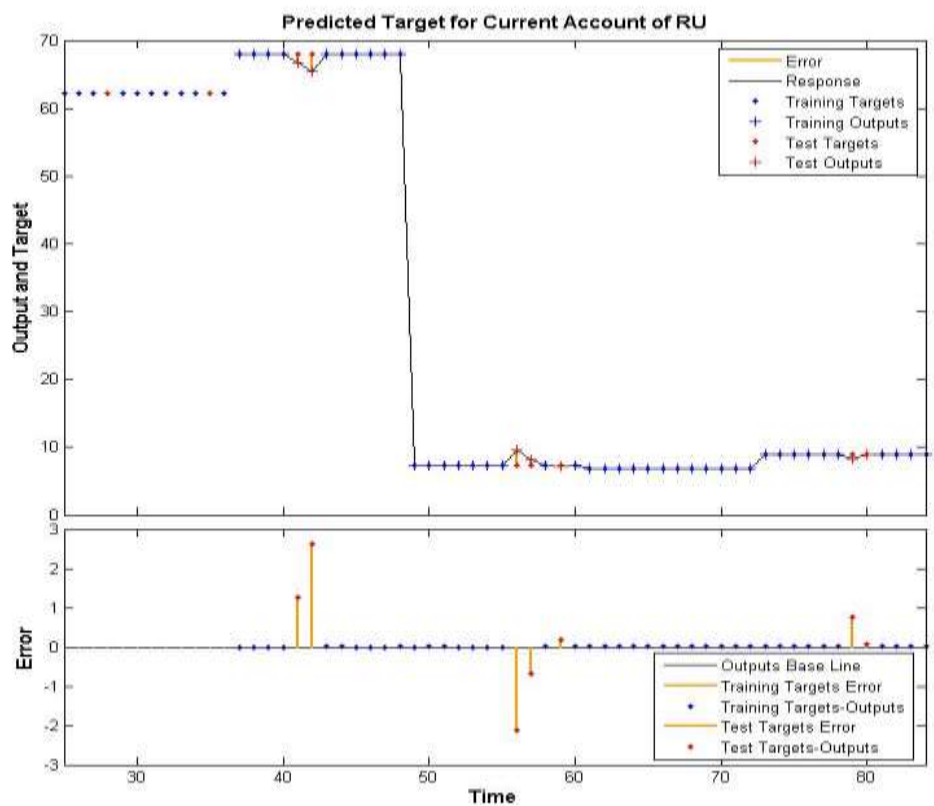

Source: RFSSS; CBRF; RFT; FTSR; MFRF; MLSSRF; RFCS; MEDRF; Moscow Exchange; IMEMO RAS; VEB; HSER; BIS; IMF; WB; CEIC Database; author's calculation

Notes: After training the specific NARX neural network based on abundant & comprehensive exogenous inputs, the upper subgraph displays this neural network's predicted targets, actual official targets, and errors versus time, while the lower subgraph adopts this neural network's predicted targets as the benchmark and shows the gaps between actual official targets and predicted targets versus time. The time on the horizontal axis corresponds to each month from January 2014 to December 2020, which is also the time span of the exogenous inputs series (from January 2013 to December 2020) minus the initial twelve months as the initial time delays for target prediction, while there are no predicted targets in the initial several periods, due to the training target missing caused by the data missing in the actual official targets series (only from January 2016 to December 2020). Both subgraphs indicate which time points were selected for neural network training and network performance testing.

Figure 71. Specific NARX neural network's prediction of target for current account of RU based on abundant & comprehensive exogenous inputs

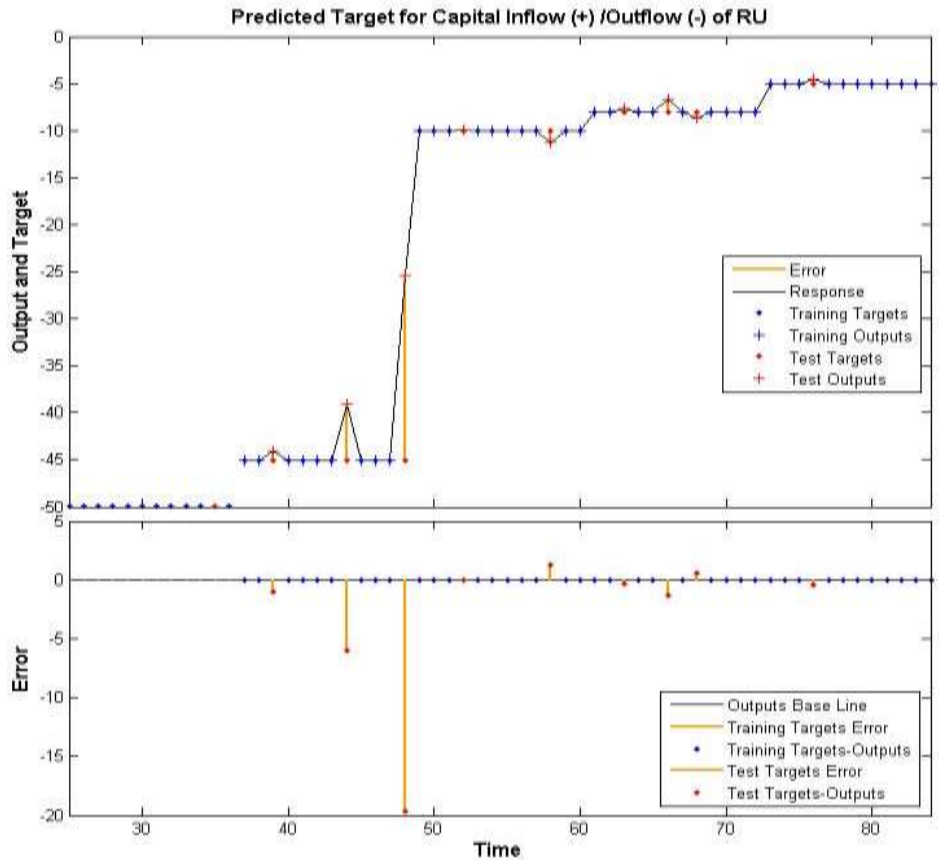

Source: CBRF; author's calculation

Notes: After training the specific NARX neural network based on limited & most relevant exogenous inputs, the upper subgraph displays this neural network's predicted targets, actual official targets, and errors versus time, while the lower subgraph adopts this neural network's predicted targets as the benchmark and shows the gaps between actual official targets and predicted targets versus time. The time on the horizontal axis corresponds to each month from January 2014 to December 2020, which is also the time span of the exogenous inputs series (from January 2013 to December 2020) minus the initial twelve months as the initial time delays for target prediction, while there are no predicted targets in the initial several periods, due to the training target missing caused by the data missing in the actual official targets series (only from January 2016 to December 2020). Both subgraphs indicate which time points were selected for neural network training and network performance testing.

Figure 72. Specific NARX neural network's prediction of target for capital inflow (+) /outflow (-) of RU based on limited & most relevant exogenous inputs

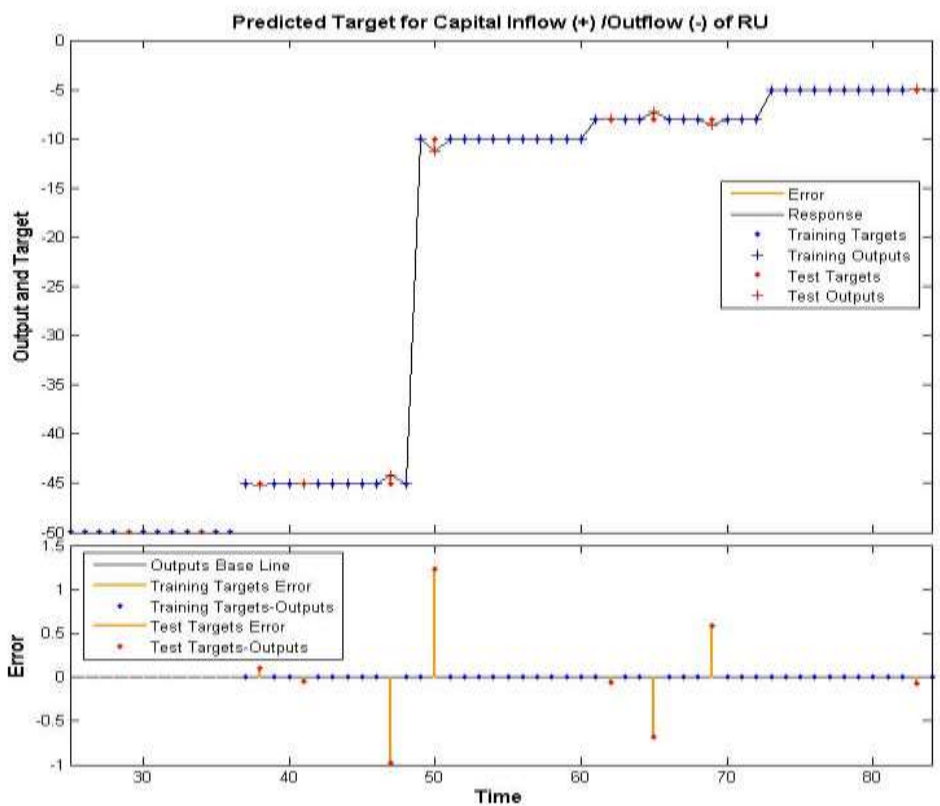

Source: RFSSS; CBRF; RFT; FTSR; MFRF; MLSSRF; RFCS; MEDRF; Moscow Exchange; IMEMO RAS; VEB; HSER; BIS; IMF; WB; CEIC Database; author's calculation

Notes: After training the specific NARX neural network based on abundant & comprehensive exogenous inputs, the upper subgraph displays this neural network's predicted targets, actual official targets, and errors versus time, while the lower subgraph adopts this neural network's predicted targets as the benchmark and shows the gaps between actual official targets and predicted targets versus time. The time on the horizontal axis corresponds to each month from January 2014 to December 2020, which is also the time span of the exogenous inputs series (from January 2013 to December 2020) minus the initial twelve months as the initial time delays for target prediction, while there are no predicted targets in the initial several periods, due to the training target missing caused by the data missing in the actual official targets series (only from January 2016 to December 2020). Both subgraphs indicate which time points were selected for neural network training and network performance testing.

Figure 73. Specific NARX neural network's prediction of target for capital inflow (+) /outflow (-) of RU based on abundant & comprehensive exogenous inputs

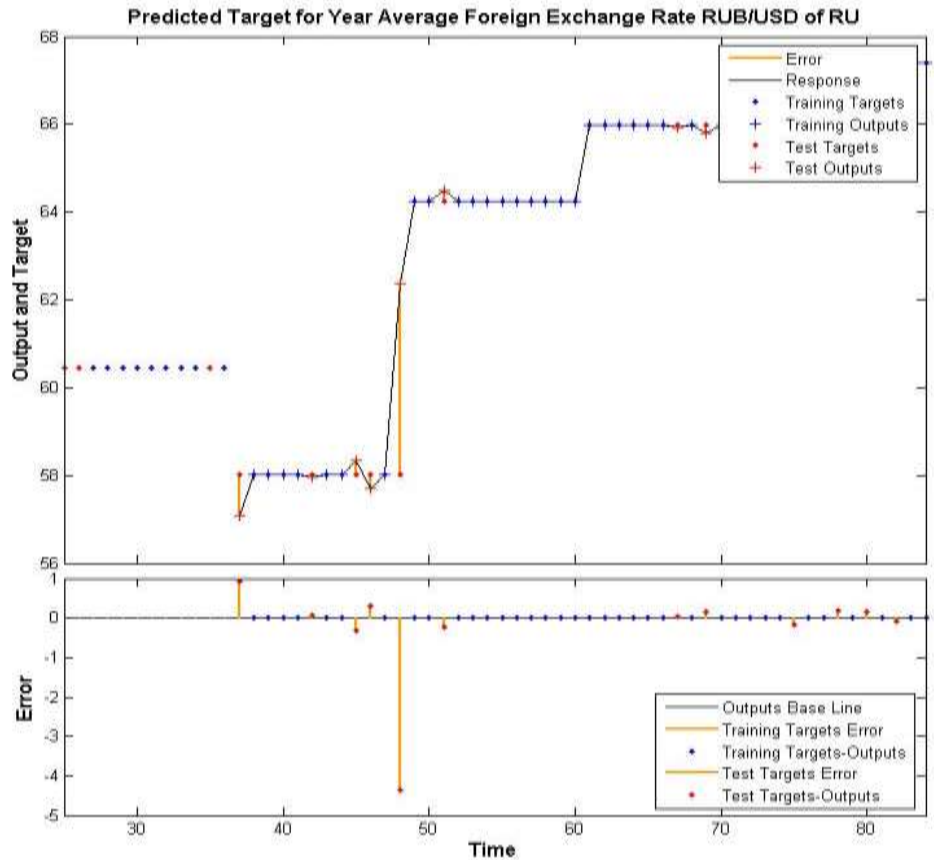

Notes: After training the specific NARX neural network based on limited & most relevant exogenous inputs, the upper subgraph displays this neural network's predicted targets, actual official targets, and errors versus time, while the lower subgraph adopts this neural network's predicted targets as the benchmark and shows the gaps between actual official targets and predicted targets versus time. The time on the horizontal axis corresponds to each month from January 2014 to December 2020, which is also the time span of the exogenous inputs series (from January 2013 to December 2020) minus the initial twelve months as the initial time delays for target prediction, while there are no predicted targets in the initial several periods, due to the training target missing caused by the data missing in the actual official targets series (only from January 2016 to December 2020). Both subgraphs indicate which time points were selected for neural network training and network performance testing.

Figure 74. Specific NARX neural network's prediction of target for year average foreign exchange rate RUB/USD of RU based on limited & most relevant exogenous inputs

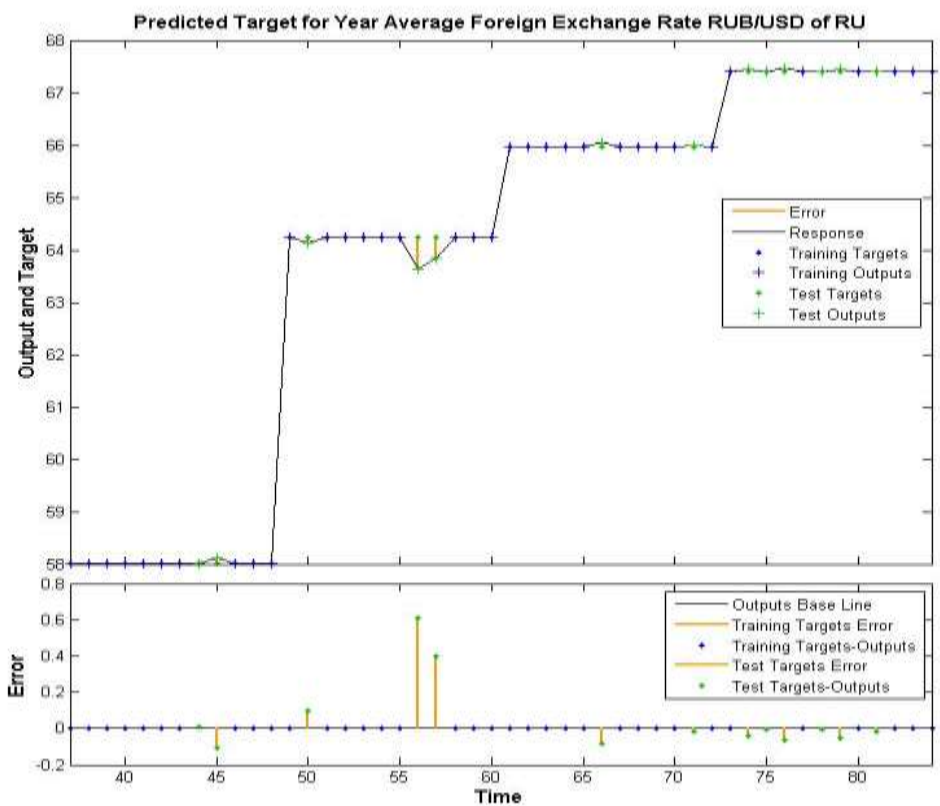

Source: RFSSS; CBRF; RFT; FTSR; MFRF; MLSSRF; RFCS; MEDRF; Moscow Exchange; IMEMO RAS; VEB; HSER; BIS; IMF; WB; CEIC Database; author's calculation

Notes: After training the specific NARX neural network based on abundant & comprehensive exogenous inputs, the upper subgraph displays this neural network's predicted targets, actual official targets, and errors versus time, while the lower subgraph adopts this neural network's predicted targets as the benchmark and shows the gaps between actual official targets and predicted targets versus time. The time on the horizontal axis corresponds to each month from January 2014 to December 2020, which is also the time span of the exogenous inputs series (from January 2013 to December 2020) minus the initial twelve months as the initial time delays for target prediction, while there are no predicted targets in the initial several periods, due to the training target missing caused by the data missing in the actual official targets series (only from January 2016 to December 2020). Both subgraphs indicate which time points were selected for neural network training and network performance testing.

Figure 75. Specific NARX neural network's prediction of target for year average foreign exchange rate RUB/USD of RU based on abundant & comprehensive exogenous inputs

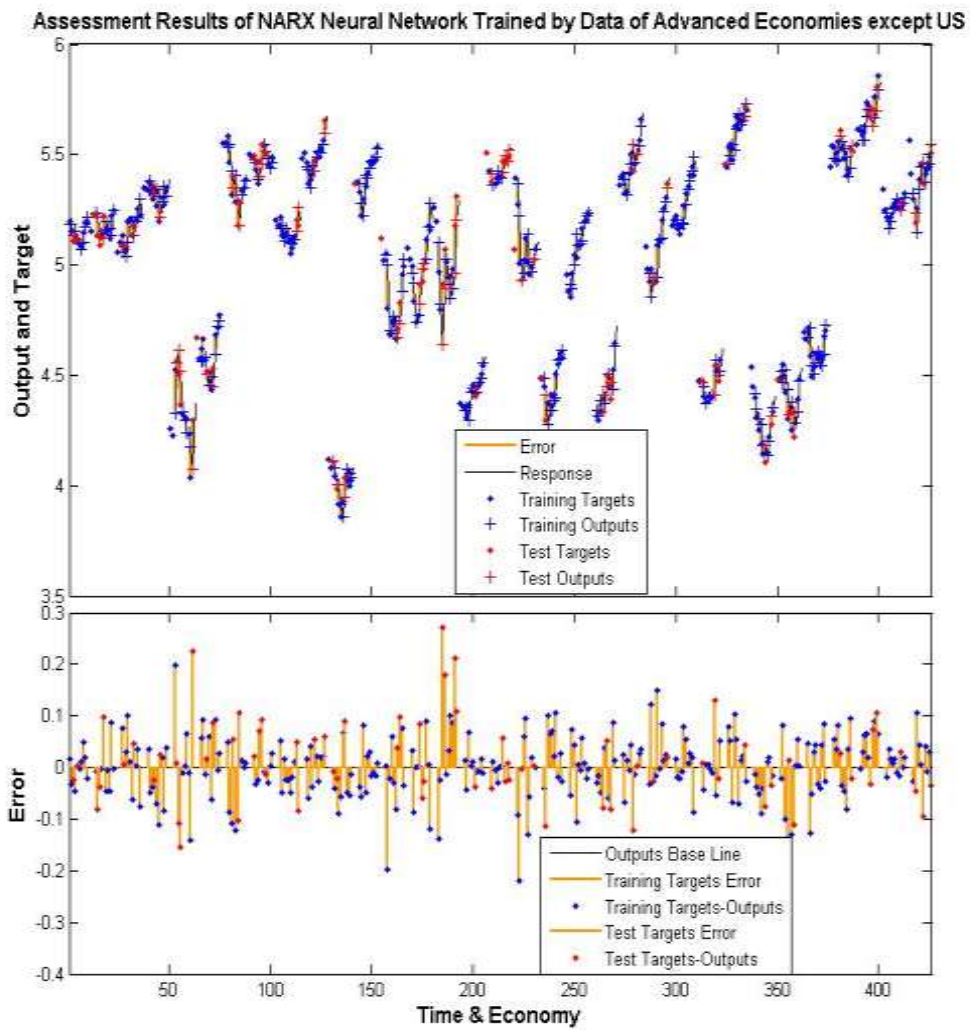

Source: World Economic Forum (WEF); IMF; WB; author's calculation

Notes: After the specific NARX neural network is trained by data of advanced economies except US, the upper subgraph displays this neural network's global competitiveness assessment results for each sample economy, actual values of global competitiveness index for each sample economy, and errors versus time & economy, each continuous curve segment from left to right belongs to Australia, Austria, Belgium, Canada, Cyprus, Czech Republic, Denmark, Finland, France, Germany, Greece, Hong Kong SAR (China), Iceland, Ireland, Israel, Italy, Japan, South Korea, Lithuania, Luxembourg, Malta, Netherlands, New Zealand, Norway, Portugal, Singapore, Slovakia, Slovenia, Spain, Sweden, Switzerland, Taiwan (China), and United Kingdom. The lower subgraph adopts this neural network's global competitiveness assessment results as the benchmark and shows the gaps between actual values of global competitiveness index and global

competitiveness assessment results versus time & economy. The time & economy on the horizontal axis corresponds to each year from 2008 to 2017 (the time span of the exogenous inputs series minus the initial two years as the initial time delays for assessment) for each sample economy. Both subgraphs indicate which time points were selected for neural network training and network performance testing.

Figure 76. Each sample economy's global competitiveness assessment results from specific NARX neural network trained by data of advanced economies except US

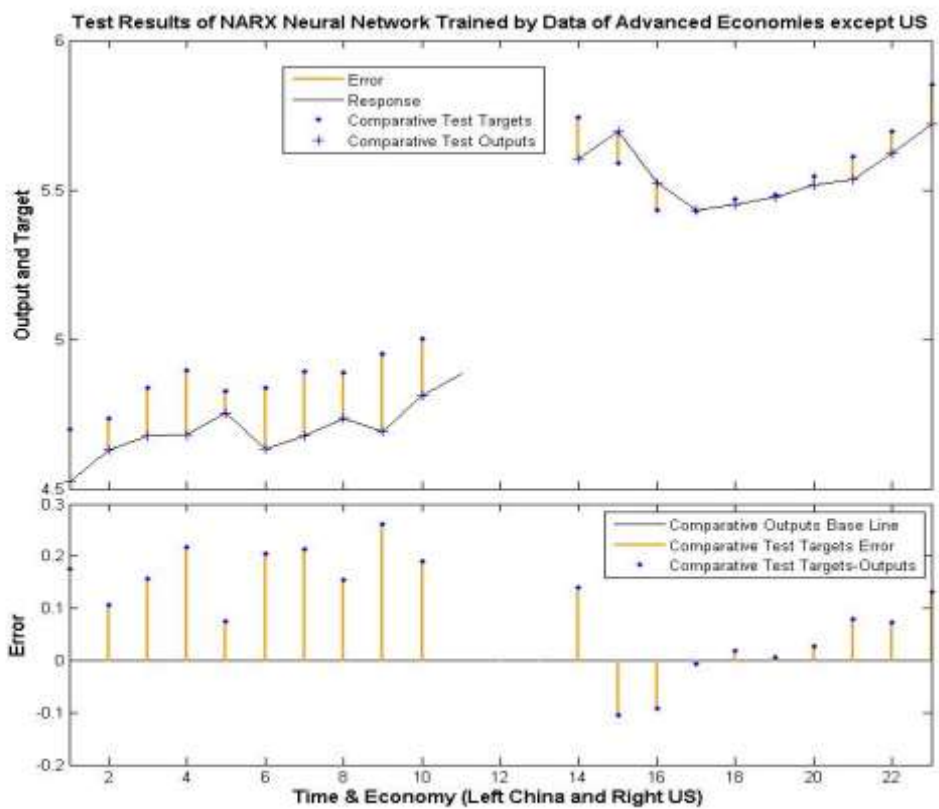

Source: WEF; IMF; WB; author's calculation

Notes: This figure applies the specific NARX neural network trained by data of advanced economies except US to data of China and US, in order to output global competitiveness assessment results for these two test economies. The upper subgraph displays this neural network's global competitiveness assessment results for China and US, actual values of global competitiveness index for China and US, and errors versus time & economy (left China and right US), while the lower subgraph adopts this neural network's global competitiveness assessment results as the benchmark and shows the gaps between actual values of global competitiveness index and global competitiveness assessment results versus time & economy (left China and right US). The time & economy on the horizontal axis corresponds to each year from 2008 to 2017 (the time span of the exogenous inputs series minus the initial two years as the initial time delays for assessment) for each economy (left China and right US).

Figure 77. Test results for China and US from specific NARX neural network trained by data of advanced economies except US

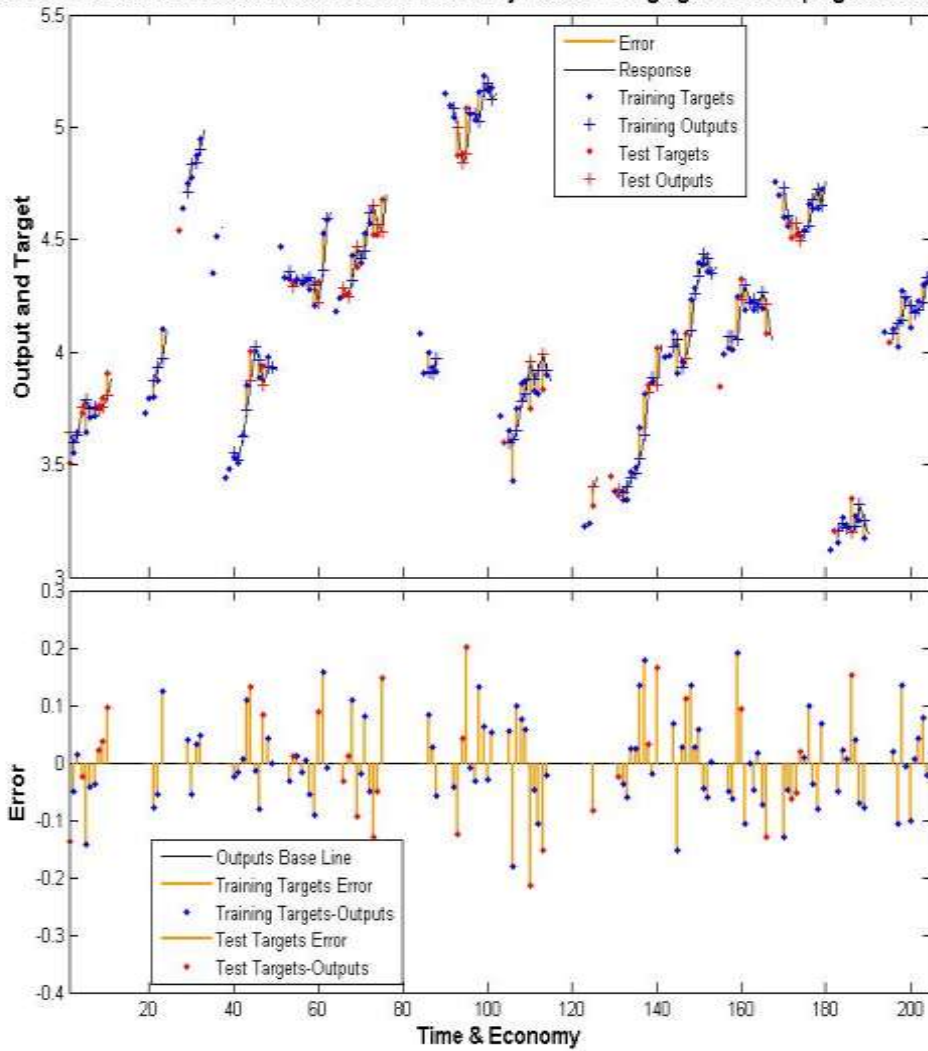

Notes: After the specific NARX neural network is trained by data of emerging and developing Asia except China, the upper subgraph displays this neural network's global competitiveness assessment results for each sample economy, actual values of global competitiveness index for each sample economy, and errors versus time & economy, each continuous curve segment from left to right belongs to Bangladesh, Bhutan, Brunei, Cambodia, India, Indonesia, Laos, Malaysia, Mongolia, Myanmar, Nepal, Philippines, Sri Lanka, Thailand, Timor-Leste, and Vietnam. The lower subgraph adopts this neural network's global competitiveness assessment results as the benchmark and shows the gaps between actual values of global competitiveness index and global competitiveness assessment results versus time & economy. The time & economy on the

horizontal axis corresponds to each year from 2008 to 2017 (the time span of the exogenous inputs series minus the initial two years as the initial time delays for assessment) for each sample economy. Both subgraphs indicate which time points were selected for neural network training and network performance testing.

Figure 78. Each sample economy's global competitiveness assessment results from specific NARX neural network trained by data of emerging and developing Asia except China

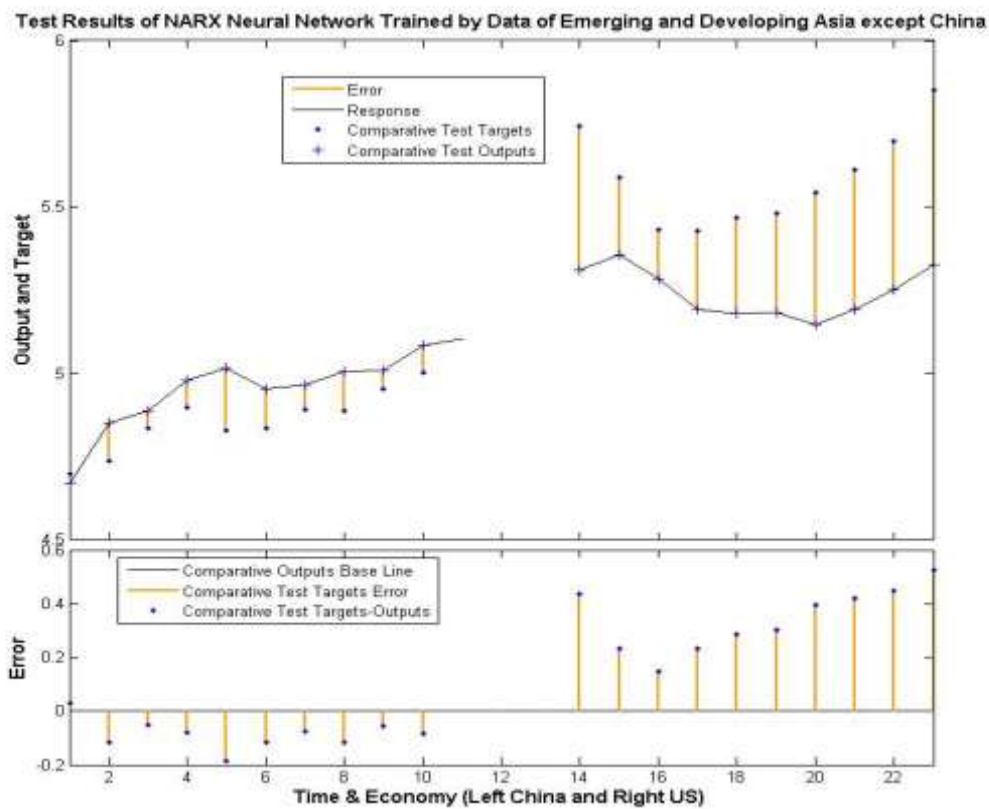

Source: WEF; IMF; WB; author's calculation

Notes: This figure applies the specific NARX neural network trained by data of emerging and developing Asia except China to data of China and US, in order to output global competitiveness assessment results for these two test economies. The upper subgraph displays this neural network's global competitiveness assessment results for China and US, actual values of global competitiveness index for China and US, and errors versus time & economy (left China and right US), while the lower subgraph adopts this neural network's global competitiveness assessment results as the benchmark and shows the gaps between actual values of global competitiveness index and global competitiveness assessment results versus time & economy (left China and right US). The time & economy on the horizontal axis corresponds to each year from 2008 to 2017 (the time span of the exogenous inputs series minus the initial two years as the initial time delays for assessment) for each economy (left China and right US).

Figure 79. Test results for China and US from specific NARX neural network trained by data of emerging and developing Asia except China

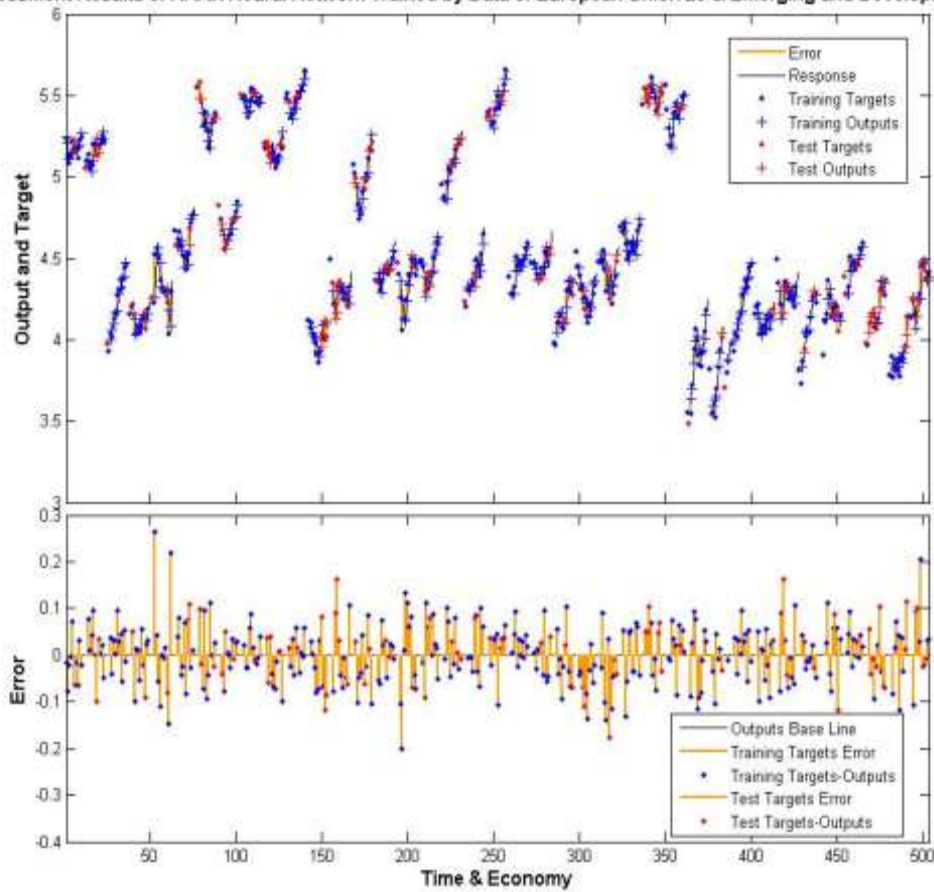

Notes: After the specific NARX neural network is trained by data of European Union 28 & emerging and developing Europe, the upper subgraph displays this neural network's global competitiveness assessment results for each sample economy, actual values of global competitiveness index for each sample economy, and errors versus time & economy, each continuous curve segment from left to right belongs to European Union 28 economies involving Austria, Belgium, Bulgaria, Croatia, Cyprus, Czech Republic, Denmark, Estonia, Finland, France, Germany, Greece, Hungary, Ireland, Italy, Latvia, Lithuania, Luxembourg, Malta, Netherlands, Poland, Portugal, Romania, Slovakia, Slovenia, Spain, Sweden, and United Kingdom (before 2019), plus emerging and developing European economies including Albania, Bosnia and Herzegovina, Bulgaria, Croatia, Hungary, North Macedonia, Montenegro, Poland, Romania, Serbia, and Turkey. The lower subgraph adopts this neural network's global competitiveness assessment results as the benchmark and shows the gaps between actual values of global

competitiveness index and global competitiveness assessment results versus time & economy. The time & economy on the horizontal axis corresponds to each year from 2008 to 2017 (the time span of the exogenous inputs series minus the initial two years as the initial time delays for assessment) for each sample economy. Both subgraphs indicate which time points were selected for neural network training and network performance testing.

Figure 80. Each sample economy's global competitiveness assessment results from specific NARX neural network trained by data of European Union 28 & emerging and developing Europe

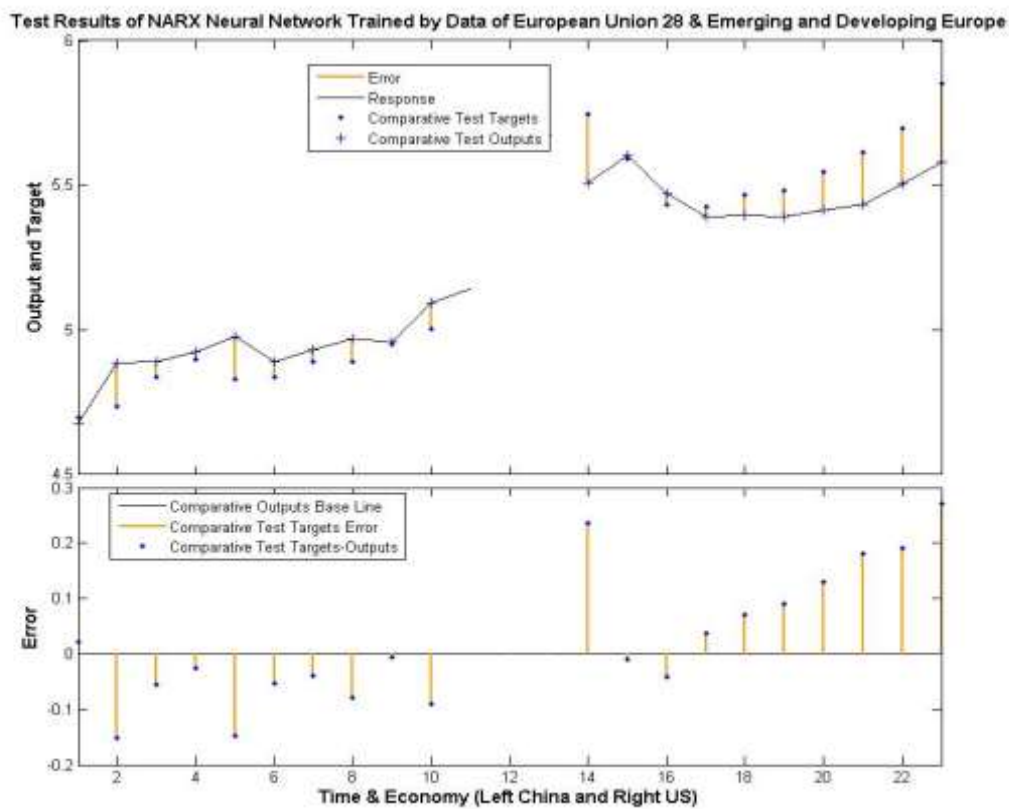

Source: WEF; IMF; WB; author's calculation

Notes: This figure applies the specific NARX neural network trained by data of European Union 28 & emerging and developing Europe to data of China and US, in order to output global competitiveness assessment results for these two test economies. The upper subgraph displays this neural network's global competitiveness assessment results for China and US, actual values of global competitiveness index for China and US, and errors versus time & economy (left China and right US), while the lower subgraph adopts this neural network's global competitiveness assessment results as the benchmark and shows the gaps between actual values of global competitiveness index and global competitiveness assessment results versus time & economy (left China and right US). The time & economy on the horizontal axis corresponds to each year from 2008 to 2017 (the time span of the exogenous inputs series minus the initial two years as the initial time delays for assessment) for each economy (left China and right US).

Figure 81. Test results for China and US from specific NARX neural network trained by data of European Union 28 & emerging and developing Europe